\documentclass[12pt,a4paper]{article}
\usepackage{amsmath}
\usepackage{bbold}
\usepackage{wasysym}
\usepackage{hyperref}
\usepackage{graphics}
\usepackage{url}
\usepackage{axodraw}
\usepackage{graphicx}

\newcommand{\nc}{\newcommand}
\nc{\beq}{\begin{equation}}
\nc{\eeq}{\end{equation}}
\nc{\bea}{\begin{eqnarray}}
\nc{\eea}{\end{eqnarray}}
\nc{\nn}{\nonumber}
\nc{\bi}{\begin{itemize}} 
\nc{\ei}{\end{itemize}}
\nc{\ib}{\item [$\bullet$]}
\nc{\cd}{\cdot}
\nc{\cds}{\cdots}
\nc{\pr}{\prime}
\nc{\tz}{\tilde{z}}
\nc{\msbar}{\overline {\textrm{MS}}}
\def\eg{{\sl e.g.}~}
\nc{\veps}{\varepsilon}
\nc{\as}{\alpha_s}
\nc{\med}{\medskip}

\def\eg{{\sl e.g.}~}
\def\ie{{\sl i.e.}~}
\def\cf{{\sl c.f.}~}
\def\fig{Fig.~}
\def\Fig{Fig.~}
\def\Figs{Figs.~}

\def\eq{Eq.~}
\def\eqs{Eqs.~}

\newcommand{\LA}{\mathrm{A}}
\newcommand{\LB}{\mathrm{B}}

\newcommand{\La}{\mathrm{a}}
\newcommand{\Lb}{\mathrm{b}}

\newbox\charbox
\newbox\slabox
\def\s#1{{      
        \setbox\charbox=\hbox{$#1$}
        \setbox\slabox=\hbox{$/$}
        \dimen\charbox=\ht\slabox
        \advance\dimen\charbox by -\dp\slabox
        \advance\dimen\charbox by -\ht\charbox
        \advance\dimen\charbox by \dp\charbox
        \divide\dimen\charbox by 2
        \raise-\dimen\charbox\hbox to \wd\charbox{\hss/\hss}
        \llap{$#1$}
}}
\newlength{\nseparation}
\newenvironment{nfigure}[1]
        {\begin{figure}[#1]\hrule\vspace{\nseparation}\par}
        {\vspace{\nseparation}\par \hrule \end{figure}}

\nc{\lag}{\cal L }
\nc{\matx}{\left|\cal {M}\right|^2}
\nc{\lqcd}{\Lambda_\textrm{QCD}}
\nc{\really}{\stackrel{!}{=}}

\newcommand{\eqn}{equation}

\newcommand{\wt}{\widetilde}

\newcommand{\lb}{\left(}
\newcommand{\rb}{\right)}

\newcommand{\al}{\alpha}
\newcommand{\M}{\mathcal{M}}
\newcommand{\mO}{\mathcal{O}}

\newcommand{\vareps}{\varepsilon}
\newcommand{\D}{\mathcal{D}}

\newcommand{\ph}{\hat{p}}

\newcommand{\prefaceps}{\lb\frac{4\,\pi\,\mu^{2}}{2\,p_{a}\cdot p_{b}}\rb^{\vareps}}

\begin{document}

\thispagestyle{empty}
\def\thefootnote{\fnsymbol{footnote}}
\setcounter{footnote}{1}
\null
\hfill TTK-10-59
\vskip 0cm
\begin{center}

{\Large \boldmath{\bf
    An alternative subtraction scheme for next-to-leading order QCD calculations}
    \par} \vskip 2.5em {\large
{\sc Cheng-Han Chung, Michael Kr\"amer}\\[1ex]
{\normalsize \it Institute for Theoretical Particle Physics and Cosmology,\\
RWTH Aachen University, D-52056 Aachen, Germany}\\[2ex]
{\sc Tania Robens}\\[1ex]
{\normalsize \it SUPA, School of Physics and Astronomy,\\ 
University of Glasgow, Glasgow, G12 8QQ,  Scotland, UK}}
\par \vskip 2em
\end{center}\par

\noindent{\bf Abstract:}\\[0.25em]
\noindent We propose a new subtraction scheme for next-to-leading
order QCD calculations. Our scheme is based on the momentum mapping
and on the splitting functions derived in the context of an improved
parton shower formulation. Compared to standard schemes, the new
scheme features a significantly smaller number of subtraction terms
and facilitates the matching of NLO calculations with parton showers
including quantum interference. We provide formulae for the momentum
mapping and the subtraction terms, and present a detailed comparison
with the Catani-Seymour dipole subtraction for a variety of $2\to 2$
scattering processes.
\par
\null
\setcounter{page}{0}
\clearpage
\def\thefootnote{\arabic{footnote}}
\setcounter{footnote}{0}

\section{Introduction}

The Standard Model (SM) of particle physics has been tested and
confirmed by a large variety of experimental data
\cite{Nakamura:2010zzi}.  With the start of data taking at the Large
Hadron Collider (LHC) in 2009~\cite{LHCstart}, the exploration of the
properties and predictions of the SM is being extended to
center-of-mass energies of several TeV. To further consolidate the
Standard Model is an important effort per se; however, a detailed
knowledge of SM-induced backgrounds is equally important in searches
for physics beyond the Standard Model. For both tasks, leading-order
(LO) QCD calculations of SM processes are not sufficient: the
theoretical uncertainty at LO is substantial, and next-to-leading
order (NLO) corrections to LHC processes are in general large and have
to be included to match experimental accuracies (for a recent review
on higher-order corrections for hadron collider processes see
\cite{Binoth:2010ra}).  More specifically, the analysis and
interpretation of experimental signatures at the LHC require
theoretical predictions for differential distributions or cross
sections with cuts on kinematic variables. NLO calculations should
thus be set up in form of fully differential parton-level Monte Carlo
programs.  In recent years, a lot of effort has gone into semi- or
fully-automated inclusion of higher-order corrections in such tools
for processes at hadron colliders.  Most of these, as \eg MFCM
\cite{mcfm}, NLOJet++ \cite{nlojetpp}, VBFNLO \cite{Arnold:2008rz},
Rocket \cite{Ellis:2005zh}, Blackhat/Sherpa \cite{Berger:2009zg,
  Berger:2009ep}, GOLEM/Whizard \cite{Binoth:2009rv}, GOLEM/Madgraph
\cite{Binoth:2009rv}, and Helac \cite{Kanaki:2000ey,
  Bevilacqua:2009zn, Bevilacqua:2010ve}, are now able to provide NLO
QCD predictions for several processes at parton level.  Finally, a
leading logarithmic description of multiple soft and collinear parton
emissions in the form of parton showers should be combined with the
NLO partonic predictions; this has been realised in a variety of
different approaches~\cite{Collins:2001fm, Frixione:2002ik,
  Kramer:2003jk, Soper:2003ya, Nason:2004rx, Nagy:2005aa,
  Bauer:2006mk, Giele:2007di, Lavesson:2008ah}, and mature program
packages like MC@NLO \cite{Frixione:2002ik, Torrielli:2010aw,
  Frixione:2010ra, Frixione:2010wd} and POWHEG \cite{Frixione:2007vw,
  Alioli:2010xd,Hoche:2010pf} have been developed for LHC physics.

An important calculational tool for the implementation of NLO QCD
corrections in Monte Carlo style programs are subtraction
schemes~\cite{Ellis:1980wv, Frixione:1995ms,
  Catani:1996vz,Dittmaier:1999mb, Catani:2002hc, Weinzierl:2003ra},
which facilitate the treatment of infrared and collinear divergences
originating from different phase-space contributions.  Subtraction
schemes introduce local counterterms which mimic the behaviour of the
real-emission matrix element in the singular limits. The difference
between the local counterterms and the real-emission matrix element is
thus finite and can be integrated numerically. The infrared and
collinear singularities are isolated by integrating the subtraction
terms over the singular regions within a given regularisation scheme,
\eg dimensional regularisation. The soft and collinear singularities
then cancel when the integrated subtraction terms are added to the
virtual cross section. After standard UV-renormalisation, all
contributions to the NLO cross section are finite, and the further
phase-space integrations can be performed numerically by means of
Monte Carlo techniques.

Various general NLO subtraction schemes have been proposed (see
\cite{Frixione:1995ms, Catani:1996vz,Dittmaier:1999mb,
  Weinzierl:2003ra}), and several (semi)automated implementations are
available \cite{Gleisberg:2007md, Czakon:2009ss,Hasegawa:2009tx,
  Frederix:2009yq, Frederix:2010cj, Hoche:2010pf}. The schemes differ
in the phase-space mapping which relates real-emission and
leading-order kinematics and which is needed to define the subtraction
terms, and in the finite parts of the subtraction.  Unfortunately, the
schemes developed so far suffer from a rapidly rising number of
momentum mappings needed to evaluate the subtraction terms, which
basically scales like $N^{3}$ for a leading order $2\rightarrow N$
process.\footnote{We note that recently a constant scaling behaviour
  for $N$-gluon final states has been achieved within the Madgraph
  framework \cite{Frederix:2009yq}; however, this scaling behaviour
  relies on the symmetrisation of the matrix element and is therefore
  applicable independent of the subtraction scheme and its mapping
  prescription.  Our argument concerns the general combinatoric number
  of mappings and the resulting scaling behaviour if no further
  process-specific simplifications are applied.}  With an increasing
number of final-state particles this scaling leads to a rapidly rising
number of momentum mappings and subsequent re-evaluations of matrix
elements. In this paper, we therefore propose a subtraction scheme
where the number of mappings scales like $N^{2}$, thereby reducing the
number of matrix element evaluations by a factor $N$. This scaling is
achieved by the use of a mapping prescription which, for emissions
from final-state particles, redistributes the momenta to all
non-emitting final state particles simultaneously, thereby leading to
one unique momentum mapping per emitter. Furthermore, the subtraction
terms in this scheme are derived from splitting functions which have
been proposed in the context of a parton shower including interference
effects \cite{Nagy:2007ty,Nagy:2008ns,Nagy:2008eq}.  When NLO
parton-level calculations and parton showers are combined, specific
counterterms have to be added in order to avoid double counting of
contributions which are included in both the NLO calculation and the
parton shower \cite{Frixione:2002ik, Kramer:2005hw}. The use of the
shower splitting functions as subtraction terms reduces the number of
these coun-terterms \cite{Nagy:2005aa} and therefore facilitates the
combination of shower and NLO calculations.  First results obtained
with the new subtraction scheme have been published in
\cite{Robens:2010zr}.

This paper is organised as follows. After a brief revision of the
generic features of NLO subtraction schemes in Section 2, we shall
present the setup of our new scheme in Section 3. We discuss the
momentum mapping, the factorisation of the matrix element in the soft
and collinear limits, and present the subtraction terms and their
integrated counterparts for scattering processes with at most two
particles in the final state. In Section 4, we apply our scheme to
well-known collider processes at NLO: single-W production, dijet
production at lepton colliders, gluon-induced Higgs production and
Higgs decay into gluons, and deep inelastic scattering. We show that,
for these processes, the results obtained using our subtraction
prescription agree with the results obtained within the commonly used
Catani-Seymour dipole subtraction scheme~\cite{Catani:1996vz}.  We
conclude in Section 5. Some additional useful formulae are listed in
the Appendix.

\section{General structure of NLO cross sections and subtraction
  schemes}\label{sec:gen_struct}
In this section, we shall briefly review the general structure of
subtraction schemes for NLO cross-section calculations at colliders
and set up our notation, following closely the notation established in
the context of the Catani-Seymour dipole subtraction
scheme~\cite{Catani:1996vz}.

We consider a generic jet cross-section $\sigma$ with
\begin{eqnarray}
\sigma &= & \sigma^{\text{LO}}+\sigma^{\text{NLO}}\nonumber \\
& = & \int_m d\sigma^B + \int_m d\sigma^V + \int_{m+1}d\sigma^R\,,
\end{eqnarray}
where $\sigma^{\text{B}}$, $\sigma^{\text{V}}$, and
$\sigma^{\text{R}}$ denote the LO, virtual and real-emission
contributions, respectively. There are $m$ partons in the final state
for the LO and virtual cross sections, and $m+1$ partons for the
real-emission contribution. After UV-renormalisation, the virtual and
real-emission cross sections each contain infrared and collinear
singularities, which we regularise using dimensional regularisation,
\ie we work in $d\,=\,4 - 2\vareps$ dimensions so that the
singularities appear as $1/\vareps^2$ (soft and collinear) and
$1/\vareps$ (soft or collinear) poles.  These poles cancel in the sum
of virtual and real contributions, but the individual pieces are
divergent and can thus not be integrated numerically in four
dimensions.

In subtraction schemes one constructs local counterterms which match
the behaviour of the real-emission matrix element in each soft and
collinear region. Subtracting these counterterms from the
real-emission matrix elements and adding back the corresponding
one-particle integrated counterparts to the virtual contribution
results in finite integrands for both the virtual correction
($m$-particle phase space) and the real contribution ($m+1$-particle
phase space): 
\bea
\label{countertermfinite85}
\sigma^{\text{NLO}}&=&
\underset{\textrm {finite} }
{\underbrace{\int_{m}\,d\sigma^V+\int_{m+1}\,d\sigma^A}}   +
\underset{ \textrm {finite} }
{\underbrace{\int_{m+1}\left[ d\sigma^R-d\sigma^A\right]}}               \nn\\
&=&
\int_{m}\left[d\sigma^V+\int_1\,d\sigma^A\right]_{\vareps=0}
+
\int_{m+1}\left[
  d\sigma^R_{\vareps=0}-d\sigma^A_{\vareps=0}\right]\,.  
\eea
The construction of the local counterterms, collectively denoted by
$d\sigma^A$ in \eq (\ref{countertermfinite85}), relies on the
factorisation of the real-emission matrix element in the singular (\ie
soft and collinear) limits (\fig\ref{Dipolefactorizationprocedure_CS})
\cite{Altarelli:1977zs, Bassetto:1984ik,Dokshitzer:1991wu}:
\begin{nfigure}{tbp} 
\vspace*{0.5cm}
\SetScale{1}
\centerline{\unitlength 1pt
\vspace*{0.5cm}
   \begin{picture}(530,100)(0,0)
  \ArrowLine(100,40)(140,70)
  \ArrowLine(100,40)(140, 10)
  \ArrowLine(100,40)(150,60)
   \ArrowLine(100,40)(150,40)
  \DashLine(100,40)(60, 40){6}
  \GCirc(100,40){13}{0.5}
    \Text ( 90,70)[l]{${\cal M}_{m+1}$}
     \Text ( 145,75)[l]{1}
     \Text ( 145,5)[l]{$j$}
      \Text ( 155,63)[l]{$\ell$}
     \Text ( 155,40)[l]{$j$}
   \Text (185 ,21.5)[l]{\small $\ell$ and $j$ collinear }
   \Text (185 ,8)[l]{\small and/or j soft}
      \Text ( 185,40)[l]{$\longrightarrow$}
      \Text ( 225,40)[l]{\large${\sum}_{\ell}$}
  \put(125, 25){$\vdots$\normalsize }
  \DashLine(310,40)(270,40){6}
  \ArrowLine(310,40)( 350, 70)
   \ArrowLine(310,40)( 350, 10)
   \ArrowLine(310,40)( 350, 55)
  \GCirc( 310,40){13}{0.5}
   \Text ( 300,70)[l]{${\cal M}_{m}$}
   \put(345, 25){$\vdots$\normalsize }
     \Line(405,55)( 380, 55)
     \ArrowLine(405,55)( 435, 70)
     \ArrowLine(405,55)( 435, 40)
     \GCirc(405,55){10}{0.5}
     \Text ( 355,75)[l]{1}
     \Text ( 355,5)[l]{$m$}
     \Text ( 355,55)[l]{$\ell$}
      \Text ( 370,55)[l]{$\otimes$}
       \Text ( 440,75)[l]{$\ell$}
     \Text ( 440,35)[l]{$j$}
     \Text (400,76)[l]{${v}_{\ell}$}
  \end{picture}}
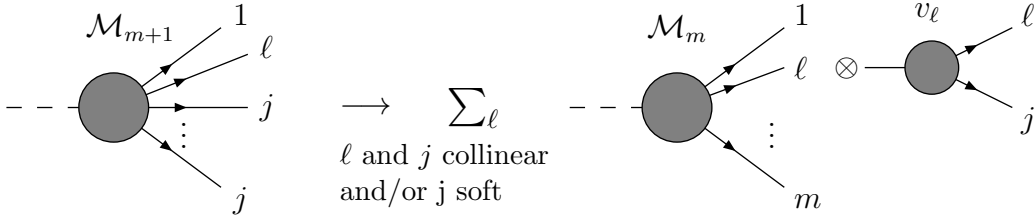
\caption{Soft/collinear factorisation: when the partons $\ell$ and $j$
  become collinear and/or parton $j$ becomes soft, the ($j$)-parton
  matrix element factorises into a sum over $m$-parton matrix elements
  times a singular factor ${v}_{\ell}$.}
\label{Dipolefactorizationprocedure_CS}
\vspace*{0.5cm}
\end{nfigure}
\beq
\label{Factorizationm1toDim}
{\cal M}_{m+1}(\{\hat p\}_{m+1}) \longrightarrow \sum_{\ell}
v_{\ell}(\{\hat p\}_{m+1})\otimes {\cal M}_m(\{p\}_m)\,, 
\eeq 
where ${\cal M}_{m+1}$ and ${\cal M}_m$ denote $(m+1)$- and $m$-parton
matrix elements, respectively, the $v_{\ell}$ are generalised
splitting functions containing the singularity structure of the $m+1$
matrix element, and the symbol $\otimes$ represents properly defined
phase-space, spin and colour convolutions.  (See
Section~\ref{sec:splttings} for a more detailed discussion of the
$(m+1))$-parton matrix element factorisation.) The momenta in the
$(m+1)$- and $m$-parton phase spaces are denoted by $\{\hat p\}_{m+1}$
and $\{p\}_m$, respectively. As ${\cal M}_{m+1}$ and ${\cal M}_{m}$
are defined in terms of $(m+1)$- and $m$-parton phase spaces, a
mapping $\{\hat p\}_{m+1} \to \{p\}_m$ needs to be introduced.  The
mapping must be such that four-momentum conservation as well as the
on-shell condition for all external particles are satisfied for both
the $\{\hat p\}_{m+1}$ and the $\{p\}_m$ momentum configurations.
Different subtraction schemes differ in the definition of the
generalised splitting functions $v_{\ell}$ and in the mapping from
$(m+1)$- to $m$-parton phase space.  The momentum mapping and the
construction of the $v_{\ell}$ for our subtraction scheme will be
discussed in detail in Sections~\ref{sec:mommap}, \ref{sec:splttings}
and \ref{sec:subtr}.
    
Squaring the generalised splitting functions $v_{\ell}$ and summing
over all different singular parton splittings $\ell \to \ell+j$ yields
the subtraction terms, symbolically written as
\beq
\label{eq:dA}
d\sigma^A = \sum_{\ell} {\cal D}_{\ell} \otimes d\sigma^B\,,
\eeq 
where ${\cal D}_{\ell}\propto \left|v_{\ell}\right|^2$, $d\sigma^B$
denotes the LO cross section, and the symbol $\otimes$ represents
properly defined phase-space, spin and colour convolutions as above.

Integrating the subtraction term $d\sigma^A$ over the one-parton
unresolved phase space, $d\xi_p$, yields an infrared- and
collinear-singular contribution
\beq
\sum_{\ell} \left[\int\,d\xi_p {\cal D}_{\ell}\right] \otimes d\sigma^B\, = \sum_{\ell} {\cal V}_{\ell} \otimes d\sigma^B\,,
\eeq
which needs to be combined with the virtual cross section to yield a
finite NLO cross section
\beq
\label{eq:nlofin}
\sigma^{\text{NLO}} =
\int_{m}\Big[ d\sigma^V+\sum_{\ell}{\cal V}_{\ell}\otimes d\sigma^B\Big]  +
\int_{m+1}\Big[ d\sigma^R-\sum_{\ell}{\cal D}_{\ell}\otimes d\sigma^B\Big]\,.
\eeq
In this form, the NLO cross section can be integrated numerically over
phase space using Monte Carlo methods.

The generalisation of Eq.~(\ref{eq:nlofin}) to hadron collisions is
straightforward and requires the inclusion of a further counterterm to
absorb initial-state collinear singularities into a re-definition of
the parton-distribution functions.  Finally, we emphasise that the jet
cross-section $\sigma$ has to be defined in a infrared-safe way by the
inclusion of a jet-function $F_J$, which satisfies $F_J^{(m+1)}\,\to
\,F_J^{(m)}$ in the collinear and infrared limits. Both, the
factorisation of initial-state collinear singularities and the
inclusion of a jet-function are standard and are included in the more
detailed version of our final formulae presented in
Section~\ref{sec:finform}.

\section{Alternative subtraction scheme: setup}\label{sec:setup}
In the scheme proposed in this paper, the NLO subtraction terms are
derived from the splitting functions introduced in the formulation of
a parton shower with quantum interference
\cite{Nagy:2007ty,Nagy:2008ns, Nagy:2008eq}, and the momentum mapping
$\{\hat p\}_{m+1} \to \{p\}_m$ used to define the subtraction terms
corresponds to the inverse of the mapping introduced in the definition
of the parton shower.

In Section~\ref{sec:mommap} we shall first describe the momentum
mappings, following closely Ref.~\cite{Nagy:2007ty}. The essential
feature of our scheme is that we use a global mapping in which all of
the partons participate, \cf Ref.~\cite{Somogyi:2006cz}.  The
factorisation of the real-emission matrix element in the singular
limits and the construction of the subtraction terms is then discussed
in Section~\ref{sec:splttings}.  The subtraction terms needed to
address the NLO collider processes considered in this paper and their
integrated counterparts are presented in Section~\ref{sec:subtr}.
Section~\ref{sec:finform} finally collects all necessary formulae.
While our scheme can be formulated in a completely general way, here
we restrict ourselves to the discussion of processes with massless QCD
partons.

\subsection{Momentum mapping} \label{sec:mommap} As before, the
$(m+1)$-parton phase-space four-vectors are denoted by
$\ph_{1},\ph_{2},.. .$ and $m$-parton phase-space four-vectors by
$p_{1},\,p_{2},\,...$. Indices $a,\,b$ are used to label initial-state
particles.

\subsubsection{Final-state emissions}\label{sec:finmap}

Let us first treat the splitting of final-state partons. We shall
briefly review the momentum mapping used in the formulation of the
parton shower in Ref.~\cite{Nagy:2007ty} and then define the inverse
transformation needed for the subtraction scheme.

\subsubsection*{Mapping in the parton shower}
We begin with a $m$-parton configuration with momenta $\{p\}_m$, where
one of the partons with label $\ell \in \{1,\ldots m\}$ splits. After
the splitting we have a $(m+1)$-parton configuration with momenta
$\{\hat p\}_{m+1}$. We label the daughter partons by indices $\ell$
and $j$, \ie we consider a splitting $p_\ell \to \hat{p}_{\ell} +
\hat{p}_{j}$.\footnote{Ref.~\cite{Nagy:2007ty} uses instead of the
  label "$j$" the label "$m+1$" for one of the daughter partons.} For
an exactly collinear splitting or the emission of a soft gluon with
momentum $\hat{p}_{j} = 0$ one has $p_\ell = \hat{p}_{\ell} +
\hat{p}_{j}$, but away from the collinear/soft limit $p_\ell \neq
\hat{p}_{\ell} + \hat{p}_{j}$ in general. Thus, in order to satisfy
four-momentum conservation and to have all external partons before and
after the splitting on their mass-shell, we need to take some momentum
from the spectator partons which are not involved in the splitting. In
contrast to the Catani-Seymour scheme \cite{Catani:1996vz}, where the
momenta of most of the partons are left unchanged, we use a mapping
where all final state partons participate.
 
Let us now first provide the formulae which specify the momentum
mapping used in the parton shower of Ref.~\cite{Nagy:2007ty}. We
choose to leave the momenta of the initial -state partons unchanged,
\beq\label{eq:padef}
p_a\,=\,\hat p_a,\quad p_b\,=\,\hat p_b\,,
\eeq
and introduce the variable $Q$ to denote the total momentum of the
final-state partons in $m$-parton phase space:
\beq\label{eq:qdef}
Q\, \equiv\, \sum_{n = 1}^m\, p_n \,=\, p_a + p_b\,.
\eeq
Since the the momenta of the incoming partons remain the same, $Q=
\hat p_a +\hat p_b$. We now define
\beq
\label{a_ell_parameter}
a_\ell \,=\, \frac{ Q^2}{2\, p_\ell\cdot Q}\,. 
\eeq
Note that $a_\ell \ge 1$. The total momenta of the two daughter
partons $\hat p_\ell$ and $\hat p_j$ are then parametrised as a linear
combination of $p_\ell$ and $Q$ according to
\beq
\label{totalmomentumfinalstatepellpj}
 P_\ell \,= \, \hat p_\ell+\hat p_j\,=\, \lambda\,p_\ell + \frac{1-\lambda+y}{2 a_\ell}\,  Q \,.
\eeq
The two parameters $\lambda, y$ in \eq
(\ref{totalmomentumfinalstatepellpj}) can be be determined from
energy-momentum conservation as
\beq\label{eq:ydef}
\lambda\,=\,\sqrt{\left(1+y\right)^2-4\,a_\ell\,y} \quad \mbox{and}\quad y = \frac{P_\ell^2}{2\, p_\ell\cdot Q}\,.
\eeq
The parameter $y$ is a measure for the virtuality of the splitting
with a maximum value
\beq	
\label{ymax}
y_\textrm{max}\,=\,\left( \sqrt{a_\ell}-\sqrt{a_\ell-1}\,\right)^{2}
\eeq
corresponding to $\lambda = 0$. 

In our scheme, the mapping includes all final-state particles, and
their momenta before and after the splitting are related by a Lorentz
transformation
\beq
\label{hatpLambdap}
\hat p^{\mu}_{n}\,=\,\Lambda (\hat{K}, K)^\mu{}_\nu \, p_n^\nu ,\quad
n\notin\{\ell,j\}\,.  
\eeq 
Here $K$ is the total momentum of the final-state spectators before
the splitting
\bea K&=& Q - p_\ell \,,
\eea and
$\hat{K}$ is the total momentum of the final-state spectators after
the splitting
\bea \hat{K}&=& Q - P_\ell\,.
\eea 
Since each final-state spectator is changed by a Lorentz transformation, 
we have 
\beq
\hat{K}^\mu \,=\, \Lambda (\hat{K},K)^\mu{}_\nu \, K^\nu 
\eeq 
with the Lorentz transformation \cite{Nagy:2007ty}
\begin{eqnarray}\label{eq:LTini}
&&\Lambda(\hat K,K)^{\mu}_{\;\;\nu} \,=\,g^{\mu}_{\;\;\nu}\,-\,\frac{2\,(\hat K+K)^{\mu}\,(\hat K+K)_{\nu}}{(\hat K+K)^{2}}\,
+\,\frac{2\,\hat{K}^{\mu}\,K_{\nu}}{K^{2}}\,.
\end{eqnarray}
In this paper we shall focus on processes with only up to two massless
partons in the final state; then, $a_\ell\,=\,1$ and $K^2 \,=\, 0$. In
this case, an alternative representation of the Lorentz transformation
has to be introduced which is well defined when $K^2 \,=\, 0$:
\beq
\label{Lorentz_transformation_final_state_alternative}
\Lambda(\hat{K}, K)^\mu{}_\nu
\,=\, g^\mu{}_\nu + \left( \frac{K\cdot n}{\hat{K}\cdot n} - 1\right) \,n^\mu\, \bar n_{\nu} + 
\left( \frac{\hat{K}\cdot n}{ K\cdot n} - 1\right) \,\bar n^\mu\,  n_{\nu}\,,
\eeq
where $n$ and $\bar n$ are light-like vectors in the $Q$-$p_\ell$ plane
with $n \cdot \bar n \,=\, 1$ and $(p_\ell\cdot n/p_\ell\cdot\bar n) <
(Q\cdot n/Q\cdot \bar n).$

Let us briefly comment on the flavour structure of the splitting. The
flavours $f \in \{g,u,\bar u, d,\bar d, \ldots\}$ of the spectator
partons remain unchanged
\beq
\hat f_n\,=\, f_n,\quad n\notin \{\ell, j\}\,,
\eeq
while the flavour of the mother parton $f_\ell$ obeys   
\beq
\hat f_\ell+\hat f_j \,=\,f_\ell\,,
\eeq
where we use the notation of adding flavours as in $q+g = q$,
$\bar{q}+q = g$ etc. Thus if the mother parton $\ell$ is a
quark/antiquark, then $(\hat f_\ell, \hat f_j) = (q/\bar q, g)$, and
if the mother parton $\ell$ is a gluon, then $(\hat f_\ell, \hat f_j)
= (g, g)$, which corresponds to $g\to g\,g$ splitting, or any choice
of quark/antiquark flavours $(\hat f_\ell, \hat f_j) = (q, \bar q)$,
which corresponds to $g\to q\, \bar q$ splitting.

\subsubsection*{Mapping in the subtraction scheme}
There is an inverse of the transformation \eq(\ref{hatpLambdap}),
which maps the $(m+1)$-parton momenta to the $m$-parton momenta. It is
this inverse transformation which is needed to determine the
subtraction terms in a NLO calculation. We thus start with a $\{\hat
p\}_{m+1}$ configuration and determine the $\{p\}_{m}$ configuration
by combining two final state partons $\hat{p}_{\ell} + \hat{p}_{j} \to
p_{\ell}$.

One can determine $p_\ell$ by rearranging
\eq(\ref{totalmomentumfinalstatepellpj})
\beq\label{eq:fin_map}
p_\ell\,=\,\frac{1}{\lambda}\,(\hat p_\ell + \hat p_j)-\frac{1 - \lambda + y}{2\, \lambda\, a_\ell}\, Q\,, 
\eeq
where we now have to express the momentum $Q$ and the parameters
$a_\ell, \lambda, y$ defined in Eqs.~(\ref{eq:qdef}),
(\ref{a_ell_parameter}) and (\ref{eq:ydef}), respectively, in terms of
the $\{\hat p\}_{m+1}$.  From Eqs.~(\ref{eq:padef}) and
(\ref{eq:qdef}) we have
\beq
Q\, = \, \hat p_a + \hat p_b \,=\,  \sum_{n = 1}^{m+1}\, \hat p_n \,, 
\eeq
while $y$ and $a_\ell$ can alternatively be written as
\beq\label{eq:yinverse}
y = \frac{P_\ell^2}{2\, P_\ell\cdot Q - P_\ell^2}  \quad \mbox{and} \quad 
a_\ell \,=\, \frac{ Q^2}{2\, P_\ell\cdot Q - P_\ell^2}\,, 
\eeq
with $P_\ell \,= \, \hat p_\ell+\hat p_j$
(\ref{totalmomentumfinalstatepellpj}).  With $y, a_\ell$ as given in
Eq.~(\ref{eq:yinverse}), the parameter $\lambda$ follows from
Eq.~(\ref{eq:ydef}).

From $K\,=\, Q \,-\, p_\ell $ and $\hat{K}\,=\, Q - P_\ell $ we can
finally obtain the Lorentz transformation which takes the spectator
partons from the $(m+1)$-parton to the $m$-parton phase space:
\beq
\label{pLambdahatp}
 p_n^\mu\,=\,\Lambda (K,\hat{K})^\mu{}_\nu \,\hat p^{\nu}_{n} ,\quad n\notin\{\ell,a,b\}\,,
\eeq
where $\Lambda (K,\hat{K})^\mu{}_\nu$ is given by \eq(\ref{eq:LTini})
with $\hat K$ and $K$ interchanged.  For $a_{\ell}=1$, \ie $\{p\}_m =
\{p_\ell, p_m\}$, the mapping is particularly simple and reads
\begin{\eqn}\label{eq:fin_map_a1}
p_{\ell}\,=\,\frac{1}{1-y}\,\left(\ph_{\ell}+\ph_{j}-y\,Q\right),\;p_m\,=\,\frac{\ph_{m}}{1-y}\,. 
\end{\eqn}

The transformation of the flavours is similar to the case of parton
splitting.  The flavour of the mother parton $f_\ell$ is given by
\beq
f_\ell\,=\, \hat f_\ell+\hat f_j \,,
\eeq 
with the rule of adding flavours, $q+g=q$ and $q+\bar q =g$, while 
the flavours of the spectators remain unchanged 
\beq
f_n\,=\,\hat f_n, \quad n\notin \{\ell, j\}\,. 
\eeq

\subsubsection{Initial-state emission}\label{sec:inimap}

Let us now turn to the description of the splitting of an
initial-state parton into an initial- and final-state parton, and the
combination of an initial-state and a final-state parton into an
initial-state parton. As before we first review the momentum mapping
defined for the parton shower evolution in Ref.~\cite{Nagy:2007ty} and
then present the inverse mapping needed for the NLO subtraction
scheme.

\subsubsection*{Mapping in the parton shower}

We take the initial state partons to be on-shell with zero transverse
momentum, \ie
\beq
p_a^2\,=\,p_b^2\,=\,\hat p_a^2\,=\,\hat p_b^2\,=\,0
\eeq
and, in general,  
\bea
p_a &\!\!=\!\!&\eta_a\,p_A,\quad p_b\,=\,\eta_b\,p_B\,;\nn\\
\hat p_a &\!\! = \!\!& \hat\eta_a\,p_A,\quad \hat p_b\,=\,\hat \eta_b\,p_B\,.
\eea
Here $p_A$ and $p_B$ are the momenta of the incoming hadrons, which
are taken massless $p_A^2 = p_B^2=0$ and $2\, p_A\cdot p_B= s$, where
$s$ denotes the hadronic center-of-mass energy. The $\eta_a,\eta_b$
are the usual momentum fractions of the hadrons $A$ and $B$ carried by
the partons $a$ and $b$, respectively.

We now consider the splitting of an initial-state parton, say parton
$a$, into a new initial-state parton and a final-state parton, $p_a
\to \hat p_a + \hat p_j$. Here, the splitting is to be understood in
the sense of backward evolution, \ie the evolution going forward in
time is $\hat p_a \to p_a + \hat p_j$, where parton $a$ enters the
hard interaction. Given $\{p\}_m$ and $\hat p_j$ we need to define how
to obtain $\{\hat p\}_{m+1}$. As in the case of final-state splitting,
we cannot have on-shell partons with $\hat p_a = p_a + \hat p_j$ away
from the soft/collinear limit, so we need a momentum mapping which
redistributes some momenta from the spectator partons.

To define the momentum mapping we first choose to keep the momentum 
fraction of parton $b$ unchanged
\beq 
\hat \eta_b\,=\,\eta_b\,.
\eeq 
The momentum fraction $\hat \eta_a$ will be determined from the
momentum $\hat p_j$. To proceed, first consider the momenta of the
final-state spectator partons after the splitting. They are related to
the momenta before the splitting by a Lorentz transformation
\beq\label{Lorentztransformation_initialstate335}
\hat p^{\mu}_{n}\,=\,\Lambda (\hat{K},  K)^\mu{}_\nu \, p_n^\nu  , \quad n\in\{1,\cdots,m\} \quad\textrm{and}\quad 
    n\neq  j \,.
\eeq
Here $\Lambda (\hat{K}, K)^\mu{}_\nu $ is the transformation specified
in \eq(\ref{eq:LTini}), and $K$ and $\hat K$ are the total momenta of
the final state spectators before and after the splitting,
\bea
K & = & p_a + p_b\nn \\
\hat{K} & = & \hat p_a +  p_b - \hat p_j  \,=\,\hat Q -\hat p_j, \quad \hat Q\,=\,\hat p_a +  p_b\,.
\eea
To determine $\hat \eta_a$ one uses the fact that $K$ and $\hat K$ are
related through the Lorentz transformation \eq (\ref{eq:LTini}),
$\hat{K}^\mu = \Lambda (\hat{K},K)^\mu{}_\nu \, K^\nu$, so that $\hat
K^2 = K^2$. This condition determines $\hat \eta_a$ as
\beq
\label{hateta_a}
\hat \eta_a \,=\, \frac{\eta_a\,\eta_b\, s +2\,\eta_b\,p_B\cdot\hat
  p_j} {\eta_b\,s-2\,p_A\cdot\hat p_j} \le 1\,.
\eeq 
\noindent
As discussed for the final-state splitting in Sec.~\ref{sec:finmap},
the flavours of the spectator partons remain unchanged
\beq
\hat f_n\,=\,f_n,\quad n\notin \{a, j\}\,,
\eeq
while the flavours of the daughter partons $\hat f_a$ and $\hat f_j$ obey   
\beq
\hat f_a+\hat f_j \,=\,f_a\,. 
\eeq

\subsubsection*{Mapping in the subtraction scheme}
To obtain the inverse mapping which takes us from $\{\hat p\}_{m+1}$
to $\{p\}_m$ and which specifies the NLO subtraction terms, we first
determine $\eta_a$ from the condition $\hat K^2 = K^2$ as
\beq \label{eq:eta_aFS}
\eta_a \,=\, 
\frac{ \hat \eta_a\,\eta_b\, s  - 2\left( \hat \eta_a\, p_A\cdot\hat p_j + \eta_b\, p_B\cdot\hat p_j \right)}{\eta_b\,s} 
\eeq
which leads to
\begin{\eqn}
p_{a}\,=\,\lb 1-\frac{\ph_{j}\cdot\hat{Q}}{\ph_{a}\cdot p_{b}} \rb\,\ph_{a}\,.
\end{\eqn}
As before, we choose to set $\eta_b = \hat \eta_b$. The final-state
spectator partons $\{p\}_m$ are determined through the inverse of
\eq(\ref{Lorentztransformation_initialstate335}):
\beq \label{eq:ini_rest_trafo}
 p_n^\mu\,=\,\Lambda (K,\hat{K})^\mu{}_\nu \,\hat p^{\nu}_{n} , \quad n\in\{1,\cdots,m\} \quad\textrm{and}\quad 
    n\neq  j \,,
\eeq
where $\Lambda (K,\hat{K})^\mu{}_\nu$ is given by \eq(\ref{eq:LTini})
with $K\leftrightarrow \hat K$. The total momentum of the final-state
partons in the $m$-particle configuration can be calculated from $K =
p_a + p_b = \eta_a p_A + \eta_b p_B$ using \eq(\ref{eq:eta_aFS}).

The flavours of the partons obey the usual relations
\bea
f_n & = & \hat f_n, \quad n\notin \{a, j\}\nn \\
f_a &=& \hat f_a+\hat f_j\,.
\eea

\subsection{Factorisation in the soft and collinear limits}\label{sec:splttings}
We now proceed to derive the subtractions terms ${\cal D}_{\ell}$ in
\eq(\ref{eq:dA}) from the factorisation of the matrix-element in the
soft and collinear limits, \cf \eq(\ref{Factorizationm1toDim}). If
partons $\ell$ and $j$ are collinear, we have
\beq
| {\cal M}(\{\hat p, \hat f\}_{m+1})\rangle \,=\, | {\cal M}_{\ell}(\{\hat p, \hat f\}_{m+1})\rangle\,,
\eeq
where the partial amplitude $| {\cal M}_{\ell}(\{\hat p, \hat
f\}_{m+1})\rangle$ can be expressed in terms of a $m$-parton amplitude
times a splitting function as specified below. Here we follow the
notation of Refs.~\cite{Catani:1996vz, Nagy:2007ty}, where a QCD
amplitude is written as a vector $|{\cal M}\rangle$ in colour and spin
space.  In the case that parton $j$ becomes soft, the full amplitude
is given by a sum of the partial amplitudes,
\beq
| {\cal M}(\{\hat p, \hat f\}_{m+1})\rangle \,=\, \sum_{\ell} | {\cal M}_{\ell}(\{\hat p, \hat f\}_{m+1})\rangle\,.
\eeq
In the soft and collinear limits, the partial amplitudes $| {\cal
  M}_{\ell}(\{\hat p, \hat f\}_{m+1})\rangle$ take the factorised form
\beq
\label{QCDFactorizationm1tVm}
| {\cal M}_{\ell}(\{\hat p, \hat f\}_{m+1})\rangle \,=\,t^\dagger_{\ell}(f_\ell \to \hat f_\ell + \hat
f_{j})\,V^\dagger_{\ell}(\{\hat p, \hat f\}_{m+1})\,| {\cal M}(\{ p, f\}_{m})\rangle\,,
\eeq 
where $V^\dagger_{\ell}(\{\hat p, \hat f\}_{m+1})$ is an operator
acting on the spin part of the amplitude, while the operator
$t^\dagger_{\ell}(f_\ell \to \hat f_\ell + \hat f_{j})$ acts on the
colour part.

The $m$-parton amplitude $| {\cal M}(\{ p, f\}_{m})\rangle $ is
evaluated at momenta and flavours $\{p, f\}_{m}$ determined from
$\{\hat p, \hat f\}_{m+1}$ according to the transformations specified
in Sections \ref{sec:finmap} and \ref{sec:inimap} for final- and
initial-state emitters, respectively.  The matrix elements of the
spin-dependent splitting operator $V^\dagger_{\ell}$ can be expressed
as
\begin{alignat}{53}
\label{eq:defVij}
\langle \{\hat s\}_{m+1} | V^\dagger_{\ell}(\{\hat p, \hat f\}_{m+1}) | \{s\}_m\rangle \,=\, \left(\prod_{n\notin\{\ell,j\}}
      \delta_{\hat s_n,s_n}\right)\, v_{\ell} (\{\hat p, \hat f\}_{m+1},\hat s_{j},\hat s_{\ell},s_\ell) \,.
\end{alignat}
The splitting amplitudes $v_{\ell}$ in \eq(\ref{eq:defVij}) have been
derived in Ref.\cite{Nagy:2007ty} from QCD vertices. In general, they
depend on the type of partons involved in the splitting. The results
are collected in Appendix~\ref{app:splittings}.  In the special case
that parton $j$ is a gluon and that $\hat p_j$ is soft, or soft and
collinear with $\hat p_{\ell}$, the splitting amplitudes simplify and
are given by the eikonal approximation,
\beq\label{eq:veik}
v_\ell^{\textrm {eikonal}} (\{\hat p, \hat f\}_{m+1},\hat s_{j},\hat s_{\ell},s_\ell) \,=\,
   \sqrt{4\pi\as}\,\delta_{\hat s_\ell, s_\ell} \,
   \frac{\veps (\hat p_j,\hat s_j,\hat Q)^*\cd \hat p_\ell}
   {\hat p_j\cd \hat p_\ell}\,,
\eeq
where $\hat Q = Q$ is the total momentum of the final state partons,
see \eq(\ref{eq:qdef}).

To construct the NLO subtraction terms, we consider the square of the
amplitude in the singular limits, $\langle {\cal M}(\{\hat p, \hat
f\}_{m+1}) | {\cal M}(\{\hat p, \hat f\}_{m+1})\rangle$.  There are
two kinds of contributions: the direct terms which correspond to the
amplitude for a parton $\ell$ to split times the complex conjugate
amplitude for the same parton to split, and the interference terms
where parton $j$ is emitted from parton $\ell$ in the amplitude and
from parton $k\neq \ell$ in the complex conjugate amplitude. The
direct terms contain singularities when partons $\ell$ and $j$ are
collinear, and when parton $j$ is soft but not necessarily collinear
with parton $\ell$. Since our definition of the amplitude in the
singular limits is based on the use of physical polarisations for the
final-state partons (see \eq(\ref{transverseprojectiontensor}) below),
the interference terms do not exhibit purely collinear singularities.
They do, however, contain soft and soft/collinear singularities.

Let us first consider the direct terms , which are products of a
splitting amplitude $v_\ell$ times a complex conjugate splitting
amplitude $v^*_\ell$. Summing over the daughter parton spins and
averaging over the mother parton spins, leads to the spin-averaged
direct splitting functions $\overline{W}_{\ell\ell}$ which will form
part of the subtraction terms
\beq \overline{W}_{\ell\ell}\,\equiv\,
\frac12 \sum_{\hat{s}_\ell, \hat{s}_j, s_\ell}|v_\ell (\{\hat p, \hat
f\}_{m+1},\hat s_{j},\hat s_{\ell},s_\ell)|^2 \,.
\eeq 
In $d$ dimensions, when the mother parton is a gluon, the averaging
factor becomes $1/(2(1-\vareps))$.  An additional symmetry factor
$1/2$ has to be introduced for a final state $g\to g+ g$ splitting. In
the soft-gluon limit, $\overline{W}_{\ell\ell}$ can easily be obtained
from \eq(\ref{eq:veik}),
\beq
\label{spinaveragedsplittingfunctions_Wll_eikonal}
\overline{W}_{\ell\ell}^{\textrm {eikonal}}\,=\, 4\,\pi\,\as\, 
\frac{\hat p_\ell\cd  D(\hat p_{j},\hat Q) \cd  \hat p_\ell}
     { (\hat p_{j}\cd \hat p_\ell)^2 }\,,
\eeq
where flavour-dependent averaging factors are already taken into
account.  The transverse projection tensor $D_{\mu\nu}$ in
\eq(\ref{spinaveragedsplittingfunctions_Wll_eikonal}) is given by
\beq
\label{transverseprojectiontensor}
D_{\mu\nu}(\hat p_j,\hat Q) \,=\,- g_{\mu\nu} + \frac{\hat p_j^\mu\, \hat Q^\nu + \hat Q^\mu\, \hat p_j^\nu}
{\hat p_j\cd \hat Q}- \frac{\hat Q^2\, \hat p_j^\mu\, \hat p_j^\nu}{(\hat p_j\cd \hat Q)^2}\,.
\eeq
It will be convenient to define a dimensionless function $F$:
\beq
F\,=\,\frac{\hat p_\ell\cdot\hat p_j}{4\,\pi\,\as}\, \overline W_{\ell\ell}\,,
\eeq
and we then have
\beq
F_{\textrm{eik}} \,\equiv \,  \frac{\hat p_\ell\cdot\hat p_j}{4\,\pi\,\as}\, \overline W_{\ell\ell}^{\textrm{eikonal}} \,=\,
\frac{\hat p_\ell\cdot  D(\hat p_j, \hat Q)\cdot\hat p_\ell}{\hat p_\ell\cdot\hat p_j}\,=\, \frac{2\,\hat p_\ell\cdot Q}
{\hat p_j\cdot Q}-\frac{ Q^2\,\hat p_\ell\cdot\hat p_j}
{(\hat p_j\cdot Q)^2}\,.
\eeq
\noindent
For the direct terms, the colour part of the squared amplitude factorises 
\bea
\lefteqn{\langle {\cal M}(\{\hat p, \hat f\}_{m+1}) | {\cal M}(\{\hat p, \hat f\}_{m+1})\rangle }\nn\\
&& \sim 
C(\hat{f}_{\ell},\hat{f}_{j}) \langle {\cal M}_\ell(\{p, f\}_{m} | V_{\ell}(\{\hat p, \hat f\}_{m+1}) V^\dagger_{\ell}(\{\hat p, \hat f\}_{m+1}) | {\cal M}_\ell(\{p, f\}_{m})\rangle
\eea 
with overall colour factors 
\begin{eqnarray}
C(\hat{f}_{\ell},\hat{f}_{j})\,=\,
\begin{cases}
C_{F}& (\hat{f}_{\ell},\hat{f}_{j})\,=\,(q,g), (g,q)\,,\\
C_{A}& (\hat{f}_{\ell},\hat{f}_{j})\,=\,(g,g)\,,\\
T_{R}& (\hat{f}_{\ell},\hat{f}_{j})\,=\,(q,\bar{q})\,.
\end{cases}
\end{eqnarray}

We now turn to the interference terms, where parton $j$ is emitted
from parton $\ell$ in the amplitude and from parton $k\neq \ell$ in
the complex conjugate amplitude
(\fig\ref{amplitudesquare_Softdiagram_qqg}).
\begin{nfigure}{tbp} 
\vspace*{0.7cm}
\SetScale{1}
\centerline{\unitlength 1pt
\vspace*{0.7cm}
   \begin{picture}(630,100)(0,0)
  \CArc(320,10)(50,45,135)
  \CArc(320,70)(50,-135,-45)
  \CArc(320,23)(50,20,160)
  \CArc(320,57)(50,-160,-20)
   \put(323, 62){$\vdots$\normalsize }
   \put(323,  9){$\vdots$\normalsize }
  \Line(280,40)(240,65)
  \Line(280,40)(240,15)
  \Line(360,40)(400,65)
  \Line(360,40)(400,15)
  \GCirc( 280,40){13}{0.5}
  \GCirc(360,40){13}{0.5}
   \DashLine(320,-20)(320,105){6}
     \put(298, 36){$\vdots$ }
   \put(338,36){$\vdots$ }
   \Text ( 205,85)[l]{ $| {\cal M}_\ell(\{\hat p, \hat f\}_{m+1})>$}
   \Text ( 350,85)[l]{ $      < {\cal M}_k(\{\hat p, \hat f\}_{m+1}) |$}
      \Text ( 325,50)[l]{$j$}
      \Text ( 310,66)[l]{$\ell$}
      \Text ( 310,14)[l]{$k$}
     \LongArrow( 295,57)( 305, 62)
  \LongArrow( 295,22)( 305, 17)
  \LongArrow( 326,41)( 335, 31)
  \Gluon(305.6,58)(334.4,22){2}{6}
  \end{picture}}
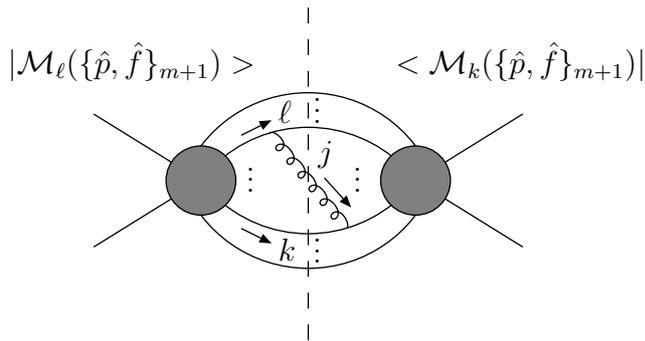
\caption{Interference diagram: parton $j$ is emitted from parton
  $\ell$ in the scattering amplitude and parton $j$ is emitted from
  parton $k$ in the complex-conjugate scattering amplitude. }
 \label{amplitudesquare_Softdiagram_qqg}
\vspace*{0.7cm}
\end{nfigure}
As mentioned above, the interference terms contain the soft and
soft/collinear singularities, and the splitting amplitudes are given
by the eikonal approximation.  The interference splitting function
thus becomes
\beq\label{eq:wlkdef}
{W}_{\ell k}\, = \,
v_\ell^{\textrm {eikonal}}(\{\hat p, \hat f\}_{m+1}, \hat s_j, \hat s_\ell, s_\ell)\,
   v_k^{\textrm {eikonal}}(\{\hat p, \hat f\}_{m+1}, \hat s_j, \hat s_k,    s_k)^* \,
    \delta_{\hat s_\ell, s_\ell} \,
    \delta_{\hat s_k, s_k}\,.
\eeq
There is an ambiguity in the momentum mapping to be used in
\eq(\ref{eq:wlkdef}).  One could associate ${W}_{\ell k}$ with the
splitting of parton $\ell$ and the momentum mapping for that
splitting, or with the splitting of parton $k$ and the corresponding
momentum mapping, or one could use an average of both. To define the
average, we introduce weight factors~\cite{Nagy:2008ns} which
redistribute the splitting function to the two possible mappings
\beq
W_{\ell k}\,\longrightarrow\,A_{\ell k}\,W^{(\ell)}_{\ell k}+A_{k \ell}W_{\ell k}^{(k)}\,.
\eeq
Here $W^{(\ell)}_{\ell k}$ and $W_{\ell k}^{(k)}$ denote splitting
functions with the mapping corresponding to the splitting of parton
$\ell$ and $k$, respectively. Furthermore we have
\beq
A_{\ell k}(\{\hat p\}_{m+1})\,+\,A_{k \ell}(\{\hat p\}_{m+1})\,=\,1
\eeq
for any fixed momenta $\{\hat p\}_{m+1}$. The conceptually simplest
choice $A_{\ell k} = A_{k \ell} = 1/2$ has been adopted in
Ref.~\cite{Nagy:2007ty}.  However, we would like to set up our NLO
scheme in a way that facilitates matching with a parton shower
including quantum and spin interferences.  Thus, as advocated
Ref.~\cite{Nagy:2008eq}, we will adopt weight functions which have
more favourable properties when used in the formulation of parton
showers.

Summing over the two graphs with interference of gluons emitted from
partons $\ell$ and $k$ we obtain a term
\bea
W_{\ell k,k\ell} &\!\!= \!\!&W_{\ell k}\,t^{\dagger}_{\ell}\,\otimes t_{k}\,+\,W_{k \ell}\,t^{\dagger}_{k}\,\otimes t_{\ell}\nn\\
&\!\! = \!\!&A_{\ell k}\,\left[ W^{(\ell)}_{\ell k}\,t^{\dagger}_{\ell}\otimes\,t_{k} 
\,+\, W^{(\ell)}_{k \ell}\,t^{\dagger}_{k}\otimes\,t_{\ell}\right]\nn\\
&+& A_{k\ell}\,\left[ W^{(k)}_{\ell k}\,t^{\dagger}_{\ell}\otimes\,t_{k} \,+\, W^{(k)}_{k\ell}\,t^{\dagger}_{k}\otimes\,t_{\ell}\right]\,.
\eea
Let us consider the part proportional to $A_{\ell k}$. The
spin-averaged splitting function $\propto A_{\ell k}$ can be
simplified to
\beq
\frac{1}{2}\,\left[t^{\dagger}_{k}\,\otimes\,t_{\ell}+ t^{\dagger}_{\ell}\,\otimes\,t_{k}\right]\,\overline{W}_{\ell k}\,,
\eeq
where 
\begin{equation}
\label{Wellk2842010}
\overline W_{\ell k} \,=\, 4\pi\as\, 2 A_{\ell k}
\frac{\hat p_\ell\cd D(\hat p_{j}, \hat Q) \cd  \hat p_k}
{\hat p_{j}\cd \hat p_\ell \, \hat p_{j}\cd \hat p_k}\,.
\end{equation}

We now combine $\overline W_{\ell k}$ with the direct splitting
function squared $\overline{W}_{\ell \ell}$, which has the same
momentum mapping and which comes with a colour factor
$t^{\dagger}_{\ell}\,\otimes\,t_{\ell}$.  Invariance of the matrix
element under colour rotations implies \cite{Nagy:2008ns}:
\beq
t^{\dagger}_{\ell}\,\otimes\,t_{\ell}\,=\,-\sum_{k\,\neq\,\ell}\,\frac{1}{2}\,\left[t^{\dagger}_{k}\,\otimes\,t_{\ell}+ t^{\dagger}_{\ell}\,\otimes\,t_{k}\right]\,,
\eeq
and the complete contribution obeying one particular mapping is then
given by
\beq
-\frac{1}{2}\,\left[t^{\dagger}_{k}\,\otimes\,t_{\ell}+ t^{\dagger}_{\ell}\,\otimes\,t_{k}\right]\,\left[\overline{W}_{\ell \ell}-\overline{W}_{\ell k} \right]\,.
\eeq

Following \cite{Nagy:2008ns}, we now add and subtract the soft-gluon
approximation to the direct splitting function
$\overline{W}_{\ell\ell}$ according to
\beq
\label{splitsplittingfunctions335}
\overline{W}_{\ell\ell}- \overline{W}_{\ell k} \,=\,\left( \overline{W}_{\ell\ell} - \overline{W}_{\ell\ell}^{\textrm {eikonal}} \right)
+ \left(\overline{W}_{\ell\ell}^{\textrm {eikonal}}  - \overline{W}_{\ell k}\right)\,,
\eeq
where $\overline{W}_{\ell\ell}^{\textrm {eikonal}}$ is given in
\eq(\ref{spinaveragedsplittingfunctions_Wll_eikonal}).  The first term
in \eq(\ref{splitsplittingfunctions335}) only has a collinear
singularity, while the soft and the soft/collinear singularities are
contained in the second term $\left(\overline{W}_{\ell\ell}^{\textrm
    {eikonal}} - \overline{W}_{\ell k}\right)$. This second term can
be written as~\cite{Nagy:2008eq}
\beq
\overline{W}_{\ell\ell}^{\textrm {eikonal}}  - \overline{W}_{\ell k}\,=\,
4\pi\as\,A'_{\ell k} \,\frac{-\hat{P}_{\ell k}^{2}} {(\hat p_j  \cd \hat p_\ell \,\hat p_j  \cd \hat p_k )^{2}}\,,
\eeq
where $ \hat{P}_{\ell k} \,=\,\hat p_j \cd \hat p_\ell \,\hat p_k -
\hat p_j \cd \hat p_k \,\hat p_\ell$, and $A'_{\ell k}$ is defined in
\cite{Nagy:2008eq} in terms of $A_{\ell k}$.  Several choices for
$A'_{\ell k}$ have been proposed in Ref.~\cite{Nagy:2008eq}; all
results given here have been obtained using
\beq
\label{Nagy2008eq712}
A'_{\ell k} (\{\hat p\}_{m+1})\,=\,\frac{\hat p_j  \cd \hat p_k\, \hat p_\ell\cd \hat Q}
{\hat p_j  \cd \hat p_k\, \hat p_\ell\cd \hat Q +\hat p_j  \cd \hat p_\ell\, \hat p_k\cd \hat Q}\,,
\eeq
specified in \eq (7.12) of Ref.~\cite{Nagy:2008eq}. As argued in
\cite{Nagy:2008eq}, the choice in \eq (\ref{Nagy2008eq712}) has
various favourable features when used in the formulation of a parton
shower.

The general form of the interference spin-averaged splitting function
is then given by
\beq
\label{interferencespinaveragedsplittingfunction}
\Delta W \,=\, \overline{W}_{\ell\ell}^{\textrm {eikonal}}  - \overline{W}_{\ell k} \,=\,4\pi\al_s
\frac{2\, \hat p_\ell\cdot\hat p_k\, \hat p_\ell\cdot\hat Q } 
{\hat p_\ell\cdot\hat p_j\,
\left(\hat p_j\cdot\hat p_k\, \hat p_\ell\cdot\hat Q+\hat p_\ell\cdot\hat p_j\,\hat p_k\cdot\hat Q \right)}\,.
\eeq
The only singularity in
\eq(\ref{interferencespinaveragedsplittingfunction}) arises from the
factor $\hat p_\ell\cd \hat p_j $ in the denominator; the interference
term is constructed such that it vanishes for $ \hat p_j\cd \hat p_k
\,\rightarrow\,0$.  Note that the interference term $\Delta W$ only
needs to be considered if the emitted parton $j$ is a gluon. If parton
$j$ is a quark or antiquark, this term vanishes.

\subsection{Subtraction terms}\label{sec:subtr}
The subtraction terms ${\cal D}_{\ell}$ in \eq(\ref{eq:dA}) are
constructed from the splitting functions that describe the
$(m+1)$-parton matrix-element squared in the soft and collinear
limits, as derived in Section~\ref{sec:splttings}.  Results for the
splitting functions have already been presented in
Ref.~\cite{Nagy:2008ns}.\footnote{Note that there is an error in \eq
  (2.38) of \cite{Nagy:2008ns}, which should read
\begin{\eqn*}
F\,=\,F_{\text{eikonal}}+\frac{x-y}{1-x}\,.
\end{\eqn*}} Integrating the subtraction terms over the one-parton
unresolved phase space yields an infrared- and collinear-singular
contribution which needs to be combined with the virtual cross section
to yield a finite NLO cross section. In this section we present
results for the subtraction terms ${\cal D}_{\ell}$ and their
integrated counterparts ${\cal V}_{\ell}$. We first consider the
collinear contributions proportional to $\overline{W}_{\ell\ell} -
\overline{W}_{\ell\ell}^{\textrm {eikonal}}$ and then discuss the soft
singularities contained in the interference terms proportional to
$\overline{W}_{\ell\ell}^{\textrm {eikonal}} - \overline{W}_{\ell k}$,
see \eq(\ref{splitsplittingfunctions335}).

In the formulae presented below, we leave out a common factor
$4\pi\al_{s}$ in the expressions for the squares $v_{\ell}^2$ of the
splitting amplitudes; however, the expressions for the subtraction
terms ${\cal D}_{\ell}$ and the integrated subtraction terms
$\mathcal{V}_{\ell}$ contain all factors, so that they can directly be
used in the final formulae presented in Section~\ref{sec:finform}
below.  Some of the integrals have been calculated using the
Mathematica package HypExp \cite{Huber:2005yg,Huber:2007dx}.

\subsubsection{Final-state collinear subtractions}\label{sec:finsq}

In this paper we consider processes with only up to two massless
partons in the final state, so that $a_{\ell} = 1$ (\cf
\eq(\ref{a_ell_parameter})).  We use the labelling
$\D_{f_{\ell}\hat{f}_{\ell}\hat{f}_{j}}$ and
$\mathcal{V}_{f_{\ell}\hat{f}_{\ell}\hat{f}_{j}}$ for a process with
the splitting $p_{\ell}\,\rightarrow\,\ph_{\ell}+\ph_{j}$. For
final-state collinear splittings, the subtraction terms can be
expressed through the variables
\begin{eqnarray}
&&y\,=\,\frac{\ph_{\ell}\cdot \ph_{j}}{p_{\ell}\cdot Q} \quad \mbox{and}\quad 
z\,=\,\frac{\ph_{j}\cdot n_{\ell}}{P_{\ell}\cdot n_{\ell}}\,,
\end{eqnarray}
with
\beq
P_{\ell}\,=\,\ph_{\ell}+\ph_{j},\;n_{\ell}\,=\,\frac{1}{1-y}\,
\lb Q-P_{\ell} \rb\,=\,p_{m},\;p_{\ell}\cdot Q\,=\,P_{\ell}\cdot Q-\ph_{\ell}\cdot\ph_{j}\,,
\eeq
where $p_{m}$ is the non-emitting final-state parton in the process.

The integration of the subtraction terms over the one-parton
unresolved phase space makes use of the phase-space factorisation
\begin{\eqn}
\left[d\{\ph,\hat{f}\}_{m+1}\right]g(\{\ph,\hat{f}\}_{m+1})\,=\,
\left[d\{p,f\}_{m}\right]\,d\xi_{p}g(\{\ph,\hat{f}\}_{m+1})\,,
\end{\eqn}
where $g(\{\hat p, \hat f\}_{m+1})$ is an arbitrary function. The
definition of the unresolved one-parton integration measure is
\begin{alignat}{5}
d\xi_p&=
dy\,\theta(y_\textrm{min}< y  <
y_\textrm{max})\,\lambda^{d-3}\,\frac{p_{\ell}\cdot Q}{\pi}\,
\frac{d^d\hat p_{\ell}}{(2\,\pi)^{d}}\,
2\,\pi\,\delta^{+}(\hat p_{\ell}^{2} )\,
\frac{d^{d}\hat p_{j}}{(2\,\pi)^{d}}\,2\,\pi\,\delta^{+}(\hat p_{j}^{2} )\, 
\notag \\
&\times
(2\,\pi)^{d}\,\delta^{(d)}\left(\hat p_\ell + \hat p_{j}- \lambda\, p_\ell - \frac{1 - \lambda  + y}{2\,a_\ell}\, Q
\right).
\label{eq:fin_meas}
\end{alignat}
Here $y_\textrm{min} = 0$ for massless partons and $y_\textrm{max}$ is
given by \eq(\ref{ymax}). For $a_\ell\,=\,1$, \eq(\ref{eq:fin_meas})
reduces to
\begin{alignat}{5}
d\xi_p^{a_\ell=1}&=
dy\,\theta\lb y(1-y)\rb\,(1-y)^{d-3}\,\frac{p_{\ell}\cdot Q}{\pi}\,\frac{d^d\hat p_{\ell}}{(2\,\pi)^{d}}\,
2\,\pi\,\delta^{+}(\hat p_{\ell}^{2} )\,
\frac{d^{d}\hat p_{j}}{(2\,\pi)^{d}}\,2\,\pi\,\delta^{+}(\hat p_{j}^{2} )\, 
\notag \\
&\times
(2\,\pi)^{d}\,\delta^{(d)}\left(\hat p_\ell + \hat p_{j}- (1-y) p_\ell - y Q
\right).
\label{eq:meas_fin}
\end{alignat}

\subsubsection*{Subtraction terms and integrals}
\subsubsection*{qqg, $\bar{\text{q}}\bar{\text{q}}$g}
The squared splitting amplitude for final state $qqg$ couplings in the
case of massless quarks is given by
\begin{\eqn}\label{eq:qqgfinsq}
v_{qqg}^{2}-v_{\text{eik}}^{2}\,=\,\frac{2}{y\,(p_{\ell}Q)}\,\left\{\frac{y}{1-y}\,F_\text{eik}\,+\,(1-\vareps)\,z\right\}\,,
\end{\eqn}
where
\begin{\eqn}\label{eq:feikfin}
F_\text{eik}\,=\,2\,\,\lb\,-1\,+\,\frac{1+y}{y+z\,(1-y)}\,-\,\frac{y}{(y+z\,(1-y))^{2}}\rb.
\end{\eqn}
Thus we have for the subtraction term
\begin{\eqn}\label{eq:dqqg_fin}
\D^{\text{coll}}_{qqg}\,=\,\frac{4\pi\al_{s}}{2}\,C_{F}\,\lb v_{qqg}^{2}-v_{\text{eik}}^{2} \rb.
\end{\eqn}
The collinear part of the integrated
subtraction term is given by
\begin{eqnarray}\label{eq:vqqg_fin}
\mathcal{V}_{qqg}^{\text{coll}}
&=&\frac{4\pi\al_{s}}{2}\mu^{2\vareps}\,C_{F}\,\int\,d\xi_{p}\,
\left[v_{qqg}^{2}-v^{2}_\text{eik}\right] \nn\\
&=&
\frac{\al_{s}}{4\,\pi}\,C_{F}\,\frac{1}{\Gamma(1-\vareps)}\,
\lb\frac{2\,\pi\,\mu^{2}}{p_{\ell}Q} \rb^{\vareps}\,
\lb-\frac{1}{\vareps}\,-14\,+\,\frac{4}{3}\,\pi^{2}\rb.
\end{eqnarray}
\subsubsection*{gq$\bar{\text{q}}$, g$\bar{\text{q}}$q}
The $gq\bar{q}$ splitting function for massless quarks is given by
\begin{\eqn}
v_{gq\bar{q}}^{2}\,=\,\frac{2}{y\,p_{\ell}Q}\,\lb 1-\vareps-2\,z\,(1-z) \rb.
\end{\eqn}
Averaging over the helicity of the incoming particles gives an
additional factor $1/(2\,(1-\vareps))$, and we obtain for the
subtraction term
\begin{\eqn}\label{eq:dgqq_fin}
\D_{gqq}\,=\,\frac{4\pi\al_{s}}{2\,(1-\vareps)}\,T_{R}\,v_{gq\bar{q}}^{2}\,.
\end{\eqn}
Integrating this over the unresolved one-parton phase-space, we obtain
\begin{eqnarray}\label{eq:vgqq_fin}
\mathcal{V}_{gqq}\,=\,\frac{4\pi\al_{s}}{2\,(1-\vareps)}\mu^{2\vareps}\,T_{R}\,
\int\,d\xi_{p}\,v_{gq\bar{q}}^{2}\,=\,\frac{T_{R}}{\pi}\,
\frac{\al_{s}}{\Gamma(1-\vareps)}\,\lb\frac{2\,\pi\,\mu^{2}}{p_{\ell} 
\cdot Q}\rb^{\vareps}\,\left[-\frac{1}{3\,\vareps}\,-\,\frac{8}{9}\right]\,.
\end{eqnarray}
\subsubsection*{ggg}
The total (unaveraged) splitting amplitude squared is given by
\begin{\eqn}
v_{ggg}^{2}\,=\,\frac{1}{2\,(\ph_{\ell} \cdot
  \ph_{j})^{2}}\left\{(d-2)\,
\left[\ph_{\ell}\cdot D_{j}\cdot \ph_{\ell}+\ph_{j}\cdot D_{\ell}\cdot
  \ph_{j}\right]\,
-\,k_{\perp}^{2}\mbox{Tr}\left[D_{\ell}\cdot D_{j}
  \right] \right\},
\end{\eqn} 
with
\begin{eqnarray}
\ph_{\ell}\cdot D_{j}\cdot \ph_{\ell}&=&\frac{2\,y\,p_{\ell} \cdot
  Q}{y+z\,(1-y)}\,
\left[1\,-\,z\,(1-y)\,-\,\frac{y}{y+z\,(1-y)}
  \right],\nn\\
\ph_{j}\cdot D_{\ell}\cdot \ph_{j}&=&\frac{2\,y\,p_{\ell} \cdot
  Q}{1-z\,(1-y)}\,
\left[y\,+\,z\,(1-y)\,-\,\frac{y}{1-z\,(1-y)}
  \right],\nn\\
k_{\perp}^{2}&=&-2\,y\,z\,(1-z)\,p_{\ell} \cdot Q,\nn\\
\mbox{Tr}\left[D_{\ell}\cdot D_{j}\right]&=&d-2-2\,\Delta+\Delta^{2}
\end{eqnarray}
and
\begin{\eqn}
\Delta\,=\,\frac{\hat{Q}^{2}\,(\ph_{\ell} \cdot \ph_{j})}{(\ph_{\ell}
  \cdot 
\hat{Q})\,(\ph_{j} \cdot \hat{Q})}\,=\,\frac{2\,y}{(y+z\,(1-y))\,(1-z\,(1-y))}\,.
\end{\eqn}
Instead of using this as a subtraction term, however, we proceed in a
different way \cite{Nagy:2008ns}: in order to well separate the
singularities in the triple-gluon final state, we will use a slightly
modified splitting function, where all soft divergences originating
from particle $\hat{\ell}$ are transferred to the subtraction term
where $\ph_{\ell}$ and $\ph_{j}$ are interchanged.  This can be
achieved by subtracting the term
\begin{\eqn}
v_{ggg,\text{sub}}^{2}\,=\,v_{2}^{2}-v_{3}^{2}\,=\,\frac{d-2}{2\,(\ph_{\ell}
  \cdot 
\ph_{j})^{2}}\,\left[\ph_{\ell}\cdot D_{j}\cdot \ph_{\ell}-\ph_{j}\cdot D_{\ell}\cdot\ph_{j} \right],
\end{\eqn}
where $v_{2,3}$ are defined corresponding to \eqs (2.40)-(2.42) in
\cite{Nagy:2008ns}.  In the end, we obtain
\begin{\eqn}
\tilde{v}_{ggg}^{2}\,=\,v^{2}_{ggg}+v_{ggg,\text{sub}}^{2}\,=\,\frac{1}{2\,(\ph_{\ell}
  \cdot \ph_{j})^{2}}\left\{2\,(d-2)\,\,\ph_{\ell}\cdot D_{j}\cdot
  \ph_{\ell}\,
-\,k_{\perp}^{2}\mbox{Tr}\left[D_{\ell}\cdot D_{j}
  \right] \right\}\,,
\end{\eqn}
which is the unintegrated subtraction term for each gluon emission.
The first part is the unaveraged eikonal splitting function; if we
combine this with the interference term, we have
\begin{\eqn}
\tilde{v}_{ggg}^{2}-v^{2}_\text{eik}\,=\,-\frac{k_{\perp}^{2}}{2\,(\ph_{\ell}
  \cdot \ph_{j})^{2}}\,\mbox{Tr}\left[D_{\ell} \cdot D_{j}\right]\,=
\,\frac{z\,(1-z)}{y\,(p_{\ell}\,\hat{Q})}\,\left[d-2-2\,\Delta+\Delta^{2}\right]\,,
\end{\eqn}
and the collinear subtraction term reads
\begin{\eqn}\label{eq:dggg_fin}
D^{\text{coll}}_{ggg}\,=\,\frac{4\,\pi\,\al_{s}}{2\,(1-\vareps)}\,C_{A}\,\lb
\tilde{v}_{ggg}^{2}-v^{2}_\text{eik} \rb.
\end{\eqn}
Integrating and taking all averaging factors into account, we obtain
\begin{eqnarray}\label{eq:vggg_fin}
\mathcal{V}_{ggg}^{\text{coll}}&=&\mu^{2\vareps}\,\frac{4\,\pi\,\al_{s}}{2\,(1-\vareps)}\,C_{A}
\,\int\,d\xi_{p}\,\lb\tilde{v}_{ggg}^{2}-v^{2}_\text{eik}\right)\nn\\
&=&\lb\frac{2\,\pi\,\mu^{2}}{p_{\ell} \cdot
  \hat{Q}}\rb^{\vareps}\,\frac{1}{\Gamma(1-\vareps)}
\,\frac{\al_{s}}{2\pi}\,C_{A}\,
\left[-\frac{1}{6\,\vareps}\,+\,\frac{55}{18}\,-\frac{3}{8}\pi^{2}\right]\,.
\end{eqnarray}

\subsubsection{Initial-state collinear subtractions}\label{sec:inisq}
In the following, the parent parton will always be $\ph_{a}$; for
processes with two initial-state partons, the corresponding formulae
for the other incoming parton can be obtained from interchanging
$a\,\leftrightarrow\,b$. We label the processes according to backward
evolution; \eg g$\bar{\text{q}}$\,q denotes a process where the gluon
participates in the hard interaction. Therefore, this splitting
function needs to be applied for $g+X\,\rightarrow\,Y$ in the
$m$-particle and $q+X\,\rightarrow\,Y+q$ in the $(m+1)$-particle phase
space. See \Fig \ref{Aninitialstate_qqg_splitting} for an illustration
for qqg splitting.
\begin{nfigure}{tbp} 
\vspace*{0.5cm}
\SetScale{1}
\centerline{\unitlength 1pt
\vspace*{0.5cm}
\begin{picture}(315,105)
 \Line (90, 50)(170,50)
 \Gluon(122, 50)(235,105) {2}{9}
 \Line  (90,0)(170,40)
 \Gluon(170,50)(260,75){2}{9}
 \Line (185,40)(260,45)
 \Gluon(160,40)(260,0){2}{8}
  \Text ( 90,40)[l]{$\hat p_a$}
  \Text ( 200,105)[l]{$\hat p_j$}
  \Text ( 135,40)[l]{$ p_a$}
    \GCirc( 170,40){22}{0.5}
  \end{picture}
}
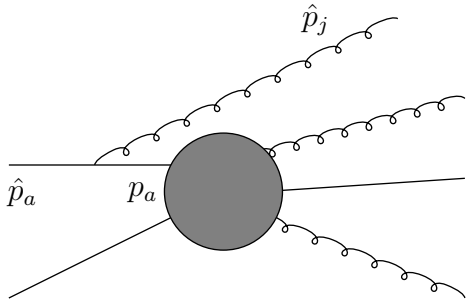
\caption{Example for the initial state splitting
  $q(p_{a})q(\ph_{a})g(\ph_{j})$, which has to be applied for
  processes with $q+X\,\rightarrow\,Y$ in the $m$-particle and
  $q+X\,\rightarrow\,g+Y$ in the $(m+1)$-particle phase space. Our
  notation follows backward evolution, with the first flavour denoting
  the parton which participates in the hard interaction.}
\label{Aninitialstate_qqg_splitting}
\vspace*{0.5cm}
\end{nfigure}

We will use the following variables to describe in the initial-state
subtraction terms:
\begin{\eqn}
x\,=\,1-2\,\frac{\ph_{j} \cdot \hat{Q}}{\hat{Q}^{2}} \quad \mbox{and} \quad y\,=\,2\,\frac{\ph_{a} \cdot \ph_{j}}{\hat{Q}^{2}}\,,
\end{\eqn}
and we also define $y'\,=\, y/(1-x)$.

The momentum $\ph_{j}$ can be expressed in terms of these variables by
introducing a Sudakov parametrisation
\begin{\eqn}\label{eq:pjsud}
\ph_{j}\,=\,\al\,p_{a}\,+\,\beta\,p_{b}\,-k_{\perp},
\end{\eqn}
where  
 \begin{\eqn}\label{eq:albe_sud}
\al\,=\,\frac{1-x-y}{x} \quad \mbox{and} \quad \beta\,=\,y.
\end{\eqn}
From $\ph_{j}^{2}=0$ it follows that
$k_{\perp}^{2}\,=\,-2\,\al\,\beta\,p_{a}\cdot\,p_{b}$.  If we define
the $z-$axis by the direction of the incoming beams, we can
furthermore specify
\begin{\eqn}\label{eq:kperp}
k_\perp\,=\,-|k_\perp|\,\lb\begin{array}{c}0\\1-2\,v\\2\,\sqrt{v\,(1-v)}\\0 \end{array}\rb,
\end{\eqn} 
where $v$ parametrises the additional angle in the interference terms;
$\ph_{j}$ is then completely parametrised in terms of
$x,\,y',\,v,\,2p_{a}\cdot p_{b}$, and we can easily reconstruct
\begin{\eqn}\label{eq:khat_ini_m}
\hat{K}\,=\,\frac{1}{x}p_{a}+p_{b}-\ph_{j}
\end{\eqn}
required for the inverse transformation given in \eq
(\ref{Lorentztransformation_initialstate335}).

\noindent
The phase space factorises according to
\begin{\eqn}
\left[d\{\ph,\hat{f}\}_{m+1}\right]g(\{\ph,\hat{f}\}_{m+1})\,=\,\left[d\{p,f\}_{m}\right]\,d\xi_{p}g(\{\ph,\hat{f}\}_{m+1})\,,
\end{\eqn}
with the $d$-dimensional integration measure
\begin{\eqn}\label{eq:ini_meas}
d\xi_{p}\,=\,\frac{d^{d}\ph_{j}}{(2\,\pi)^{d-1}}\,\delta_{+}\lb\ph_{j}^{2}\rb.
\end{\eqn}
A more explicit form of the integration measure is given in Appendix \ref{app:ints}.

\subsubsection*{Subtraction terms and integrals}
\subsubsection*{$\bar{\text{q}}\bar{\text{q}}\text{g}, \text{qqg}$} 
The unaveraged $qqg$ splitting function is given by
\begin{\eqn}\label{qqgsplitting}
v_{\bar{q}\bar{q}g}^{2}-v_{\text{eik}}^{2}\,=\,\frac{(1-x-y)}{x\,y\,\ph_{a}\cdot\,\ph_{b}}\,(d-2)\,.
\end{\eqn}
Including all prefactors, we obtain
\begin{\eqn}\label{eq:dqqg_ini}
\D^{\text{coll}}_{q\bar{q}g}\,=\,\frac{4\,\pi\,\al_{s}}{2}\,C_{F}\,\lb v_{\bar{q}\bar{q}g}^{2}-v_{\text{eik}}^{2} \rb
\end{\eqn}
and
\begin{eqnarray}\label{eq:vqqg_ini}
\lefteqn{\mathcal{V}^{\text{coll}}_{qqg}\,=\,\frac{4\,\pi\,\al_{s}}{2}\,C_{F}\,
\int\,\mu^{2\,\vareps}\,d\xi_{p}\,\lb v_{\bar{q}\bar{q}g}^{2}-v^{2}_{\text{eik}}\rb,}\nonumber\\
&&\hspace*{-2mm} =
\frac{\al_{s}}{2\,\pi}\,C_{F}\,\frac{1}{\Gamma(1-\vareps)}\,
\lb\frac{4\,\pi\,\mu^{2}}{2\,p_{a}\cdot
  p_{b}}\rb^{\vareps}\,\int^{1}_{0}\,dx\,\frac{1-x}{x}\,
\left[ -\frac{1}{\vareps}-\,\ln\,x\,+\,2\,\ln\,(1-x) \right]\,.
\end{eqnarray}
\subsubsection*{g$\bar{\text{q}}$q, gq$\bar{\text{q}}$}
The $g\bar{q}q$ splitting function for massless quarks is given by
\begin{\eqn}
v_{g\bar{q}q}^{2}\,=\,\frac{(d-2)}{y\, \ph_{a} \cdot \ph_{b}}\,+\,\frac{4\,(1-(x+y))}{\ph_{a} \cdot \ph_{b}\,y\,(x+y)^{2}}\,,
\end{\eqn}
and we have
\begin{\eqn}\label{eq:dgqq_ini}
\D_{g\bar{q}q}\,=\,\frac{4\,\pi\,\al_{s}}{2}\,C_{F}v^{2}_{g\bar{q}q}\,.
\end{\eqn}
Integrating, we obtain 
\begin{eqnarray}\label{eq:vgqq_ini}
\mathcal{V}_{g\bar{q}q} &=
&\frac{4\,\pi\,\al_{s}}{2}\,C_{F}\,\mu^{2\vareps}\,
\int\,d\xi_{p}\,v_{g\bar{q}q}^{2}\nn\\
&=&
\frac{1}{\Gamma(1-\vareps)}\,\frac{\al_{s}}{2\pi}\,C_{F}\,\prefaceps\,
\int^{1}_{0}\,dx\,\left[-\frac{1}{\vareps}\frac{1}{x}\,
\lb\frac{1+\,(1-x)^{2}}{x}\rb\,+\,g(x,\vareps^{0})\right]\,,\nn\\&&
\end{eqnarray}
with
\begin{eqnarray}
g(x,\vareps^{0})&=&\frac{x^{2}-2\,(1-x)}{x^{2}}\,-\,\ln\,x\,+\,\frac{2\,\ln(1-x)}{x^{2}}\,\lb (1-x)^{2}+1\rb.
\end{eqnarray}
\subsubsection*{qgq, $\bar{\text{q}}$g$\bar{\text{q}}$} 
The $gq\bar{q}$ splitting function for massless quarks is given by
\begin{\eqn}
v_{qgq}^{2}\,=\,\frac{(d-2)}{x\,y\,\ph_{a} \cdot \ph_{b}}\,+\,\frac{4\,(x+y)}{x\,y\,\ph_{a} \cdot \ph_{b}}\,\lb (x+y)-1\rb,
\end{\eqn}
and we obtain
\begin{\eqn}\label{eq:dqgq_ini}
\D_{qgq}\,=\,\frac{4\,\pi\,\al_{s}}{2\,(1-\vareps)}\,T_{R}\,v_{qgq}^{2}\,.
\end{\eqn}
The integrated splitting function reads
\begin{eqnarray}\label{eq:vqgq_ini}
  \mathcal{V}_{qgq}
  &=&\frac{4\,\pi\,\al_{s}}{2\,(1-\vareps)}\,T_{R}\,\mu^{2\vareps}\,
\int\,d\xi_{p}\,v_{qgq}^{2}\nn\\
  &=&\frac{1}{\Gamma(1-\vareps)}\,\frac{\al_{s}}{2\pi}\,T_{R}\,\,\prefaceps\,\int^{1}_{0}\,
\frac{dx}{x}\,\left[-\frac{1}{\vareps}\,\lb x^{2}\,+\,(1-x)^{2}\rb\,+\,g(x,\vareps^{0})\right],\nn\\
\end{eqnarray}
with
\begin{\eqn}
g(x,\vareps^{0})\,=\,(1-x)\,(5\,x-1)\,+\lb\,2\,\ln(1-x)\,-\,\ln\,x\rb\,\left[x^{2}+(1-x)^{2}\right]\,.
\end{\eqn}
\subsubsection*{ggg}
For initial states, the $ggg$ splitting function is given by
\begin{eqnarray}
v_{ggg}^{2}-v_{\text{eik}}^{2}&=&\frac{2\,(1-x-y)}{y\,\,\ph_{a} \cdot \ph_{b}}\,(d-2)\lb 1\,+\,\frac{1}{(x+y)^{2}}\rb
\,-\,\frac{4\,(1-x-y)}{\ph_{a} \cdot \ph_{b}\,(1-x)\,(x+y)}\,,\nn\\&&
\end{eqnarray}
and
\begin{\eqn}\label{eq:dggg_ini}
\D^{\text{coll}}_{ggg}\,=\,\frac{4\,\pi\,\al_{s}}{2\,(1-\vareps)}\,C_{A}\,\lb v_{ggg}^{2}-v_{\text{eik}}^{2}\rb.
\end{\eqn}
After the integration, we obtain
\begin{eqnarray}\label{eq:vggg_ini}
\lefteqn{\mathcal{V}^{\text{coll}}_{ggg}\,=\,\frac{4\,\pi\,\al_{s}}{2\,(1-\vareps)}\,C_{A}\,\mu^{2\vareps}\,
\int\,d\xi_{p}\,\lb v_{ggg}^{2}-v_{\text{eik}}\rb\,=\,
\frac{1}{\Gamma(1-\vareps)}\,\frac{\al_{s}}{2\pi}\,C_{A}\,\,\prefaceps\,}\nonumber\\
&\times&\int^{1}_{0}\,dx\,2\,(1-x)\,\left[\frac{1+x^{2}}{x^{2}}
  \lb-\frac{1}{\vareps}\,+\,2\,\ln\,(1-x)\rb\,
\,+\,\frac{x\,(2-x)}{(1-x)^{2}}\,\ln\,x\right.\nn\\
&&\hspace*{32mm}\left. +\,\lb \frac{x}{1-x}-\frac{1}{x^{2}}  \rb\right]\,.
\end{eqnarray}

\subsubsection{Soft and soft/collinear subtractions}\label{sec:soft}
We now turn to the discussion of the soft and soft/collinear
singularities which arise when the emitted parton is a gluon and which
are contained in $\overline{W}_{\ell\ell}^{\textrm {eikonal}} -
\overline{W}_{\ell k}$, as given in
\eq(\ref{interferencespinaveragedsplittingfunction}). There are
contributions from initial--initial-, initial--final-,
final--initial-, and final--final-state interference terms, which we
now discuss in turn.

\subsubsection*{Initial--initial-state interference}
In terms of the variables $x$ and $y$ defined in Section
\ref{sec:inisq}, the initial--initial-state interference term is given
by
\begin{\eqn}
\frac{1}{4\,\pi\,\al_{s}}\Delta W_{ab}\,=\,\frac{2}{(1-x)\,y\,\ph_{a}\cdot\,p_{b}}\,,
\end{\eqn}
so that 
\begin{\eqn}\label{eq:Dif_iniini}
\D^{\text{if}}_{ab}\,=\,C_{i}\,\Delta W_{ab}\,.
\end{\eqn}
For the simple scattering processes we consider in this paper, the
colour algebra factorises and leads to the colour factors
$C_{i}\,=\,C_{F}\,(C_{A})$ for the qqg (ggg) splitting functions.

Integrating this over the phase space of the unresolved parton, we
obtain
\begin{eqnarray}\label{eq:Vif_iniini}
\lefteqn{\mathcal{V}^{\text{if}}_{ab}\,=\,\mu^{2\,\vareps} \int\,d\xi_{p}\Delta
  W_{ab}\,=\,\lb\frac{4\,\pi\,\mu^{2}}{2\,p_{a}\cdot
    p_{b}}\rb^{\vareps}\,
\frac{\al_{s}}{2\,\pi}\,\frac{1}{\Gamma(1-\vareps)}\,C_{i}}\nonumber\\
&&\times\int^{1}_{0}\,dx\,\left[\frac{1}{\vareps^{2}}\,\delta(1-x)-\,\frac{2}{\vareps}\,\frac{1}{(1-x)_{+}}
-\frac{\pi^{2}}{6}\,\delta(1-x)\,-2\,\frac{\ln\,x}{(1-x)_{+}}\,\right.\nn\\
&&\hspace*{1.5cm}\left.+\,4\,\lb\frac{\ln(1-x)}{1-x}\rb_{+}\right]\,.
\end{eqnarray}

\subsubsection*{Initial--final-state interference}
For the initial--final-state interference, we obtain
\begin{\eqn}
\frac{1}{4\,\pi\,\al_{s}}\Delta W_{ak}\,=\,\frac{2\,(\ph_{a} \cdot
  \ph_{k})\,\hat{Q}^{2}}{(\ph_{a} \cdot \ph_{j})\,
\lb(\ph_{j} \cdot \ph_{k})\hat{s}\,+\,2\,(\ph_{a} \cdot \ph_{j})\,(\ph_{k} \cdot \hat{Q})\rb}
\end{\eqn}
and
\begin{\eqn}\label{eq:Dif_inifin}
\D^{\text{if}}_{ak}\,=\,C_{i}\,\Delta W_{ak}.
\end{\eqn}
Using the initial-state integration measure given by \eq (\ref{eq:ini_meas}), we have
\begin{eqnarray}\label{eq:Vif_inifin}
  \lefteqn{\mathcal{V}^{\text{if}}_{ak}\,=\, \mu^{2\,\vareps}
    \int\,d\xi_{p}(\Delta\,W_{ak})\,=\,
\lb\frac{4\,\pi\,\mu^{2}}{2 p_{a} \cdot p_{b}\,}\rb^{\vareps}\,\frac{\al_{s}}{2\,\pi}\,\frac{1}{\Gamma(1-\vareps)}\,C_{i}}\nn\\
  &&\times
  \,\int^{1}_{0}\,dx\,\Bigg\{\frac{1}{\vareps^{2}}\,\delta(1-x)\,-\,\frac{1}{\vareps}\,
\left[ \frac{2}{(1-x)_{+}}\,+\,\delta(1-x)\,\ln\tilde{z}_{0} \right]\nonumber\\
  &&+\,4\lb\frac{\ln(1-x)}{1-x}\rb_{+}\,-\,\frac{2\,\ln\,x}{(1-x)_{+}}\,+\,4\,\delta(1-x)
\,(\ln\,\tilde{z}_{0})\,(\ln\,2)\,+\,\frac{2}{\pi}\,I_\text{fin}(x,\tilde{z})\Bigg\},\nonumber\\
  &&
\end{eqnarray}
with
\begin{eqnarray}\label{eq:ifin2}
\lefteqn{I_\text{fin}(x,\tilde{z})\,=\,\pi\,\delta(1-x)\,\times}\nonumber\\
&&\Bigg\{\int^{1}_{0}\,dy'\,\lb\frac{\tilde{z}_{0}}{y'\,\sqrt{4\,y'^{2}\,(1-\tilde{z}_{0})
+\tilde{z}_{0}^{2}}}\,\ln\,\left[\frac{2\,\sqrt{4\,y'^{2}\,(1-\tilde{z}_{0})+\tilde{z}_{0}^{2}}\,
\sqrt{(1-y')}}{2\,y'+\tilde{z}_{0}-2\,y'\,\tilde{z}_{0}+ \sqrt{4\,y'^{2}\,(1-\tilde{z}_{0})+\tilde{z}_{0}^{2}}}\right]\rb\,\nonumber\\
&&\,-\ln\,4\,\,\ln\,\tilde{z}_{0}\,+\,\frac{1}{4}\,\lb
\text{Li}_{2}\lb 1-\tilde{z}_{0}\rb
+\ln^{2}\,\tilde{z}_{0}\rb\,\Bigg\}\nonumber\\
&&+\,\frac{1}{(1-x)_{+}}\,\int^{1}_{0}\,\frac{dy'}{y'}\,\left[\int^{1}_{0}\,\frac{dv}{\sqrt{v\,(1-v)}}\,
\frac{\tilde{z}}{N(x,y',\tilde{z},v)}-1\right]
\end{eqnarray}
and
\begin{eqnarray}
&&\tilde{z}\,=\,\frac{p_{a} \cdot \ph_{k}}{x\ph_{k}\cdot
  \hat{Q}},\;\tilde{z}_{0}\,=\,\frac{p_{a} 
\cdot p_{k}}{p_{k}\cdot Q},\;N\,=\,\frac{\ph_{j} \cdot \ph_{k}}{\ph_{k}\cdot \hat{Q}}\,\frac{1}{1-x}\,+\,y'.
\end{eqnarray}
Note that $\tilde{z}$ implicitly depends on $x$, $y$ and $v$ through
momentum mapping. Furthermore, $\ph_{k}$ has to be obtained from
$p_{k}$ using the transformation specified in \eq
(\ref{Lorentztransformation_initialstate335}) and with $\hat{K}$ given
by \eq (\ref{eq:khat_ini_m}); $\ph_{j}$ is given by \eqs
(\ref{eq:pjsud}), (\ref{eq:albe_sud}), and (\ref{eq:kperp}).
$\tilde{z}\,=\,0$ corresponds to a singularity in the $m$-particle
phase space; this singularity should be excluded by an appropriate
infrared-safe jet function. As before, $C_{i}\,=\,C_{F}\,(C_{A})$ in
the qqg (ggg) splitting function.

\subsubsection*{Final--initial-state and final--final-state interference}
The final--initial and final--final-state interference terms have the
same structure, with the only difference of $\ph_{k}\,=\,\ph_{\ell}$
(final-state particle) and $\ph_{k}\,=\,\ph_{a}$ (initial-state
particle) for the initial- and final-state integrals, respectively.

For interference terms with a final-state emitter and final-state
momentum mapping, we obtain
\begin{\eqn}
\frac{1}{4\,\pi\,\al_{s}}\Delta\,W\,=\,\frac{2\,(\ph_{\ell} \cdot
  \ph_{k})\,(\ph_{\ell} 
\cdot \hat{Q})}{(\ph_{\ell} \cdot \ph_{j})\,\lb(\ph_{j} \cdot
\ph_{k})\,(\ph_{\ell} 
\cdot \hat{Q})+(\ph_{\ell} \cdot \ph_{j})(\ph_{k} \cdot \hat{Q})\rb}
\end{\eqn}
and
\begin{\eqn}\label{eq:Dif_fin}
\D^{\text{if}}\,=\,C_{i}\,\Delta W\,.
\end{\eqn}
These are the general expressions for interference terms using the
final-state mappings; if the spectator is an initial-state particle,
$\ph_{k}\,=\ph_{a}$.

We obtain for the integrated interference term with final-state
spectators
\begin{eqnarray}\label{eq:vif_finfin}
\mathcal{V}^{\text{if}}&=&\mu^{2\,\vareps}\,\int\,d\xi_{p}\,(\Delta\,W)\,=\,
\lb\frac{2\,\mu^{2}\,\pi}{p_{\ell}\cdot
  Q}\rb^{\vareps}\frac{\al_{s}}{\pi}\,\frac{1}{\Gamma(1-\vareps)}
\,C_{i}\,\lb\frac{1}{2\,\vareps^{2}}\,+\,\frac{1}{\vareps}\,-\,\frac{\pi^{2}}{4}\,+\,3\,\rb,\nn\\ 
\end{eqnarray}
with $C_{i}\,=\,C_{F}\,(C_{A})$ in the qqg (ggg) splitting function.
 
For interference terms where the spectator is the initial-state parton
$\ph_{a}$, we obtain for the integrated subtraction term
\begin{eqnarray}\label{eq:vif_finini}
\mathcal{V}^{\text{if}}&=&\mu^{2\,\vareps}\,\int\,d\xi_{p}\,(\Delta\,W)\,=\,
\lb\frac{2\,\mu^{2}\,\pi}{p_{\ell}\cdot Q}\rb^{\vareps}\frac{\al_{s}}{\pi}\,\frac{1}{\Gamma(1-\vareps)}\,C_{i}\nn\\
&&\times\left\{\frac{1}{2\,\vareps^{2}}\,+\,\frac{1}{\vareps}\,\left[1\,+\,\frac{1}{2}\,\ln
    \lb \tilde{a}_{0}+1\rb\right]\,\right.\nonumber\\
&&\left. \hspace*{5mm}-\,\frac{\pi^{2}}{6}\,+\,3\,-\,2\,\ln\,2\,\ln\lb
  \tilde{a}_{0}
+1 \rb\,+\,\frac{1}{\pi}\left[I^{(b)}_\text{fin}(\tilde{a}_{0})\,+\,I^{(c)}_\text{fin}(\tilde{a})\right]\right\},
\end{eqnarray}
with
\begin{eqnarray}\label{eq:ibfin}
I^{(b)}_\text{fin}(\tilde{a}_{0}) &=& \frac{\pi}{2}\Bigg[\int^{1}_{0}\,\frac{du}{u}\,\Bigg\{2\ln2\,
+\,\frac{1}{\sqrt{1+4\,\tilde{a}_{0}(1+\tilde{a}_{0})\,u^{2}}}\nn\\
&&\hspace*{20mm}\left.\times\,\ln\left[\frac{(1-u)}{\lb 1+2\,\tilde{a}_{0}\,u\,+\,\sqrt{1+4\,\tilde{a}_{0}\,(1+\tilde{a}_{0})\,u^{2}}
         \rb^{2}}
    \right]\right\}\nonumber\\
&&\hspace*{5mm}+\,2\,\ln\,2\,\ln\lb 1+\tilde{a}_{0}\rb\,+\,\frac{1}{2}\,\ln^{2}\,\lb
    1+\tilde{a}_{0}\rb\,+\,\frac{5}{2}\text{Li}_{2}\,\lb\frac{\tilde{a}_{0}}{\tilde{a}_{0}+1}\rb \nn\\
    &&\hspace*{5mm}-\,\frac{1}{2}\text{Li}_{2}\,\left[\lb\frac{\tilde{a}_{0}}{\tilde{a}_{0}+1}\rb^{2}\right]\Bigg],\\
&&\nn\\    
I^{(c)}_\text{fin}(\tilde{a})&=&\pi\int^{1}_{0}\,\frac{du}{u}\,\int^{1}_{0}\,\frac{dx}{x}\,
\Bigg[\frac{x \lb 1-x+u\,x\,\left[(1-u\,x)\,\tilde{a}\,+\,2 \right]\rb}{k(u,x,\tilde{a})}\nn\\
&&\hspace*{30mm}-\frac{1}{\sqrt{1+4\,\tilde{a}_{0}\,u^{2}\,(1+\tilde{a}_{0})}}\Bigg]\,.
\end{eqnarray}
We have introduced 
\begin{eqnarray}
k^{2}(x,u,\tilde{a}) &=& 
\left[(1+ux\,-x)(z-\tilde{z})\,+\,ux\,\lb(1-ux)\,\tilde{a}+1 \rb \right]^{2}\nn\\
&&+\,4\,u\,x\,\tilde{z}\,(1-z)\lb 1+u\,x-x \rb\,\lb (1-ux)\,\tilde{a}+1 \rb
\end{eqnarray}
and 
\begin{\eqn}
z\,=\,\frac{x\,(1-u)}{1-ux},\;z'\,=\,u\,x\,\tilde{a},\;\tilde{a}\,=\,\frac{p_{a}
  \cdot n_{\ell}}{p_{a} \cdot p_{\ell}\,+\,y\,p_{a} \cdot
  n_{\ell}},\;\tilde{a}_{0}\,=\,
\tilde{a}(y=0)\,=\,\frac{p_{a} \cdot n_{\ell}}{p_{a} \cdot p_{\ell}}\,.
\end{\eqn}
Note that the treatment of interference terms significantly differs
from \cite{Catani:1996vz}; here, our choice of momentum mapping leads
to more complicated integrated interference terms, the finite parts of
which we choose to evaluate numerically.

\subsection{Final expressions}\label{sec:finform}
Let us finally collect the formulae that are needed to address the
scattering processes considered in this paper.  The NLO parton level
cross section for any collider process is given by the sum of
$\sigma^{\text{LO}}_{ab}$ and $\sigma^{\text{NLO}}_{ab}$, with
\begin{alignat}{53}
\sigma^{\text{LO}}_{ab}&= \int_m d\sigma^B_{ab}(p_a,p_b)\notag \\
\sigma^{\text{NLO}}_{ab}&=\int_{m+1}d\sigma^R_{ab}(\ph_a,\ph_b)+
\int_{m}d\sigma^V_{ab}(p_a,p_b)+\int_{m}d\sigma^C_{ab}(p_a,p_b,\mu_F^2)\,.
\end{alignat}
The hadronic cross section is obtained from the parton level cross
section by convoluting with parton distribution functions.  The
collinear counterterms $\int_{m}d\sigma^C_{ab}(p_a,p_b,\mu_F^2)$ are
needed to absorb initial-state collinear singularities into a
re-definition of the parton-distribution functions; in the $\msbar$
scheme, they are given by
\begin{alignat}{53}
\label{universalcollinearcounterterm_general}
\int_md\sigma^C_{ab}(p_a,p_b,\mu_F^2)&= \frac{\as}{2 \pi}\,
\frac{1}{\Gamma(1-\vareps)}\,
\sum_{c}\int_0^1 dx\int_m d\sigma_{cb}^{B}(xp_a, p_b)\,\frac{1}{\vareps}\,
\left( \frac{4 \pi \mu^2}{\mu_F^2} \right)^{\vareps}\, P^{ac}(x)   \notag \\
&+ \frac{\as}{2 \pi}\,
\frac{1}{\Gamma(1-\vareps)}\,
\sum_{c}\int_0^1 dx\int_m d\sigma_{ac}^{B}(p_a, xp_b)\,\frac{1}{\vareps}\,
\left( \frac{4 \pi \mu^2}{\mu_F^2} \right)^{\vareps}\, P^{bc}(x)  \,. 
\end{alignat}
Here the $ P^{ab}(x)$ are the Altarelli-Parisi kernels in four
dimensions \cite{Altarelli:1977zs}; their explicit form is given in Appendix \ref{sec:app_splittings}. 
 We then obtain for the parton-level NLO contribution
\begin{alignat}{53}
\label{eq:sig_nlo}
\sigma_{ab}^{\text{NLO}}(p_a,p_b,\mu_F^2)&= \int_{m+1} \left[
  d\sigma_{ab}^{R}(\ph_a,\ph_b) -
  d\sigma_{ab}^{A}(\ph_a,\ph_b) \right]                    \notag \\
&+ \int_{m}\, \left[\int d\sigma^{V}_{ab}(p_a,p_b)+
  \int_1d\sigma^{A}_{ab}(\ph_a,\ph_b)+d\sigma_{ab}^{C}(p_a,p_b,\mu_F^2)\right]_{\vareps=0},
\end{alignat}
where $\int_1 d\sigma^A_{ab}+ d\sigma^C_{ab}$ can be written as
\begin{alignat}{53}\label{IKP2842010}
&\hspace*{-10mm}\int_{m}  \left[ \int_{1} d\sigma^A_{ab}(\ph_a,\ph_b)+ d\sigma^C_{ab}(p_a,p_b,\mu_F^2) \right]    \notag \\
=&
\int_m  d\sigma_{ab}^{B}(p_a,p_b) \otimes { I}(\vareps)
+ \int_0^1 dx \int_m  d\sigma_{ab}^{B}(x\ph_a,p_b)\otimes
\left[ { K}^a(x\,\ph_a) +   { P}(x,\mu_F^2)  \right]                             \notag \\
+&  \int_0^1 dx \int_m d\sigma_{ab}^{B}(\ph_a,x\ph_b)\otimes
\left[ { K}^b(x\ph_b) + { P}(x,\mu_F^2)  \right]\,. 
\end{alignat}
This equation defines the insertion operators $I(\vareps),\,K(x),\,
P(x;\mu_{F})$ at the cross section level, where we follow the standard
notation introduced in Ref.~\cite{Catani:1996vz}.
\eq(\ref{IKP2842010}) can be divided into two parts: the first part is
the universal insertion operator ${ I}(\vareps)$, which contains the
complete singularity structure of the virtual contribution and has LO
kinematics. The second part consists of the finite pieces that are
left over after absorbing the initial-state collinear singularities
into a redefinition of the parton distribution functions at NLO.  It
involves an additional one dimensional integration over the momentum
fraction $x$ of an incoming parton with the LO cross sections.

Obviously, the cross sections and all observables have to be defined
in an infrared-safe way through the introduction of jet functions; one
should thus replace $\sigma^{\text{LO}}$ and
$\sigma^{\text{\text{NLO}}}$ by the jet cross sections
\begin{alignat}{53}
\sigma^{ LO}&= \int dPS_m (p_1,\cdots,p_m)\, \left| {\cal M}_{m} (p_1,\cdots,p_m)\right|^2\, F_J^{(m)} (p_1,\cdots,p_m)            \notag \\
\sigma^{ NLO}&=\int dPS_{m+1} (\ph_1,\cdots,\ph_{m+1})\,\left| {\cal M}_{m+1} (\ph_1,\cdots,\ph_{m+1})\right|^2 \,
F_J^{(m+1)} (\ph_1,\cdots,\ph_{m+1})     \notag \\
&+ \int dPS_{m} (p_1,\cdots,p_m)\,\left| {\cal M}_{m} (p_1,\cdots,p_m)\right|^2_{\textrm{one-loop}}\, 
F_J^{(m)} (p_1,\cdots,p_m) \,.
\label{eq:withf}
\end{alignat}
In general, the jet function may
contain $\theta$-functions (which define cuts and corresponding cross
sections) and $\delta$-functions (which define differential cross
sections). Infrared safety now requires that
\begin{eqnarray}
&&\hspace*{-5mm}F_J^{(m+1)} (p_1, \cdots ,p_j=\lambda\,q, \cdots ,p_{m+1} ) \to F_J^{(m)} (p_1,\cdots ,p_{m+1} ) 
                       \quad {\textrm {if}}\quad \lambda \to 0                                    \nn    \\
&&\hspace*{-5mm}F_J^{(m+1)} (p_1,..,p_i,..,p_j,..,p_{m+1} ) \to F_J^{(m)} (p_1,..,p,..,p_{m+1} ) 
  \quad {\textrm {if}}\quad   p_i\to z p,\,p_j\to (1-z)p                                      \nn  \\
&&\hspace*{-5mm}F_J^{(m)} (p_1,\cdots ,p_m ) \to 0   \quad{\textrm {if}}\quad  p_i \cdot p_j \to 0  
\label{jetobservableinfraredandcollinearsafe}
\end{eqnarray}
The first two conditions of
\eq(\ref{jetobservableinfraredandcollinearsafe}) define the essential
property of the jet function that the jet observable has to be
infrared and collinear safe for any number $m$ of partons in the final
state, \ie to any order in QCD perturbation theory. The last condition
of \eq(\ref{jetobservableinfraredandcollinearsafe}) guarantees that
the Born-level cross section is well defined. To summarise, we require
that \beq F_J^{(m+1)}\,\to \,F_J^{(m)} \eeq in the singular limits.

For processes where one or both incoming particles are leptons, the
collinear counterterms are set to zero and the parton distribution
functions are replaced by $\delta$-distributions, \ie
$f_{i/I}^{\text{ew}}\,=\,\delta(1-\eta_{i})$.\footnote{This of course
  only holds for the discussion of higher-order effects from strong
  interactions; for electroweak processes, the inclusion of structure
  functions can not be neglected; \cf \cite{Kalinowski:2008fk} and
  references therein.}

In the following, we discuss the specific form of
$d\sigma^{A}_{ab}(p_a,p_b)$ which corresponds to the subtraction term
in the real-emission contribution of the process, as well as the
integrated $d$-dimensional counterterm
$\int_1d\sigma^{A}_{ab}(p_a,p_b)$. In general, the subtraction term
can be split in maximally four contributions for processes with
maximally two final-state particles in the leading-order contribution;
in our scheme, each of these contributions requires exactly one
momentum mapping. We then have
\begin{\eqn}\label{eq:master_sub}
d\sigma^{A}_{ab}(\ph_a,\ph_b)\,=\,d\sigma^{A,a}_{ab}(\ph_a,\ph_b)+d\sigma^{A,b}_{ab}(\ph_a,\ph_b)
+d\sigma^{A,k_{1}}_{ab}(\ph_a,\ph_b)+d\sigma^{A,k_{2}}_{ab}(\ph_a,\ph_b)\,,
\end{\eqn}
where $k_{1,2}$ now label the momenta of the outgoing particles in
$m$-particle phase space. The associated real-emission contributions
can be obtained from considering possible splittings in flavour space:
the sum over flavours in each of the above contributions is such that
$(f_{a},f_{k_{i}})\,=(\hat{f}_{a}+\hat{f}_{k_{j}},\hat{f}_{k_{i}})$
for all combinations, \ie we only consider flavour mapping where
$f_{a}\,=\,\hat{f}_{a}+\hat{f}_{k_{j}}$ is physically allowed. We will
denote this by a delta function in flavour-space mapping
$\delta_{a;\hat{a},\hat{j}}$. From initial-state splittings, we then
have for the single contribution with fixed initial and final-state
flavours (in the following, we omit the jet functions for notational
reasons; however, full expressions should always be read according to
\eq (\ref{eq:withf}) where all jet functions are included):
\begin{eqnarray}\label{eq:counter_ini}
d\sigma^{A,a}_{ab}(\hat{p}_a,\hat{p}_b)&=&\frac{N_{m+1}}{\Phi_{m+1}}
\sum_{i=1,2,3}\Bigg\{
\left[\D_{gqq}(\ph_{i})\delta_{g;q,q_{i}}+\D_{ggg}(\ph_{i})
\delta_{g;g,g_{i}}\,\right]|\M_{\text{Born},g}|^{2}(x\,\ph_{a},p_{b};p_{n})\nn\\
&&\,+\left[ \D_{qgq}(\ph_{i})\delta_{q;g,q_{i}}\,+\,\D_{qqg}(\ph_{i})
\delta_{q;q,g_{i}}(\ph_{i})\right]|\M_{\text{Born},q}|^{2}(x\,\ph_{a},p_{b};p_{n})\Bigg\},
\end{eqnarray}
where $N_{m+1}$ incorporates all symmetry factors of the $m+1$ process
and $\Phi_{m+1}\,=\,2\,\hat{s}$ is the respective flux factor.
$D(\ph_{i})$ now signifies that $\ph_{j}\,=\,\ph_{i}$, and
$\M_{\text{Born},g(q)}$ corresponds to the underlying Born matrix
element for the process $p_{a}+p_{b}\,\rightarrow\,\sum_{n} p_{n} $
with an incoming gluon (quark) such that $f_{a}=g\,(q)$ and all other
flavours unchanged, and where $n\,\neq\,i$ labels the momenta of the
final state particles. The momentum mapping is here given by \eqs
(\ref{eq:eta_aFS}) and (\ref{eq:ini_rest_trafo}) respectively. Note
that, while $\D_{gqq}$ and $\D_{qgq}$ contain collinear singularities
only, $\D_{qqg}$ and $\D_{ggg}$ are split into a collinear and soft
and interference term as discussed in Section \ref{sec:soft}:
\begin{\eqn}
\D_{qqg}(\ph_{i})\,=\,\D^{\text{coll}}_{qqg}(\ph_{i})\,+\,\D^{\text{if}}(\ph_{i},\hat{p}_{b})\delta_{\hat{f}_{b},g}\,+\,\sum_{k\,\neq\,i}\D^{\text{if}}(\ph_{i},\ph_{k})\delta_{f_{k},g},
\end{\eqn}
where $D(\ph_{i},\ph_{k})$ now denotes an interference contribution
where $\ph_{k}$ acts as a spectator. An equivalent expression holds
for $\D_{ggg}(\ph_{i})$.  The second term sums over all final-state
particles with $k\,\neq\,i$ which are gluons.
$d\sigma^{A,b}_{ab}(\hat{p}_a,\hat{p}_b)$ is obtained from
$d\sigma^{A,a}_{ab}(\hat{p}_a,\hat{p}_b)$ by interchanging
$a\,\leftrightarrow\,b$ in the corresponding expressions as well as in
the flavour mapping functions. All subtraction terms given here refer
to the expressions in Section \ref{sec:inisq}, and to the expressions
for initial-initial interference terms for
$\D^{\text{if}}(\ph_{i},\hat{p}_{b})$ and initial-final interference
terms for $\D^{\text{if}}(\ph_{i},\hat{p}_{k})$ in Section
\ref{sec:soft}.

For the final-state splittings, we have a similar expression
\begin{eqnarray}\label{eq:counter_fin}
d\sigma^{A,k_{1}}_{ab}(\hat{p}_a,\hat{p}_b)&=&\frac{N_{m+1}}{\Phi_{m+1}}
\sum_{i=1,2,3}\Bigg\{
\left[\D_{gqq}(\ph_{i})\delta_{g;q,q_{i}}+\D_{ggg}(\ph_{i})
\delta_{g;g,g_{i}}\,\right]|\M_{\text{Born},g}|^{2}(p_{a},p_{b};p_{n})\nn\\
&&\,+\left[ \D_{qqg}(\ph_{i})\delta_{q;g,q_{i}}\,+\,\D_{qqg}(\ph_{i})
\delta_{q;q,g_{i}}(\ph_{i})\right]|\M_{\text{Born},q}|^{2}(p_{a},p_{b};p_{n})\Bigg\}\,,
\end{eqnarray}
where now the delta functions in flavour space are defined as
$\delta_{k_{i};\ph_{i},\ph_{j}}$, and $\M_{\text{Born},g(q)}$
corresponds to the underlying Born matrix element for the process
$p_{a}+p_{b}\,\rightarrow\, p_{n}+p_{k_{i}} $, where now
$f_{k_{i}}=g\,(q)$ and all other flavours unchanged, and where
$n\,\neq\,k_{i}$ labels the momentum of the remaining final-state
particle. As before, $\D_{qqg}$ and $\D_{ggg}$ contain both collinear
and interference terms, and we have
\begin{\eqn}
\D_{qqg}(\ph_{i})\,=\,\D^{\text{coll}}_{qqg}(\ph_{i})\,+\sum_{k=a,b}\,\D^{\text{if}}(\ph_{i},\hat{p}_{k})\delta_{\hat{f}_{k},g}\,+\,\sum_{k\,\neq\,i}\D^{\text{if}}(\ph_{i},\ph_{k})\delta_{f_{k},g}
\end{\eqn}
and the same for $\D_{ggg}$. The mapping needed to define the
subtraction terms is given by \eqs (\ref{eq:fin_map}) and
(\ref{pLambdahatp}). All subtraction terms given here refer to the
expressions in Section \ref{sec:finsq}, and the expressions for the
final-initial interference terms for
$\D^{\text{if}}(\ph_{i},\hat{p}_{a,b})$ or the final-final
interference terms for $\D^{\text{if}}(\ph_{i},\hat{p}_{k})$ in
Section \ref{sec:soft}.

The integrated counterterms are generically given as 
\begin{\eqn}
\int_{1}d\sigma^{A}_{ab}(\ph_a,\ph_b)\,=\,\int_{1}d\sigma^{A,a}_{ab}(\ph_a,\ph_b)+\int_{1}d\sigma^{A,b}_{ab}(\ph_a,\ph_b)+\int_{1}d\sigma^{A,k_{1}}_{ab}(\ph_a,\ph_b)+\int_{1}d\sigma^{A,k_{2}}_{ab}(\ph_a,\ph_b)\,.
\end{\eqn}
The collection of the integrated counterterms is then straightforward:
for each term which has been subtracted in the real-emission part, the
corresponding integrated contribution to $I,\,K,\,P$ needs to be added
to the virtual contribution as in \eq (\ref{eq:sig_nlo}).  Finally,
note that the expressions given in the following sections are derived
on the matrix-element level:
\begin{\eqn}
\int_{1}\,|\M|^{2}_{m+1}\,\rightarrow\,\int_{1} \D |\M|^{2}_{m}\,=\,\mathcal{V}|\M|^{2}_{m};
\end{\eqn}
on the cross-section level, we additionally have to take the flux and
combinatoric factors into account, such that
\begin{eqnarray}
\int_{1}d\sigma^{A}_{m+1;ab}(\ph_a,\ph_b)&=&\frac{N_{m+1}}{2\hat{s}}\,
\int_{1}\D|\M|^{2}_{m}\,=\,\frac{N_{m+1}}{2\hat{s}}\mathcal{V}|\M|^{2}_{m},\nn\\
\int_{m}\int_{1}d\sigma^{A}_{m+!;ab}&=&N_{m+1}\,\int_{m}\frac{1}{2\hat{s}}\mathcal{V}|\M|^{2}_{m}\,=\,
\frac{N_{m+1}}{N_{m}}\,x\,\mathcal{V}\int_{m}d\sigma_{m}\,.
\end{eqnarray}
Here the factors $N_{m},\,N_{m+1}$ account for possible symmetry
factors of the specific process. We then have the relation
\begin{\eqn}\label{eq:v_ikp}
\sum\,\mathcal{V}\,=\,\frac{1}{x}\,\frac{N_{m}}{N_{m+1}}\lb
I\,+\,K\,+\,P \rb
\end{\eqn}
between the integrated splitting functions $\mathcal{V}$ derived in
the previous section and the insertion operators $I,K,P$ as defined in
\eq (\ref{IKP2842010}).

\section{Application to physical processes}\label{sec:appl}
In the last section, we have presented the subtraction terms for the
real emission and their integrated counterparts for processes with up
to two particles in the final state. We will now use these expressions
in well-known processes at NLO, showing that indeed the singularity
structures of the real-emission terms and one-loop contributions is
reproduced by the squared and averaged (one-particle integrated)
splitting functions of the parton shower \cite{Nagy:2007ty} in the
singular limits. We validate our scheme by showing that the
application of our subtraction terms reproduces standard results from
the literature for all processes. In more detail, we discuss single-W
production (initial-state $qqg$ and $qgq$ splittings, initial-initial
state interference term), dijet production at lepton colliders ($qqg$
final-state splittings, final-final state interference term),
gluon-induced Higgs production in an effective theory description
($ggg$ and $gq\bar{q}$ initial-state splitting functions,
initial-initial state interference terms), Higgs decay to two gluons
($ggg$ and $gq\bar{q}$ final-state splitting functions, final-final
state interference terms) and a subprocess of deep-inelastic
scattering (DIS) ($qqg$ initial and final-state splitting functions,
initial-final and final-initial state interference terms), \cf Table
\ref{tab:funlist}. We find that the splitting functions for the parton
shower as described in \cite{Nagy:2007ty,Nagy:2008ns,Nagy:2008eq}, in
combination with the momentum mapping, are well suited as subtraction
terms in NLO-QCD calculations. The numerical evaluations of
phase-space integrals in this section have been obtained using
routines from the Cuba library \cite{Hahn:2004fe}.
\begin{table}
\begin{center}
\begin{tabular}{c||c}
Splitting function&Process\\ \hline\hline
&\\
gqq, initial state& Higgs production\\
qqg, initial state& single W, DIS\\
qgq, initial state& single W\\
ggg, initial state&Higgs production\\
initial initial interference&single W, Higgs production\\
initial final interference&DIS\\
&\\
\hline\\
gqq, final state& Higgs decay\\
qqg, final state& Dijet, DIS\\
ggg, final state&Higgs decay\\
final final interference&Higgs decay, Dijet\\
final initial interference&DIS\\
\end{tabular}
\end{center}
\caption{\label{tab:funlist} List of all splitting functions presented
  in Section \ref{sec:subtr} and test processes used for the scheme validation in Section \ref{sec:appl}. }
\end{table}

\subsection{Single-W production}\label{sec:single_W}
We start with a simple process: single-W production at a hadron
collider. The tree-level process here is given by
\begin{\eqn*}
q(p_{1})\,\bar{q}(p_{2})\,\longrightarrow\,W,
\end{\eqn*}
and real-emission processes include both quark- and gluon-induced
cases
\begin{\eqn*}
q(\ph_{1})\,\bar{q}(\ph_{2})\,\longrightarrow\,W\,g(\ph_{3}),\;g(\ph_{1})\,q(\ph_{2})\,\longrightarrow\,W\,q(\ph_{3}).
\end{\eqn*} 
This process contains both the $q\,\bar{q}\,g$ as well as
$g\,q\,\bar{q}$ initial state splittings given by \eqs
(\ref{eq:dqqg_ini}), (\ref{eq:dqgq_ini}) as well as the initial state
interference term \eq (\ref{eq:Dif_iniini}). All results for
tree-level and real-emission matrix elements are well known and have
been taken from the literature (\eg \cite{Ellis:1991qj}).
\subsubsection*{Tree-level contribution}
The squared matrix element for the tree-level process $q \bar q^\prime
\to W$ is
\beq  
\mid{\cal M}_B\mid^2=\frac{g^2}{12}\mid
V_{qq^\prime}\mid^2M_W^2\nn \eeq 
where we have averaged over initial-state particle spins and colours.
The one-particle phase space is given by
\bea 
\int dPS_1&=&2\pi\delta^{+}(s-M_W^2)\nn
\eea
with $s$ being the partonic center-of-mass energy. 
\subsubsection*{Virtual correction}
The virtual contribution in the
$\overline {\textrm{MS}} $ renormalisation scheme is 
\bea 
2\,\text{Re}\,\lb \M_{B}\M^{*}_{\text{virt}}\rb&\equiv &\mid {\cal M}_V\mid^2\nn\\
&&\hspace*{-25mm}=\, \mid {\cal M}_B\mid^2 \frac{\alpha_s}{2\pi}C_F\frac{1}{\Gamma(1-\vareps)}
\left (\frac{4\pi\mu^2}{  Q^2}  \right)^\vareps 
\left\{-\frac{2}{\vareps^2}-\frac{3}{\vareps}-8+\pi^2
+ {\cal O}(\vareps) \right\}\,,
\eea
and we have
\begin{\eqn}\label{eq:sigvirt_W}
\sigma^{\text{virt}}\,=\,\frac{\alpha_s}{2\pi}C_F\frac{1}{\Gamma(1-\vareps)}
\left (\frac{4\pi\mu^2}{  Q^2}  \right)^\vareps 
\left\{-\frac{2}{\vareps^2}-\frac{3}{\vareps}-8+\pi^2\right\}\,\sigma^{\text{LO}}.
\end{\eqn}
\subsubsection*{Real emission}
The matrix element for the quark-induced NLO real-emission process 
$q\bar q^\prime\to Wg$ is given by
\bea  
\mid{\cal M}^{q}_R\mid^2&=&\pi\as \sqrt{2}G_FM_W^2\mid V_{qq^\prime}\mid^2
\frac{32}{9}\,\frac{\hat t^2+\hat u^2+2M_W^2\hat s}{\hat t\hat u}\nn\\
&=&\frac{8}{9}g^2\pi\as\mid V_{qq^\prime}\mid^2 \frac{\hat t^2+\hat u^2+2M_W^2\hat s}{\hat t\hat u}\,,
\eea
where $g$ and $M_W$ are related to the Fermi coupling constant $G_F$
by $\frac{G_F}{\sqrt{2}}=\frac{g^2}{8M_W^2}$, and we use
$\hat{s}\,=\,(\ph_{1}+\ph_{2})^{2},\,\hat{t}\,=\,(\ph_{1}-\ph_{3})^{2},
\hat{s}+\hat{t}\,+\hat{u}\,=\,m_{W}^{2}$.  The matrix element for the
gluon-induced process can be obtained by crossing symmetry
\bea  
\mid{\cal M}^{g}_R\mid^2
&=&\frac{1}{3}g^2\pi\as\mid V_{qq^\pr}\mid^2 
\frac{\hat s^2+\hat u^2+2M_W^2\hat t}{- \hat s\hat u}\,.\nn
\eea
The two-particle phase space is 
 \beq  
 \int
dPS_2=\int \frac{d^3\bf{\ph}_{W} }{(2\pi)^3 2 \ph_W^{0}}\frac{d^3\bf{\ph}_{3}}{(2\pi)^3 2 \ph_3^{0}}(2\pi)^4
\delta^{(4)}(\hat{Q}-\ph_{W}-\ph_{3}). \nn \eeq

\subsubsection*{Subtraction terms}
In the quark-induced case, we need to consider two subtraction terms
corresponding to the gluon emission from $q$ and $\bar{q}$,
respectively, which we label ${\cal D}^{q}_1$ and ${\cal D}^{q}_2$.
Their definition is given in \eqs(\ref{eq:dqqg_ini}) and
Eq.(\ref{eq:Dif_iniini}). We have
\bea\label{qqWg} 
{\cal D}^{q}_1\,=\,\D_{qqg}+\D^{\text{if}}_{ab}&=&\frac{4\,\pi\,\al_{s}}{2}\,C_{F}\,\left[ \frac{4\,\hat{u}}{\hat{t}M^{2}_{W}}\,+\,\frac{8\,\hat{u}\hat{s}}{\hat{t}\,(\hat{u}+\hat{t})^{2}}
+\frac{16 \hat s\hat u^2}{(\hat t^2+\hat u^2)(\hat t+\hat u)^2}\right]
\eea
where we consider the case where $\ph_{\ell}\,=\ph_{1}$, \ie gluon
emission from the incoming quark.  The subtraction term ${\cal
  D}^{q}_2$ can be obtained from \eq(\ref{qqWg}) by the replacement
$\hat t \leftrightarrow \hat u$. As the Born matrix element is
constant and we use a unit jet function, no momentum mapping is
required.

After subtraction of the counterterms, the final expression for
the two-particle cross section at parton level is given by
\bea 
\label{eq:NLO2_q_W}
\sigma^{\text{\text{NLO}}\,\{2\}}_{q}
&=&\frac{1}{2\hat s}\int dPS_2 \left\{\mid{\cal M}^{q}_R\mid^2 -\left({\cal D}^{q}_1+{\cal D}^{q}_2\right)
\mid{\cal M}_B\mid^2\right\}\,=\,0.
\eea
The splitting function in the gluon-induced case $\D_{g}$ is given by \eq (\ref{eq:dqgq_ini}), and we obtain
\bea  \label{eq:NLO2_g_W}
\sigma^{\text{NLO}\,\{2\}}_{g}&=&\int_2\,\,\left[d\sigma^R_{\vareps=0}-d\sigma^A_{\vareps=0}\right]\nn\\
&=&\frac{1}{2\hat s}\int dPS_2 \left\{\mid{\cal M}^{g}_R\mid^2 -
\D_{g}\mid{\cal M}_B\mid^2\right\}\nn\\
&=&\frac{1}{2\hat s}\int dPS_2 \left\{-\frac{1}{3}g^2\pi\as \frac{(2\hat t+\hat u)}{\hat s}\right\}\,.
\eea
\subsubsection*{Integrated subtraction terms}
In the $m$-particle phase-space contribution, we now have to consider
the contributions from both quark- and gluon-induced real emission
processes; note, however, that the poles from the virtual contribution
are completely cancelled by the integrated $q\,\bar{q}\,g$ splitting
function. In this case, the corresponding integrated subtraction terms
are obtained from \eqs (\ref{eq:vqqg_ini}), (\ref{eq:Vif_iniini}).

The collinear singularity needs to be cancelled by the universal
collinear counter term as explained in Section \ref{sec:finform}. We
then obtain for the quark-induced case
\beq 
\begin{split}
\label{eq:NLO1_q_W}
\int_{1}d\sigma^{\text{NLO}\{1\}}_{q}&=\,  \int_{1}d\sigma^B_{ab}  \,( 
2\,\mathcal{V}_{qqg})
+\int_1d\sigma^C_{ab} \\
&=
\int_1  d\sigma_{ab}^{B}(p_a,p_b) \otimes {\bf I}_{q}(\vareps) 
+ \int_0^1 dx \int_1 d\sigma_{ab}^{B}(xp_a,p_b)\otimes 
 \left[ {\bf K}_{q}^a(xp_a) +   
{\bf P}_{q}(x,\mu_F^2)\right]\\
&+(a\leftrightarrow b)
 \end{split}
 \eeq
The corresponding ${\bf I}$, ${\bf K}$ and ${\bf P}$ terms are 
\bea  
{\bf
I}_{q}(\vareps)&=&\frac{\as}{2\pi}C_F\frac{1}{\Gamma(1-\vareps)}
\left(\frac{4\pi\mu^2}{Q^2}\right)^\vareps
\left(
\frac{2}{\vareps^2}+ \frac{3}{\vareps}- \frac{\pi^2}{3}+{\cal O}(\vareps) 
\right)  \nn\\
{\bf K}_{q}^a(xp_a)&=& \frac{\as}{2\pi}C_F\frac{1}{\Gamma(1-\vareps)}
\left[-(1-x)\ln x+2(1-x)\ln(1-x) + 4x\left(\frac{\ln (1-x)}{1-x}\right)_+
\right.\nn\\
 &-& \left.
 \frac{2x\ln x}{\left(1-x\right)_+}
-\left(\frac{1+x^2}{1-x}\right)_+
 \ln\left(\frac{4\pi\mu^2}{2xp_a\cdot p_b}\right)  \right]\nn\\
{\bf P}_{q}(x,\mu_F^2) &=&
\frac{\as}{2\pi}C_F\frac{1}{\Gamma(1-\vareps)}
\left(\frac{1+x^2}{1-x}\right)_+\ln
\left(\frac{4\pi\mu^2}{\mu_F^2}\right)\nn \eea
We immediately see that the singularities in $\mid {\cal M}_V\mid^2$
and $ {\bf I}(\vareps)$ cancel.

The integrated subtraction term for the gluon-induced case is given by
\eq (\ref{eq:vqgq_ini}), and combined with the corresponding collinear
counterterm, we have
\bea 
\begin{split}
\label{eq:NLO1_g_W}
\int_{1}d\sigma^{\text{NLO}\{1\}}_{g}&=\,\int_1d\sigma_{ab}^B \mathcal{V}_{qgq} +
 \int_1d\sigma_{ab}^C\\
& =\int_0^1 dx \int_1 d\sigma^B_{ab}(xp_a,p_b)\otimes
\left[{\bf K}_{g}(xp_a)+{\bf P}_{g}(x,\mu_F^2) \right]  
\end{split}
\eea
where
\bea 
 {\bf K}_{g}(xp_a)&=& \frac{\as}{2\pi}\frac{1}{\Gamma(1-\vareps)}
\left\{\frac{}{}T_R( 6x-5x^2-1)+ \left[2\ln(1-x)-\ln(x)\right]P^{gq}(x)
\right. \nn\\
&-&\left.  P^{gq}(x)
\ln\left(\frac{4\pi\mu^2}{2xp_a\cdot p_b}\right)
 \right\},\nn \\
 {\bf P}_{g}(x,\mu_F^2) &=&
\frac{\as}{2\pi}\frac{1}{\Gamma(1-\vareps)}
P^{gq}(x)
\ln\left(\frac{4\pi\mu^2}{\mu_F^2}\right),\nn \\
P^{gq}(x)&=& T_R\left[x^2+(1-x)^2\right], \qquad T_R=\frac{1}{2}\,.\nn\eea

\subsubsection*{Results}
\begin{figure}
\centering
\includegraphics[width=3.5in]{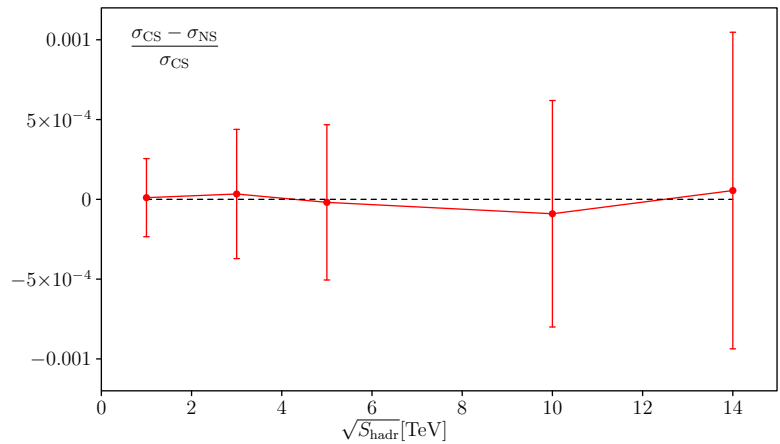}
\vspace{5mm}
\caption{\label{fig:single_w_diff} Relative difference between NLO
  corrections to single-W production using the Catani-Seymour (CS) and
  our subtraction scheme based on Nagy-Soper splitting functions (NS),
  as a function of the hadronic center of mass energy. The results
  agree at the sub-per-mil level, shown are the numerical integration
  errors. }
\end{figure}

The complete partonic cross section for the single-$W$ production at
NLO can then be obtained by combining \eqs (\ref{eq:sigvirt_W}),
(\ref{eq:NLO1_q_W}), (\ref{eq:NLO1_g_W}), (\ref{eq:NLO2_q_W}), and
(\ref{eq:NLO2_g_W}):
\begin{\eqn*}
\sigma^{\text{NLO}}\,=\,\sigma^{\text{virt}}\,+\,\sigma^{\text{NLO}\{1\}}_{q}\,+\,\sigma^{\text{NLO}\{1\}}_{g}\,+\,\sigma^{\text{NLO}\{2\}}_{q}\,+\,\sigma^{\text{NLO}\{2\}}_{g}\,.
\end{\eqn*}
Comparison with a calculation in the Catani-Seymour scheme reveals
that in the two schemes the finite terms are attributed to different
parts, but the complete contributions of course agree.  For a
comparison, we implemented both schemes in a private code; \Fig
\ref{fig:single_w_diff} shows the relative difference between the two
schemes as a function of the hadronic center of mass energy,
convoluted with parton distribution functions. The schemes are
equivalent, with agreement which is consistent with zero on sub-per-mil
level.
\begin{figure}
\centering
\includegraphics[width=3.5in]{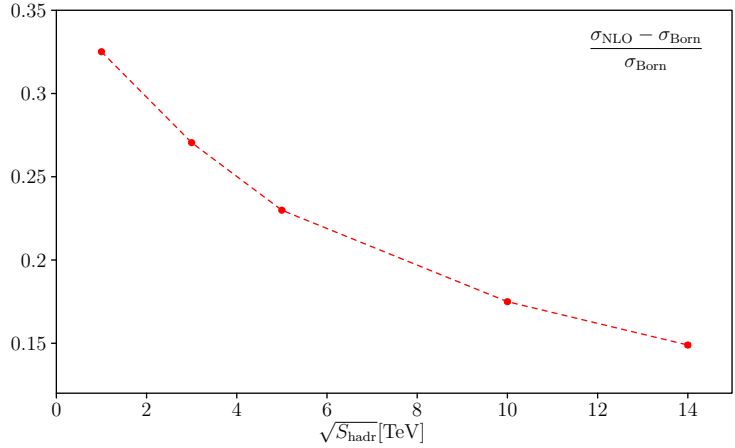}
\vspace{5mm}
\caption{\label{fig:single_w} Relative NLO correction
  $(\sigma_\text{NLO}-\sigma_\text{Born})/\sigma_\text{Born}$ to
  single-W production at the LHC as a function of the hadron center of
  mass energy. The result was obtained using the CTEQ6M parton
  distribution function~\cite{Pumplin:2002vw}.}
\end{figure} 

\newpage

\subsection{Dijet production }
We next consider dijet production at NLO. In this process, the
subtraction terms are final state $q\,q\,g$ splitting and the
final-final interference terms, given by \eqs (\ref{eq:dqqg_fin}),
(\ref{eq:Dif_fin}). The integrated counterparts can be obtained from
\eqs (\ref{eq:vqqg_fin}), (\ref{eq:vif_finfin}).  We here closely
follow the procedure in \cite{Catani:1996vz}, \ie we calculate the
prefactor which relates the NLO and LO cross section via
\bea
\sigma^{\text{NLO}}&=&\frac{3}{4} \frac{\as}{\pi}C_F \sigma^{\text{LO}}\,.
\eea
We average over the event orientation in the leading-order process; in
this case, the momentum dependence of the Born contribution vanishes.
\subsubsection*{Tree-level contribution}
We normalise the leading-order matrix element such that the phase space is given by
\bea 
\int dPS_2&=&\int dy\,\delta(1-y), \qquad y=\frac{2p_1\cdot p_2}{Q^2}\nn\\
\sigma^{\text{LO}}&=& \mid {\cal M}_2\mid^2 \int dy\,\delta(1-y)   F_J^{(2)}\nn
\eea
and $p_1, p_{2}$ are the momenta of the final state partons.
\subsubsection*{Virtual contribution}
The one-loop matrix element in the $\overline {\textrm{MS}} $
renormalisation scheme is given by
\bea 
2\,\text{Re}\lb\M_{2}\M_{\text{virt}}^{*} \rb & \equiv & |\M_{V}|^{2}\nn \\&=&\mid {\cal M}_2\mid^2 \frac{\alpha_s}{2\pi}C_F\frac{1}{\Gamma(1-\vareps)}
\left (\frac{4\pi\mu^2}{  Q^2}  \right)^\vareps 
\left\{-\frac{2}{\vareps^2}-\frac{3}{\vareps}-8+\pi^2
+ {\cal O}(\vareps) \right\}\nn
\eea
and we have
\begin{\eqn}\label{eq:sigvirt_2jets}
d\sigma^{V}\,=\,\frac{\alpha_s}{2\pi}C_F\frac{1}{\Gamma(1-\vareps)}
\left (\frac{4\pi\mu^2}{  Q^2}  \right)^\vareps 
\left\{-\frac{2}{\vareps^2}-\frac{3}{\vareps}-8+\pi^2\right\}\,d\sigma^{\text{LO}}\,.
\end{\eqn}
\subsubsection*{Real emission}
The matrix element for the NLO real-emission process 
\begin{\eqn*}
  e^+e^- \to q (\ph_{1})\,\bar q (\ph_{2})\, g(\ph_{3})
\end{\eqn*}
 in four dimensions is 
given by
\bea
\mid{\cal M}_3\mid^2&=&C_F \frac{8\pi\as}{Q^2}
\frac{x_1^2+x_2^2}{(1-x_1)(1-x_2)}
\mid{\cal M}_2\mid^2 \nn \,,
\eea
where we have used $x_i=\frac{2\hat p_i\cdot \hat Q}{\hat Q^2}$ and
$\mid{\cal M}_2\mid^2$ is the LO matrix element for the process $
e^+e^- \to q+\bar q $.  For the phase-space integration, we use
\bea
\int  dPS_3&=&\frac{Q^2}{16\pi^2}\int_0^1 dx_1dx_2 \,\Theta(x_1+x_2-1). \nn 
\eea
\subsubsection*{Subtraction terms}
The calculation of the subtracted splitting function contains two
contributions, ${\cal D}_1$ and ${\cal D}_2$, which correspond to
emission from the quark and the anti-quark, respectively. They are
given by \eqs (\ref{eq:qqgfinsq}) and (\ref{eq:Dif_fin}). We obtain
\bea\label{D1}
{\cal D}_1&=&\frac{4}{\hat Q^2} \left\{ \left(\frac{1}{x_2}\right)
\left[2\left(\frac{x_1}{x_{3}}- \frac{1-x_2}{x_{3}^2}\right)+\frac{1-x_1}{1-x_2}\right] \right.\nn\\
&&+\left.\,2 \left( \frac{1-x_{3}}{1-x_2} \right)\frac{x_1}{(1-x_1)x_1+(1-x_2)x_2} \right\} \,,
\eea
where we have used $\hat Q=\hat p_1 +\hat p_2+\hat p_3=Q,\,
x_1+x_2+x_3=2 $, and $Q^{2}\,=\,\hat Q^2$ is the square of the
center-of-mass energy. The contribution ${\cal D}_2$ can be obtained
from \eq(\ref{D1}) by the replacement $x_1 \leftrightarrow x_2$. The
final expression for the three-parton cross section is then given by
\bea 
\lefteqn{\sigma^{NLO\,\{3\}}\,=\,\int_3\,\,\left[d\sigma^R_{\vareps=0}-d\sigma^A_{\vareps=0}\right]}
\nn\\
&=\!\!\!\!&\int dPS_3\left\{\mid{\cal M}_3\mid^2 F_J^{(3)}-
\left(\frac{4\pi\as}{2}\right) C_F \left({\cal D}_1F_J^{(2)}(1)+{\cal D}_2F_J^{(2)}(2)\right)
\mid{\cal M}_2\mid^2\right\}\nn\\
&=\!\!\!\!&\left(\frac{\as}{2\pi}C_F\right)\mid{\cal M}_2\mid^2 
\int_0^1 dx_1dx_2 \,\Theta(x_1+x_2-1) \left\{
\frac{x_1^2+x_2^2}{(1-x_1)(1-x_2)} F_J^{(3)}\right.\nn\\
&-\!\!\!\!&\left. 
\left[ \left(\frac{1}{x_2}\right)
\left(2\left(\frac{x_1}{x_{3}}- \frac{1-x_2}{x_{3}^2}\right)
+\frac{1-x_1}{1-x_2}\right)  \right.\right. \nn\\
&&\hspace*{10mm} +\,\left.\left. 2 \left( \frac{1-x_{3}}{1-x_2} \right)\frac{x_1}{(1-x_1)x_1+(1-x_2)x_2} \right]F_J^{(2)}(1)
\right.\nn\\
 &-\!\!\!\!&\left. 
\left[ \left(\frac{1}{x_1}\right)
\left(2\left(\frac{x_2}{x_{3}}- \frac{1-x_1}{x_{3}^2}\right)
+\frac{1-x_2}{1-x_1}\right)  \right.\right.\nn\\
&&\hspace*{10mm}\,+\, \left.\left. 2 \left( \frac{1-x_{3}}{1-x_1} \right)\frac{x_2}{(1-x_1)x_1+(1-x_2)x_2} \right]F_J^{(2)}(2)
\right\}\,, 
\eea
where $F_{J}^{(2)}(i)$ signifies that $p_{\ell}\,=\,p_{i}$ in the
corresponding mappings in the jet functions.  For any infrared-safe
observable, $F_J^{(3)}(i)\to F_J^{(2)}$ as $x_i$ approaches 1, and the
above expression is finite. For unit jet functions, it reduces to
\bea 
\label{eq:NLO3_2jets}
\sigma^{\text{NLO}\,\{3\}}&=&
\frac{\as}{2\pi}C_F \left(\frac{23}{2}-\frac{4}{3}\pi^2\right)
\sigma^{\text{LO}}\,.
\eea 

\subsubsection*{Integrated subtraction terms}
The integrated collinear and soft subtractions are given by \eqs
(\ref{eq:vqqg_fin}) and (\ref{eq:vif_finfin}):
\bea 
2 \left(\frac{4\pi\as}{2}\right)\mu^{2\vareps}C_F\int d\xi_p 
\left(v_{qqg}^2-v_{\textrm{eik}}^2\right)&=&\nn\\&& \hspace*{-40mm}\frac{\as}{2\pi}C_F
\frac{1}{\Gamma(1-\vareps)}\left(\frac{4\pi\mu^2}
{ Q^2}\right)^\vareps \left(-\frac{1}{\vareps}-14+\frac{4\pi^2}{3}
+{\cal O(\vareps)}\right)\nn\\
2 \left(\frac{4\pi\as}{2}\right)\mu^{2\vareps}C_F\int d\xi_p 
\left(v_{\textrm{eik}}^2-v_{\textrm{soft}}^2\right)&=&\nn\\ && \hspace*{-40mm}\frac{\as}{2\pi}C_F
\frac{1}{\Gamma(1-\vareps)}\left(\frac{4\pi\mu^2}
{ Q^2}\right)^\vareps 
\left(\frac{2}{\vareps^2}+\frac{4}{\vareps}+12-\pi^2+{\cal O(\vareps)}\right)\,.
\eea
Combining these two contributions with the virtual cross section, we obtain a
finite expression for the two-parton cross section:
\bea 
\label{eq:NLO2_2jets}
\sigma^{\text{NLO}\,\{2\}}&=&\int_2 \,\,\left[d\sigma^V+\int_1 d\sigma^A\right]_{\vareps=0}\nn\\ &=&\int dPS_2\left\{ \mid {\cal M}_V\mid^2+
2 \lb \mathcal{V}_{qqg}-\mathcal{V}^{\text{if}}_{12}\rb \mid {\cal M}_2\mid^2 \right\}F_J^{(2)}\nn\\
&=&\frac{\as}{2\pi}C_F
\left(-10+\frac{4\pi^2}{3}\right)\mid {\cal M}_2\mid^2 \int dy\,\delta(1-y)   F_J^{(2)}\nn\\
&=&\frac{\as}{2\pi}C_F 
\left(-10+\frac{4}{3}\pi^2\right) \sigma^{\text{LO}}\,.
\eea
\subsubsection*{Result}
Summing \eq (\ref{eq:NLO3_2jets}) and \eq (\ref{eq:NLO2_2jets}), we obtain for unit jet functions
\bea
\sigma^{\text{NLO}}&=&\sigma^{\text{NLO}\,\{2\}}+\sigma^{\text{NLO}\,\{3\}}=\frac{3}{4} \frac{\as}{\pi}C_F 
\sigma^{\text{LO}}\nn
\eea
which agrees with the standard result in the literature.  Note that,
although the angularly-averaged tree-level matrix element is
independent of the parton momenta, non-unit jet functions need to
account for the mappings from $m+1$ to $m$ phase space as specified in
Section \ref{sec:finmap}.

\subsection{Gluon-induced Higgs production }

In this section, we discuss the NLO corrections to gluon-induced Higgs
production, where the leading-order contribution has been derived in
an effective theory approach \cite{Ellis:1975ap, Shifman:1979eb,
  Vainshtein:1980ea, Voloshin:1985tc}.  This involves $ggg$ and
$gq\bar{q}$ initial-state subtraction terms given by \eqs
(\ref{eq:dggg_ini}), (\ref{eq:dgqq_ini}) as well as the
initial-initial interference term \eq (\ref{eq:Dif_iniini}) and their
integrated counterparts \eqs (\ref{eq:vggg_ini}), (\ref{eq:vgqq_ini}),
(\ref{eq:Vif_iniini}).

The real-emission contributions in this process are \beq gg\to gH,
\quad qg\to qH, \quad q\bar q\to gH,\nn \eeq such that the full NLO
QCD cross section includes
\bea
\int d\sigma^{\text{NLO}}&=&\nn\\
&&\hspace*{-20mm}\int d\sigma_{q\bar q\to gH}+\int d\sigma_{qg\to qH}+
\int d\sigma_{gg\to gH}+\int d\sigma_{V}+\int
d\sigma_{\text{coll}}+\int d\sigma_{\text{ch}}\,+\,\int
d\sigma_{\text{eff}}\,,\nn\\&& \eea where $\int
d\sigma_{\text{ch}},\;\int d\sigma_{\text{eff}}$ are contributions
from charge renormalisation and the matching of the effective to the
full theory, respectively. In the following, we use unit jet functions
throughout the whole calculation.

\subsubsection*{Tree-level contribution}
The lowest-order cross section was first presented in
\cite{Djouadi:1991tka, Dawson:1990zj}.  From the QCD Lagrangian, we
obtain for the one-loop induced $Hgg$ coupling the following
expression for the total cross section, where the $\delta$-function
accounts for the one-particle final state:
\bea \sigma_0(\hat{s})(gg\to
H)&=&\frac{\as^2}{\pi}\frac{M_H^2}{256v^2} \left|A\right|^2
\delta\left(\hat{s}-M_H^2\right), \nn\\
\left|A\right|^2&=&\left|\sum_q\tau_q
\left(1+(1-\tau_q)f(\tau_q)\right)
\right|^2,\nn\\
\tau_q&=&\frac{4M_q^2}{M_H^2},\;
\frac{G_F}{\sqrt{2}}\,=\,\frac{g^2}{8M_W^2},\quad v^2=
\frac{4M_W^2}{g^2}=\frac{1}{\sqrt{2}G_F}= \left(246
\,\textrm{GeV}\right)^2, \nn
\eea where 
 \bea
  f(\tau_q)  &=&   \left\{  \begin{array}{cl}
           \left[\sin^{-1}\left(\sqrt{1/\tau_q} \right)\right]^2  & \quad\textrm{if}\quad
           \tau_q\geq 1 \\
          -\frac{1}{4} \left[\ln\left(\frac{1+\sqrt{1-\tau_q}}{1-\sqrt{1-\tau_q}}
          \right)-i\pi\right]^2  &\quad\textrm{if}\quad \tau_q <1\nn \\
                 \end{array}\right.
\eea 
and $\hat{s}$ is the partonic center-of-mass energy, $M_q$ the pole
mass of the heavy quark (which we assume to be the top quark), and
$M_H$ the Higgs boson mass.  In the limit that the top-quark mass is
infinitely large, $\tau_q\to\infty$, $A\to\frac 2 3 $ and
\bea\label{eq:sig04_ggH}
\sigma_0(\hat{s})(gg\to H)\to\frac{\as^2}{\pi}\frac{M_H^2}{576v^2}
\delta\left(\hat{s}-M_H^2\right).
\eea
Note that the cross section is here given in $d\,=\,4$ dimensions; the
$d\,=\,4-2\,\vareps$ dimensional cross section is related by
$\sigma_{0}^{(d)}\,=\,{\sigma_{0}}/{(1-\vareps)}$.

The cross section to ${\cal O}(\as^3)$ for $gg\to H$ in the limit
$M_{t}\gg M_H$ can be obtained from the effective Lagrangian
\bea
{\cal L}_{\textrm{eff}}=
\frac{\as}{12\pi} G_{\mu\nu}^AG^{\mu\nu A}\ln
\left(1+\frac{H}{v}\right)=\frac{\as}{12\pi v}HG_{\mu\nu}^AG^{\mu\nu A}+\cds\nn
\eea
where $G_{\mu\nu}^A$ is the gluon field strength tensor and the index
$A$ is the colour degree of freedom of the gluon field (which runs
over $1 \cdots 8$). All processes considered here and the
corresponding Feynman rules have been obtained using this effective
Lagrangian.

\subsubsection*{Virtual matrix element, charge renormalisation, and
  effective Lagrangian correction}
The one-loop matrix element in $d=4-2\vareps$ dimensions in the
$\overline {\textrm{MS}} $ renormalisation scheme is given by
\bea\label{eq:Hgg_virt}
2\,\text{Re}\lb\M_{\text{LO}}\M_{\text{virt}}^{*}\rb & \equiv & \left|{\cal M}_V\right|^2\nn\\
&=&\hspace*{-30mm}\left|{\cal M}_{\text{LO}}\right|^2
\frac{\as}{2\pi}C_A  \left(\frac{4\pi\mu^2}{M_H^2}\right )^\vareps \Gamma(1+\vareps)
\left(-\frac{1}{\vareps^2}+\frac{2}{3}\pi^2+{\cal O}(\vareps)\right )\times 2\,, 
\eea
where
\beq
\left|{\cal M}_{\text{LO}}\right|^2=\frac{\as^2M_H^4}{576\pi^2v^2}\frac{1}{\left(1-\vareps\right)}\nn
\eeq
is the $d$-dimensional leading-order squared matrix element.
For the total cross section
\bea
\sigma_{\text{virt}}\,=\,\sigma_{0}\,\frac{\as}{\pi}\,\lb \frac{4\,\pi\mu^{2}}{M_{H}^{2}} \rb^{\vareps}\,\Gamma\lb 1+\vareps \rb\,C_{A}\left[-\frac{1}{\vareps^{2}}\,-\frac{1}{\vareps}\,-\,1+\,\frac{2}{3}\pi^{2} \right],
\eea
where now $\sigma_{0}$ denotes the four-dimensional LO cross section
as given in \eq (\ref{eq:sig04_ggH}).  We also need to take charge
renormalisation into account; the charge counterterm in the $\msbar$
renormalisation scheme is (see \eg
\cite{Altarelli:2002wg,Ellis:1988vi})
\beq\label{eq:ggh_charge}
\sigma_{\textrm{ch}} \, =\, \left(4\,Z_g\right)\, \sigma^{(d)}_0
\eeq
where
\bea\label{eq:ggh_charge2}
Z_g&=&-\frac{\as}{2\,\vareps}\, \left(\frac{4\pi\mu^2}{\mu_F^2}\right)^\vareps\, b_0\, \Gamma(1+\vareps) \,
 \left(\frac{\mu_F^2}{\mu^2}\right)^\vareps ,         \nn\\
b_0&=&\frac{1}{2\,\pi} \, \left(\frac{11}{6}\,C_A-\frac{2}{3}\,n_f\, T_R\right)\,=\,\frac{1}{2\,\pi}\beta_{0}
\eea
and $n_f$ is the number of light quarks.  Finally, we have to add the
${\cal O}(\as)$ correction to the matching coefficient which relates
the effective Lagrangian to the full theory,
\bea
\sigma_{\text{eff}}\,=\,\sigma_{0}\,\frac{\as}{\pi}\,\frac{11}{2}\,.
\eea
\subsubsection*{Real emission}
The real-emission contributions relevant for this process are given by
\beq
 g(\ph_{a})g(\ph_{b})\to g(\ph_{1})H, \quad q(\ph_{a})g(\ph_{b})\to q(\ph_{1})H, \quad q(\ph_{a})\bar q(\ph_{b})\to g(\ph_{1})H\nn, 
\eeq
and the Mandelstam variables as defined as
\begin{\eqn*}
\hat{s}\,=\,\lb\ph_{a}+\ph_{b} \rb^{2},\;\hat{t}\,=\,\lb\ph_{a}-\ph_{1}\rb^{2},\;\hat{u}\,=\,M^{2}_{H}-\hat{s}-\hat{t}\,.
\end{\eqn*}
Pole cancellations occur between the purely gluon-induced real
emission process and the virtual contribution; all other processes
contain at most collinear divergences which are eliminated by the
addition of collinear counterterms.

For the real-emission matrix elements, we obtain for NLO $ gg \to gH $
in four dimensions
\bea \left|{\cal M}(gg\to
gH)\right|^2=\frac{\as^3}{v^2}\frac{32}{3\pi} \frac{M_H^8+\hat
s^4+\hat t^4+\hat u^4}{\hat s\hat t\hat u},\nn 
\eea 
which has singularities when $\hat t\to 0$ or $ \hat u\to 0$. The spin
and colour averages yield an additional factor $1/2/(1-\vareps) \times
1/2/(1-\vareps) \times 1/8 \times 1/8=1/256/(1-\vareps)^2$.

The quark-gluon induced contribution is given by
\bea \left|{\cal M}(qg\to
qH)\right|^2=-\frac{16}{9}\frac{\as^3}{\pi v^2} \frac{\hat
s^2+\hat u^2}{\hat t},\nn 
\eea 
which has a singularity when $\hat t\to 0$. The spin and colour
averages yield an additional factor $1/2 \times 1/2/(1-\vareps) \times
1/3 \times 1/8=1/96/(1-\vareps)$.

The purely quark-induced matrix element can be obtained from this by
crossing symmetry, and we have
\bea \left|{\cal M}(q\bar q\to
gH)\right|^2=\frac{16}{9}\frac{\as^3}{\pi v^2} \frac{\hat t^2+\hat
u^2}{\hat s},\nn \eea which is completely finite. The spin and colour
averages yield an additional factor $1/2 \times 1/2
\times 1/3 \times 1/3=1/36$.

\subsubsection*{Subtraction terms}
We only need to consider subtractions for the $gg$ and $qg$ induced
cases; these are given by \eqs (\ref{eq:dggg_ini}),
(\ref{eq:dgqq_ini}), and (\ref{eq:Dif_iniini}) respectively. Combining
them with the real emission matrix elements and integrating over phase
space, we obtain
\bea 
\lefteqn{\sigma^{\text{NLO},2}(gg \to gH)\,=\,\frac{1}{2\hat s}\int dPS_2\left\{
 \left|{\cal M}(gg \to gH)\right|^2-\lb D_{ggg}+D^{\text{if}}_{12} \rb\,\left|{\cal M}_{\text{LO}}\right|^2\right\}}\nn\\
 &=&\frac{1}{384\hat s}\left(\frac{\as^3}{\pi^2v^2}\right)
\left(1-\frac{M_H^2}{\hat s}\right)\times \nn \\
&&
\left\{
\frac{4\hat s  (2M_H^4-2M_H^2\hat s+\hat s^2 )}{(M_H^2-\hat s)^2}
 \ln\left(\frac{\hat s}{M_H^2}\right) 
\right. \,+\,\left. \frac{  M_H^4+34M_H^2\hat s+\hat s^2 }{3\hat s}
+\frac{4M_H^2\hat s }{M_H^2-\hat s } 
 \frac{}{}
\right\}\nn\\
&&
\eea
and
\bea \sigma^{\text{NLO},2}(qg\to
qH)&=&\frac{1}{2\hat s}\int dPS_2\left\{\left|{\cal M}(qg\to
qH)\right|^2- \D_{gqq}\left|{\cal M}_{\text{LO}}\right|^2\right\}\nn\\
&=&\frac{1}{1728\hat s}\left(\frac{\as^3}{\pi^2v^2}\right)
\left(1-\frac{M_H^2}{\hat s}\right)\left(3M_H^2+\hat s+4\hat
s\ln\left(\frac{\hat s}{M_H^2}\right)\right).\nn\\
&&\eea
As the Born contribution is constant and we are using unit jet
functions, we do not need to apply any momentum mappings.

The finite contribution from the quark-induced case is
\bea
\sigma^{\text{NLO},2}(q\bar q \to gH)&=&\frac{1}{2\hat s}\int
dPS_2\left|{\cal M}(q\bar q\to gH)\right|^2
\,=\,\frac{1}{486}\left(\frac{\as^3}{\pi^2v^2}\right)
\left(1-\frac{M_H^2}{\hat s}\right)^3.\nn\\
&&\eea

\subsubsection*{Integrated subtraction terms}
The integrated subtraction terms for the gluon- and quark-induced
cases are given in \eqs (\ref{eq:vggg_ini}), (\ref{eq:vgqq_ini}),
(\ref{eq:Vif_iniini}). After addition of the collinear counterterm, we
have for the gluon-induced case
\bea 
d\sigma^{\text{NLO}, 1}_{gg}&=&\int^{1}_{0}\,dx d\sigma_{0}(xp_a, p_b)
\left[ \mathcal{V}_{ggg}(p_{a})+\mathcal{V}^{\text{if}}_{ab}(p_{a}) \right]
\nn\\
&+&
\int^{1}_{0}\,dx d\sigma_{0}(p_a, xp_b) \left[ \mathcal{V}_{ggg}(p_{b})+\mathcal{V}^{\text{if}}_{ab}(p_{b}) \right]
+
d\sigma^C(p_a,p_b,\mu_F^2)\,+\,d\sigma_{\text{ch}}\nn\\
&=&
d\sigma_{0}(p_a,p_b)\otimes {\bf I}_{gg}(\vareps)+
\int_0^1 dx  d\sigma_{0}(xp_a,p_b)\otimes
\left[{\bf K}_{gg}^a(xp_a)+{\bf P}_{gg}(x,\mu_F^2) \right] \nn\\
&&+(a\leftrightarrow b) 
\eea
where $\mathcal{V}(p_{i})$ signifies that $p_{\ell}\,=\,p_{i}$ in the
respective integrated counterterm and where
\bea 
 {\bf I}_{gg}(\vareps) &=&
\frac{\as}{2\pi}C_A\frac{1}{\Gamma(1-\vareps)}\left(\frac{4\pi\mu^2}{Q^2}\right)
^\vareps
\left(\frac{2}{\vareps^2}-\frac{\pi^2}{3}\right), \nn\\
 {\bf K}_{gg}^a(xp_a)&=& \frac{\as}{2\pi}C_A\frac{1}{\Gamma(1-\vareps)}
\left\{A_{gg}(x, \vareps^0)\right.\nn\\
&&\left.
- 2\left(\frac{x}{(1-x)_+}+x(1-x)+\frac{1-x}{x}\right)
\ln\left(\frac{4\pi\mu^2}{2\,x\,p_a\cdot p_b}\right)
 \right\} \nn\\
&&-\,\frac{\as}{2\pi}\frac{1}{\Gamma(1-\vareps)}\,\delta(1-x)\,\lb \frac{11}{6}C_{A}-\frac{2}{3}\,n_{f}\,T_{R}\rb\,\ln \lb 4\,\pi\rb,\nn\\
 {\bf P}_{gg}(x,\mu_F^2) &=&
\frac{\as}{2\pi}\frac{1}{\Gamma(1-\vareps)} P_{gg}(x)
\ln\left(\frac{4\pi\mu^2}{\mu_F^2}\right) \nn
\eea
with
\bea 
A_{gg}(x, \vareps^0)&=&
4x\left(\frac{\ln(1-x)}{1-x}\right)_+
-2x (1-x)\ln x +4(1-x) \ln (1-x)\left(\frac{1+x^2}{x}\right) \nn\\&&
+2\left(x^2-\frac{1-x}{x}\right),\nn\\
P_{gg}(x)&= &\,2\,C_A\, \left(\frac{x}{(1-x)_+}+x(1-x)+\frac{1-x}{x}\right)
+
\delta(1-x)\left(\frac{11}{6}\,C_A-\frac{2}{3}\,n_f\,T_R\right).\nn
\eea
Integration over phase space yields
\bea
\lefteqn{\sigma^{\text{NLO,1}}_{gg}\,=\,\frac{1}{576}\,\lb\frac{\al_{s}^{3}}{v^{2}\,\pi^{2}}\rb\,C_{A}\,
\frac{1}{\Gamma(1-\vareps)}\,\times}\nn\\
&&\Bigg\{\lb\frac{4\,\pi\,\mu^{2}}{M_{H}^{2}}\rb^{\vareps}\lb
\frac{1}{\vareps^{2}}+\frac{1}{\vareps}
+1\,-\frac{\pi^{2}}{6}\rb\,\delta(1-z)
+ 4\,\left[\lb\frac{\ln(1-z)}{1-z}\rb_{+} \nn\right.\\
&&\hspace*{30mm}\left.-z\,\lb2-z\,(1-z)\rb\,\ln(1-z) \right.\bigg] \nn\\
&&+2\,\left[
  z^{3}-(1-z)\right]\,-\,2\,\left[\frac{z^{2}}{(1-z)_{+}}+z^{2}\,(1-z)+(1-z) \right]\,
\ln\frac{\mu^{2}_{F}}{\hat{s}}\nn\\
&&+\,2\,\left[\frac{z^{2}}{(1-z)_{+}}+(1-z)\right]\,\ln\,z
+\,\frac{1}{C_{A}}\,\,\delta(1-z)\lb 
\frac{11}{6}C_{A}-\frac{2}{3}n_{f}T_{R}\rb\,\ln\frac{\mu^{2}}{\mu_{F}^{2}}\Bigg\}\,,
\nn\\&&
\eea
where we introduced $z\,=\,M^{2}_{H}/\hat{s}$.
\noindent
For the contribution from the $q g$-induced case, we obtain
\bea 
\begin{split}
d\sigma^{\text{NLO}, 1}_{qg}&=\,\int_{0}^{1}\,dx  d\sigma_{ab}^B(xp_a,
p_b)
\left[\mathcal{V}_{gqq}(p_{a})+\mathcal{V}^{\text{if}}_{ab}(p_{a}) \right] +
 \sigma_{ab}^C(p_a,p_b,\mu_F^2)\nn\\
& =\int_0^1 dx d\sigma^B_{ab}(xp_a,p_b)\otimes
\left[{\bf K}_{qg}(xp_a)+{\bf P}_{qg}(x,\mu_F^2) \right],  
\end{split}
\eea
where
\bea 
 {\bf K}_{qg}(xp_a)&=& \frac{\as}{2\pi}\frac{1}{\Gamma(1-\vareps)}
\left\{A_{qg}(x, \vareps^0)C_F
-  P_{gq}(x)
\ln\left(\frac{4\pi\mu^2}{2\,x\, p_a\cdot p_b}\right)
 \right\}, \nn \\
 {\bf P}(x,\mu_F^2) &=&
\frac{\as}{2\pi}\frac{1}{\Gamma(1-\vareps)}
P_{gq}(x)
\ln\left(\frac{4\pi\mu^2}{\mu_F^2}\right),\nn \\
P_{gq}(x)&=& C_F\,  \frac{1+(1-x)^2}{x}\,.\nn
\eea
$A_{qg}(x, \vareps^0)$ is given by
\bea 
A_{qg}(x, \vareps^0)&=&\frac{x^2-2(1-x)}{x}-x\ln x+2\ln(1-x) \left(\frac{1+(1-x)^2}{x}\right). \nn
\eea
Integration over phase space then yields
\bea
\sigma^{\text{NLO}, 1}_{qg}&=&\frac{\al_{s}^{3}}{v^{2}\,\pi^{2}}\,\frac{z}{1152}
\,\left\{C_{F}\,\left[ \frac{z^{2}-2\,(1-z)}{z}-z\,\ln\,z \right]\nn\right.\\
&&\left.+\,P_{gq}(z)\,\left[ 2\,\ln\,(1-z)+\ln\lb \frac{M_{H}^{2}}{\mu_{F}^{2}} \rb \right] \right\}\,.
\eea
\subsubsection*{Result}
In this subsection, we again summarise the results we obtain from the
different subprocesses of gluon-induced Higgs production; we find
total agreement with the results in \cite{ Djouadi:1991tka,
  Dawson:1990zj}. We have
\bea
\sigma^{\text{NLO},2}_{q\bar{q}}&=&\frac{1}{486}\left(\frac{\as^3}{\pi^2v^2}\right)
\left(1-z\right)^3,\nn\\
\sigma^{\text{NLO},1}_{qg}\,+\,\sigma^{\text{NLO},2}_{qg}&=&\frac{1}{576}\left(\frac{\as^3}{\pi^2v^2}\right)\,
\left\{-(1-z)\,\lb \frac{7-3\,z}{3} \rb\nn\right.\\
&&\left. +\,\frac{1}{2}z\,P_{gq}(z)\,\left[ 1+\ln\lb 
\frac{M^{2}_{H}\,(1-z)^{2}}{z\,\mu^{2}_{F}}\rb \right]  \right\},\nn
\eea
\bea
\lefteqn{\sigma^{\text{NLO},1}_{gg}\,+\,\sigma^{\text{NLO},2}_{gg}
\,+\,\sigma_{\text{virt}}\,+\,\sigma_{\text{eff}}\,=\,}\nn\\
&&\frac{1}{576}\left(\frac{\as^3}{\pi^2v^2}\right)\,\Bigg\{\delta(1-z)\,\left[
  \frac{11}{2}+\pi^{2}\,+\,\lb 
\frac{11}{6}C_{A}-\frac{2}{3}n_{f}T_{R}\rb\,\ln\frac{\mu^{2}}{\mu_{F}^{2}}\right]\nn\\
&&\,+\,12\left[\lb\frac{\ln(1-z)}{1-z}\rb_{+}-z\,\lb2-z\,(1-z)\rb\,\ln(1-z) \right]\,-\,\frac{11}{2}\,(1-z)^{3}\nn\\
&& \hspace*{15mm}-\,z\,P_{gg}(z)\,\ln\,\frac{\mu_{F}^{2}}{\hat{s}}\Bigg\}\,.\nn
\eea

\subsection{Higgs decay to gluons }
This process is used to validate the final-state collinear splitting
functions $g\,\rightarrow\,g\,g,\,g\,\rightarrow\,q\,\bar{q}$, \eqs
(\ref{eq:dggg_fin}), (\ref{eq:dgqq_fin}), (\ref{eq:vggg_fin}),
(\ref{eq:vgqq_fin}) as well as the final-final state interference term
\eqs (\ref{eq:Dif_fin}), (\ref{eq:vif_finfin}).  Besides the virtual
correction to the vertex, both $H\,\rightarrow\,g\,g\,g$ and
$H\,\rightarrow\,g\,q\,\bar{q}$ contribute to the real radiation terms
for this process.
\subsubsection*{Leading-order contribution}
The leading-order squared and averaged matrix element for the process
$H\,\rightarrow\,g\,g$ in $d\,=\,4-2\vareps$ dimensions is given by
\beq \left|{\cal
M}_{\text{LO}}\right|^2=\frac{2}{9}\frac{\as^2M_H^4}{\pi^2v^2} (1-\vareps);\nn
 \eeq
integration over the two-particle phase space then yields 
\beq
\Gamma_{\text{LO}}(H\to gg) \, =\, \frac{G_F\, M_H^3\, \as^2}{36\, \sqrt{2}\,\pi^3}(1-\vareps). \\
\eeq
\subsubsection*{Virtual correction, charge renormalisation, and effective theory correction term}
The virtual matrix element is given by \eq (\ref{eq:Hgg_virt})
\bea
2\,\text{Re}\,\lb\M_{\text{LO}}\M^{*}_{\text{virt}}\rb & \equiv &\left|{\cal M}_V\right|^2\nn\\
&=& \left|{\cal M}_{\text{LO}}\right|^2
\frac{\as}{2\pi}C_A  \left(\frac{4\pi\mu^2}{M_H^2}\right )^\vareps \Gamma(1+\vareps)
\left(-\frac{1}{\vareps^2}+\frac{2}{3}\pi^2+{\cal O}(\vareps)\right )\times 2\nn
\eea
which gives
\beq
\Gamma_{\text{virt}}\,=\,\Gamma_{\text{LO}}\,\frac{\al_{s}}{\pi}\,C_A  \left(\frac{4\pi\mu^2}{M_H^2}\right )^\vareps \Gamma(1+\vareps)
\left(-\frac{1}{\vareps^2}+\frac{2}{3}\pi^2+{\cal O}(\vareps)\right ).
\eeq
The charge renormalisation term is given as in \eq (\ref{eq:ggh_charge})
\bea
\Gamma_{\textrm{ch}}&=&\left(4Z_g\right) \Gamma_{\text{LO}}( H\to gg)\nn\\
&=&
- \Gamma_{\text{LO}}
\frac{1}{\Gamma(1-\vareps)}
\left(\frac{4\pi\mu^2}{\mu_F^2}\right)^\vareps
\left(\frac{\as}{\pi}\right) \,\beta_0\,
\frac{1}{\vareps}
-\Gamma_{\text{LO}} \left(\frac{\as}{\pi}\right) 
\,\beta_0\,
\ln \frac{\mu_F^2}{\mu^2},\nn\\
\eea
where we used that $\Gamma(1-\vareps)\,\Gamma(1+\vareps)\,=\,1\,+\,\mO(\vareps^{2})$.
We also need to add the NLO correction to the matching coefficient
\beq
\Gamma_{\text{eff}}\,=\,\Gamma_{\text{LO}}\,\frac{11}{2}.
\eeq
\subsubsection*{Real emission}
For the real-emission contributions, both
\beq
H\,\rightarrow\,g\,g\,g,\; H\,\rightarrow\,g\,q\,\bar{q}\nn
\eeq
need to be considered.
The squared matrix element for $   H(\hat{Q})\to g(\ph_1)q(\ph_2)\bar q(\ph_3) $ is
given by \bea \left|{\cal M}(H\to gq\bar q)\right|^2
&=&\frac{16}{9}\frac{\as^3}{\pi v^2} \frac{(\ph_1+\ph_2)^4+(\ph_1+\ph_3)^4}{(\ph_2+\ph_3)^2}\nn\\
&=&\frac{16}{9}\frac{\as^3}{\pi v^2}\hat
Q^2\frac{(x_1+x_2-1)^2+(1-x_2)^2}{(1-x_1)},\nn\eea 
where we used $x_{i}\,=\,\frac{2\hat p_i\cdot \hat Q}{\hat Q^2}$ and with $\hat Q^2=Q^2=M_H^2$. The
corresponding subtraction term is given by \eq(\ref{eq:dgqq_fin})
\bea
D_{gq\,\bar{q}}&=&\D_{gqq}\left|{\cal
M}_{\text{LO}}\right|^2 \,\times\,2\,=\,\frac{16}{9}\frac{\as^3}{\pi v^2} \hat Q^2\frac{1}{1-x_1}
\left[ 1-2\,\frac{(1-x_{2})\,(1-x_{3})}{x_{1}^{2}} \right], \nn\eea
where the factor $2$ arises because each of the gluons in the leading
order contribution can split into a quark-antiquark pair. Note again
that, in the case of non-unit jet functions, this factor needs to be
replaced by
\begin{\eqn}\label{eq:2tojets}
2\,\rightarrow\,F^{(2)}(p_{g_{1}})+F^{(2)}(p_{g_{2}})
\end{\eqn}
where $F^{(2)}(p_{i})$ denotes 
\begin{\eqn*}
p_{i}\,\equiv\,p_{\ell}\,=\,\ph_{2}+\ph_{3}-\frac{y}{1-y}\,\ph_{1}
\end{\eqn*}
in the respective mappings.  Integrating over the three-particle phase
space and summing over final state quark flavours yields 
\bea
d\Gamma(H\to gq\bar q)&=&\frac{1}{2M_H}\int dPS_3\sum_q
\left(\left|{\cal M}(H\to gq\bar q)\right|^2-D_{gq\bar{q}}\right)
\,=\,\Gamma_{\text{LO}}\frac{\as}{\pi} \left(-\frac{5}{18}
  n_f\right).\nn 
\eea

For the purely gluonic decay, we obtain for $   H(\hat{Q})\to g(\ph_1)g(\ph_2)g(\ph_3) $,
\bea \lefteqn{\frac{1}{3!}\left|{\cal M}(H\to ggg)\right|^2 \,=\,}\nn\\
&&
\frac{\as^3}{\pi
v^2}\frac{32}{3}\frac{1}{3!}
\left\{
\frac{2s_{123}^2 s_{12}}{s_{13}s_{23}}+\frac{2s_{123}^2
  s_{13}}{s_{12}s_{23}}
+\frac{2s_{123}^2 s_{23}}{s_{12}s_{13}}+\frac{2s_{12} s_{13}}{s_{23}}
+\frac{2s_{12} s_{23}}{s_{13}}+\frac{2s_{13} s_{23}}{s_{12}}+8
s_{123}
\right\}
\nn
\eea
where
\bea
&&s_{ij}\,=\,\left(\ph_i+\ph_j\right)^2,\,s_{123}\,=\,\left(\ph_1+\ph_2+\ph_3\right)^2,\nn\\
&&x\,=\,\frac{s_{12}}{s_{123}}, \quad y=\frac{s_{13}}{s_{123}}, \quad 
z=\frac{s_{23}}{s_{123}}\nn\,.
\eea
In the subtraction terms, we have to sum over $\ph_{j},j\,=\,1,2,3$: 
\bea
(\ph_j,\ph_{\ell})=(\ph_1,\ph_{2}),\;
(\ph_j,\ph_{\ell})=(\ph_1,\ph_{3}),\;
(\ph_j,\ph_{\ell})=(\ph_2,\ph_{3}),\nn
\eea
where we have the transformations
\begin{\eqn*}
p_{\ell}\,=\,\ph_{\ell}+\ph_{j}-\frac{y}{1-y}\ph_{k},\,p_{k}\,=\,\frac{1}{1-y}\ph_{k}
\end{\eqn*}
and we have to account for the fact that $p_{k}\,=\,(p_{g_{1}},
p_{g_{2}})$ for each of the above scenarios.  For the soft subtraction
term, we therefore obtain \bea D_s&=&\frac{1}{3}\,\sum
\D^{\text{if}}\left[ (\ph_{\ell},\cdot \ph_{j})\right] \left|{\cal
    M}_{\text{LO}}\right|^2 \,=\, \frac{\as^3}{\pi
  v^2}\frac{32}{3}\frac{1}{3!}  \left\{ \frac{2s_{123}^2
    s_{12}}{s_{13}s_{23}}+\frac{2s_{123}^2 s_{13}}{s_{12}s_{23}}
  +\frac{2s_{123}^2 s_{23}}{s_{12}s_{13}}
\right\},\nn\\
&& 
\eea 

where $\D^{\text{if}}$ is given by \eq (\ref{eq:Dif_fin}) and the sum
runs over all possible combinations. The factor $1/3$ in the above sum
reflects the fact that the final-state gluons are
indistinguishable\footnote{In principle, the real emission matrix
  element includes all possible $3!$ combinations of
  $(\ph_{\ell},\ph_{j},\ph_{k})$; for each of these, the matrix
  factorises according to
\begin{\eqn*}
\M_{\text{real}}\,\rightarrow\,v\,\M_{\text{LO}}
\end{\eqn*}
in the singular limits.  We obtain a symmetry factor $\frac{1}{3!}$ in
front of the squared matrix element; however, the factorised splitting
function is the same for $\ph_{\ell}\,\leftrightarrow\,\ph_{k}$, which
effectively leads to a factor $\frac{2}{3!}\,=\,\frac{1}{3}$.  }, and
we implicitly included a factor $2$ in this expression which, in case
of non-unit jet functions, needs to be modified according to \eq
(\ref{eq:2tojets}).  For the collinear contribution, we have
\bea D_c&=&\frac{1}{3}\,\sum \D_{ggg}\left[ (\ph_{\ell},\ph_{j})\right]\left|{\cal
M}_{\text{LO}}\right|^{2}\nn\\
&=&\frac{4}{3}\frac{\as^3}{\pi
v^2} \hat Q^4\frac{1}{3}\left\{ \frac{4s_{12} s_{13}}{s_{23}(s_{12}+ s_{13})^2}
\left[2-\frac{4s_{23}\hat Q^2}{(s_{12}+ s_{23})(s_{13}+ s_{23})}\nn\right.\right.\\
&&\hspace*{55mm}\left.\left.+\left[\frac{2s_{23}\hat Q^2}{(s_{12}+ s_{23})(s_{13}+ s_{23})}\right]^2\right] \right. \nn\\
&&\left.
+\frac{4s_{12} s_{23}}{s_{13}(s_{12}+ s_{23})^2}
\left[2-\frac{4s_{13}\hat Q^2}{(s_{12}+ s_{13})(s_{13}+ s_{23})}
+\left[\frac{2s_{13}\hat Q^2}{(s_{12}+ s_{13})(s_{13}+ s_{23})}\right]^2\right]  \right.\nn\\
&&\left.
+\frac{4s_{13} s_{23}}{s_{12}(s_{13}+
  s_{23})^2}\left[2-\frac{4s_{12}\hat Q^2}{(s_{12}
+ s_{23})(s_{12}+ s_{13})}+\left[\frac{2s_{12}\hat Q^2}{(s_{12}+ s_{23})(s_{12}+ s_{13})}\right]^2\right] \right\},\nn
\eea
where $\D_{ggg}$ is given by \eq (\ref{eq:dggg_fin}).

Integrating over the three-particle phase space yields \bea
\Gamma^{\text{NLO,3}}_{ggg}&=&\frac{1}{2M_H}\int dPS_3
\left(\left|{\cal M}(H\to ggg)\right|^2-D_c-D_s \right)
\,=\,\Gamma_{\text{LO}}\,\frac{\as}{\pi}
\left(\frac{-214+27\pi^2}{24} \right).\nn\\
&&\eea
\subsubsection*{Integrated subtraction terms}
The two-particle phase-space contribution from the integrated $gq\bar
q$ splitting function, \eq (\ref{eq:vgqq_fin}), yields 
\beq
\Gamma^{\text{NLO},2}_{gq\bar{q}}\,=\,\Gamma_{\text{LO}}\,\frac{\as}{\pi}T_R\frac{1}{\Gamma(1-\vareps)}
\left(\frac{4\pi\mu^2}{M^{2}_{H}}\right)^\vareps
n_f\left(-\frac{2}{3\vareps}-\frac{16}{9}\right)\nn\,, \eeq where we
have summed over final-state quark flavours and used $2\,p_{\ell}\cdot
Q\,=\,M^{2}_{H}$ for $l\,=\,1,2$ being the gluon indices in the
two-particle phase space.  For $ggg$ splitting, we use \eqs
(\ref{eq:vggg_fin}) and (\ref{eq:vif_finfin}) for the integrated
splitting function contributions to obtain \bea
&&\Gamma^{\text{NLO},2}_{ggg}\,=\,
\Gamma_{\text{LO}}\,\frac{\as}{2\pi}C_A \frac{1}{\Gamma(1-\vareps)}
\left(\frac{4\pi\mu^2}{M^{2}_{H}}\right)^\vareps
\left(\frac{2}{\vareps^2}+\frac{11}{3\vareps}+\frac{163}{9}-\frac{7}{4}\pi^2\right).
\eea
\subsubsection*{Result}
Combining all contributions to the total decay width at NLO, we have
\bea
\lefteqn{\Gamma_{\text{LO}}\,+\,\Gamma^{\text{virt}}\,+\,\Gamma^{\text{NLO},2}_{g\,\bar{q}\,q}\,
+\,\Gamma^{\text{NLO},2}_{ggg}\,+\,\Gamma^{\text{NLO},3}_{g\,\bar{q}\,q}\,+\,\Gamma^{\text{NLO},3}_{ggg}\,
+\,\Gamma_{\text{ch}}\,+\,\Gamma_{\text{eff}}\,=\,}\nn\\
&=&\Gamma_{\text{LO}}\left[1+\frac{\as}{\pi}\left[
\left(\frac{95}{4}-\frac{7}{6}n_f\right)
+\left(\frac{11}{6}C_A-\frac{2}{3}n_fT_R\right)\ln \frac{\mu^2}{M_{H}^2}
\right]\right]\,,
\eea
which reproduces the result in the literature \cite{Djouadi:1995gt}.
\subsection{Deep inelastic scattering}
For a check of the (integrated) initial-final and final-initial
interference terms given by \eqs (\ref{eq:Dif_inifin}),
(\ref{eq:Dif_fin}), (\ref{eq:Vif_inifin}), and (\ref{eq:vif_finini}),
we consider a subprocess of deep-inelastic scattering:
\begin{\eqn}\label{eq:dis_ep}
e^{-}(p_{i})+q(p_{1})\,\longrightarrow\,e^{-}(p_{o})\,q(p_{2})\,
\end{\eqn}
at NLO, with a pure photon exchange in the t -channel. This suffices
to test the validity of initial-final and final-initial state
interference terms, therefore completing our check of all integrated
subtraction terms for LO processes with up to two particles in the
final state.
\subsubsection{Tree-level contribution}
\subsubsection*{Matrix element}
For the process (\ref{eq:dis_ep}), we have
\begin{\eqn}\label{eq:dis_tree}
\frac{1}{4}\,\sum\,|\M|^{2}_{\text{Born}}\,=\,2\,e^{4}\,Q^{2}_{q}\,\frac{u^{2}+s^{2}}{t^{2}},
\end{\eqn}
where
$s\,=\,(p_{i}+p_{1})^{2}\,=\,(p_{o}+p_{2})^{2},\,t\,=\,(p_{i}-p_{o})^{2}\,=\,(p_{1}-p_{2})^{2},\,u\,=\,-s-t$.
Note that the singularity $t\,=\,0$ needs to be regularised by a
proper definition of the jet function.
\subsubsection*{Phase space}
The two particle phase space is given by 
\begin{\eqn*}
d\Gamma_{2}\,=\,\lb
\prod_{i=1,2}\,\frac{d^{3}p_{i}}{(2\pi)^{3}2p_{i}^{0}}\rb (2\pi)^{4}\,\delta^{(4)}(p_{i}-(p_{1}+p_{2}))\,=\,\frac{d\Omega_{1}}{8\,(2\pi)^{2}}.
\end{\eqn*}
\subsubsection{Virtual matrix element}
The loop correction to the vertex is
\begin{\eqn}
\M_\text{virt}\,=\,\frac{C_{F}\,\al_{s}}{4\pi}\,\frac{1}{\Gamma(1-\vareps)}\,\lb\frac{4\pi\,\mu^{2}}{-p^{2}_{in}}\rb^{\vareps}\,\lb-\frac{2}{\vareps^{2}}\,-\,\frac{3}{\vareps}\,-\,8\,+\,\mO(\vareps) \rb\,\M_{\text{Born}},
\end{\eqn}
and we obtain
\begin{eqnarray}
|\M_{V}|^{2} & \equiv & 2\,Re\lb\M_{\text{Born}}\,\M_\text{virt}^{*}\rb\nn \\
&=&|\M_{\text{Born}}|^{2}\,\frac{2\,\al_{s}}{3\,\pi}\,\frac{1}{\Gamma(1-\vareps)}\,\lb\frac{4\pi\mu^{2}}{2\,p_{i} \cdot p_{o}}\rb^{\vareps}\,\left\{-\frac{2}{\vareps^{2}}\,-\frac{3}{\vareps}-8+\,\mO(\vareps)\right\}\nn \\
&=&|\M_{\text{Born}}|^{2}\,\frac{2\,\al_{s}}{3\,\pi}\,
\left\{-\frac{2}{\vareps^{2}}\,-\frac{1}{\vareps}\lb 3+2\,\ln\,K-2\,\gamma\rb -8\,-3\,\lb\ln\,K-\gamma \rb\nn\right.\\
&&\hspace*{30mm} \left. -\,\ln^{2}\,K\,+\,2\,\gamma\,\ln\,K\,+\,\frac{\pi^{2}}{6}-\gamma^{2}\right\}
\end{eqnarray}
where
$K\,=\,{2\,\pi\,\mu^{2}}/{(p_{i} \cdot p_{o})}$.
\subsubsection{Real emission}
For a check of the remaining interference terms, the only relevant
real-emission process we need to take into account is the
quark-induced channel
\begin{\eqn}\label{eq:dis_epg}
e^{-}(\ph_{i})+q(\ph_{1})\,\longrightarrow\,e^{-}(\ph_{o})\,q(\ph_{4})g\,(\ph_{3}).
\end{\eqn}
A complete discussion would also include gluon-induced processes;
however, the corresponding (integrated) subtraction terms \eqs
(\ref{eq:dgqq_ini}), (\ref{eq:vgqq_ini}) have already been discussed
in the context of single-W production in Section \ref{sec:single_W},
and are therefore not considered here.
\subsubsection*{Matrix element}
The squared and spin-averaged matrix element for the real emission process is 
\begin{eqnarray}\label{eq:disrealtot}
\frac{1}{4}\,\sum_\text{spins}\,|\M|^{2} &=&\frac{8\,\pi\,\al_{s}\,e^{4}\,Q_{q}^{2}}{(p_{1} \cdot p_{3})\,(p_{3} \cdot p_{4})\,(p_{i} \cdot p_{o})}\,C_{F}\,\left[(p_{i} \cdot p_{1})^{2}\,+\,(p_{i} \cdot p_{4})^{2}\nn\right.\\
&& \left.+\,(p_{o} \cdot p_{1})^{2}\,+\,(p_{o} \cdot p_{4})^{2} \right]\,. \nn
\end{eqnarray}
\subsubsection*{Three-particle phase space}
In our code, we follow the three-particle phase-space parametrisation
of \cite{FormCalc}. We then obtain for the outgoing
four-vectors\footnote{We are aware that a more detailed analysis could
  be optimised by a more process-specific parametrisation of phase
  space which specifically maps the singularity structures, in
  combination with a multi-channel integration as \eg VAMP
  \cite{Ohl:1998jn}; however, for our purposes this simple
  parametrisation proved sufficient.}
\begin{eqnarray}
\ph_{3} &=& |{\bf{\ph}_{3}}|\,\lb\begin{array}{c}1\\ \sin\theta\\0\\
  \cos\theta \end{array}
\rb,\;\ph_{o}\,=\,|{\bf{\ph}_{o}}|\,\lb\begin{array}{c}1\\ 
\cos\theta \cos\eta \sin\xi\,+\,\sin\theta \cos\xi\\ 
\sin\eta \sin\xi\\ 
\cos\theta \cos\xi - \sin\theta \cos \eta \sin\xi \end{array} \rb,\;\nn\\
\ph_{4} & = & \lb\begin{array}{c}
  \sqrt{s}-|{\bf{\ph}_{3}}|-|{\bf{\ph}_{o}}| \\ 
- {\bf{\ph}_{3}}- {\bf{\ph}_{o}} \end{array} \rb,
\end{eqnarray}
where
\begin{\eqn*}
\cos\xi\,=\,\frac{|{\bf{\ph_{4}}}|^{2}-{|\bf{\ph_{3}}}|^{2}-{|\bf{\ph_{o}}}|^{2}}{2\,|{\bf{\ph_{3}}}|
\,|{\bf{\ph_{o}}}|},\,\ph_{4}\,=\,\hat{Q}-\ph_{3}-\ph_{o}.
\end{\eqn*}

The three-particle phase space is
\begin{\eqn}\label{eq:3psfc}
\int\,d\Gamma_{3}\,=\,\frac{1}{8\,(2\pi)^{5}}\,\int^{\sqrt{\hat{s}}/2}_{0}\,d\ph^{0}_{3}\,
\int^{\sqrt{\hat{s}}/2}_{\sqrt{\hat{s}}/2-\ph^{0}_{3}}\,d\ph_{o}^{0}\,\int^{1}_{-1}\,d\cos\theta\,
\int^{2\pi}_{0}\,d\varphi\,\int^{2\pi}_{0}\,d\eta.
\end{\eqn}
\subsubsection{Subtraction terms}
\subsubsection*{Initial-state subtraction}
The initial-state subtraction is given by \eqs (\ref{eq:dqqg_ini}) and
(\ref{eq:Dif_inifin}):
\begin{\eqn*}
D^{1,3}\,=\,\frac{4\,\pi\,\al_{s}}{x\,y\, \ph_{1} \cdot \ph_{i}}\,\,C_{F}\lb 1-x-y+\frac{2\,\tilde{z}\,x}{v\,(1-x)+y} \rb|\M_\text{Born}(p)|^{2}
\end{\eqn*}
where $|\M_\text{Born}|^{2}$ is given by \eq (\ref{eq:dis_tree}) and
with
\begin{eqnarray}\label{eq:dis_inivars3}
&&x\,=\,\frac{\ph_{o} \cdot \ph_{4}}{\ph_{i} \cdot \ph_{1}},\;y\,=\,\frac{\ph_{1} \cdot \ph_{3}}{\ph_{1} \cdot \ph_{i}},\;\tilde{z}\,=\,\frac{\ph_{1} \cdot \ph_{4}}{\ph_{4} \cdot \hat{Q}},\;v\,=\,\frac{(\ph_{1} \cdot \ph_{i})\,(\ph_{3} \cdot \ph_{4})}{(\ph_{4} \cdot \hat{Q})\,(\ph_{3} \cdot \hat{Q})}.
\end{eqnarray}
The final-state momenta $p_{4},\,p_{o}$ are mapped according to the
corresponding transformation specified below.
\subsubsection*{Initial-state kinematics: momentum mapping}
The momentum mapping needed here has been specified in Section
\ref{sec:inimap}; for the initial-state momenta, we obtain
\begin{\eqn*}
p_{1}\,=\,x\,\ph_{1},\,p_{i}\,=\,\ph_{i}.
\end{\eqn*}
The four-vectors of the outgoing particles are transformed using
($w\,=\,o,4$)
\begin{\eqn*}
\tilde{p}^{\mu}_{w}\,=\,\Lambda^{\mu}_{\phantom{\mu}\nu}(K,\hat{K})p^{\nu}_{w},
\end{\eqn*}
where $\Lambda(K,\hat{K})$ is defined according to \eq(\ref{eq:LTini})
and with
\begin{\eqn*}
K\,=\,x\,\ph_{1}+\ph_{i},\,\hat{K}\,=\,\ph_{1}+\ph_{i}-\ph_{3}\,.
\end{\eqn*}
\subsubsection*{Final-state subtraction}
The final-state subtraction term is given by \eqs (\ref{eq:dqqg_fin})
and (\ref{eq:Dif_fin})
\begin{eqnarray*}
D^{4,3}&=&\frac{4\,\pi\,\al_{s}\,C_{F}}{y\,(\ph_{i} \cdot \ph_{1})}\,\left[\frac{y}{1-y}F_\text{eik}\,+\,z\,+\,2\,\frac{(1-v)\,(1-z\,(1-y))}{v\,\left[1-z\,(1-y) \right]+y\left[(1-y)\tilde{a}+1 \right]} \right]|\M_\text{Born}(p)|^{2}
\end{eqnarray*}
where we use
\begin{eqnarray*}
&&p_{4} \cdot \hat{Q}\,=\,\ph_{1} \cdot \ph_{i},\qquad\;y\,=\,\frac{\ph_{3} \cdot \ph_{4}}{\ph_{1} \cdot \ph_{i}},\qquad\;\tilde{n}\,=\,\frac{\ph_{o}}{(1-y)},\\
&&\;z\,=\,\frac{\ph_{3} \cdot \ph_{o}}{\ph_{3} \cdot \ph_{o}+\ph_{4} \cdot \ph_{o}},\qquad\;x\,=\,\frac{\ph_{3} \cdot \hat{Q}}{\ph_{3} \cdot \hat{Q}+\ph_{4} \cdot \hat{Q}},
\qquad \tilde{a}\,=\,\frac{1}{1-y}\,\frac{\ph_{1} \cdot \ph_{o}}{\ph_{1} \cdot \ph_{3}+\ph_{1} \cdot \ph_{4}},\\
&&v\,=\,\frac{\ph_{1} \cdot \ph_{3}}{\ph_{1} \cdot \ph_{3}+\ph_{1} \cdot \ph_{4}},\qquad
F_\text{eik}\,=\,2\,\frac{(\ph_{3} \cdot \ph_{o})\,(\ph_{4} \cdot \ph_{o})}{(\ph_{3} \cdot \hat{Q})^{2}}.
\end{eqnarray*}
The momenta in the Born matrix element need to be calculated using the
transformation defined below.
\subsubsection*{Final-state kinematics: momentum mapping}
By definition, we have
\begin{\eqn*}
p_{i}\,=\,\ph_{i},\,p_{1}\,=\,\ph_{1},\,p_{4}\,=\,\frac{1}{1-y}\left[\ph_{3}+\ph_{4}-y\,(\ph_{1}+\ph_{i})
  \right]\,.
\end{\eqn*}
Energy and momentum conservation then implies
\begin{\eqn*}
p_{o}\,=\,\ph_{1}+\ph_{i}-p_{4}\,=\,\frac{\ph_{o}}{1-y}.
\end{\eqn*}
\subsubsection{Integrated subtraction terms}
\subsubsection*{Initial-state subtraction}
The initial-state integrated subtraction term, including the
interference term, is given by \eqs (\ref{eq:vqqg_ini}) and
(\ref{eq:Vif_inifin}) respectively, and we obtain
\begin{eqnarray}\label{eq:ikini}
I^{i}(\vareps)&=&\frac{1}{\vareps^{2}}\,+\,\frac{1}{\vareps}\,\lb\frac{3}{2}\,-\,\ln\tilde{z}_{0}\rb\,+\,\frac{1}{2}\,\text{Li}_{2}\,\lb 1-\tilde{z}_{0}
\rb\,+\,\frac{1}{2}\ln^{2}\tilde{z}_{0}+\,2\,I^{0}_\text{fin}(\tilde{z}_{0}),\nonumber\\
\wt{K}^{i}(x;\vareps)&=&\frac{1}{x}\left[-\frac{1}{\vareps}\,\lb\frac{1+x^{2}}{1-x}
  \rb_{+}\,+\,(1-x)\,2\,\ln\,(1-x)-\lb\frac{1+x^{2}}{1-x}\rb_{+}\,\ln\,x \,\nn\right.\\
  && \hspace*{10mm} \left. +\,4\,x\,\lb\frac{\ln(1-x)}{1-x}\rb_{+}\right] \,+\,I_\text{fin}^{1}(x,\tilde{z})\,,
\end{eqnarray} 
where $\tilde{z}$ is defined in \eq (\ref{eq:dis_inivars3}). The
four-vectors in $\tilde{z}$ now have to be calculated from $\{p\}$
using the inverse mapping
\begin{eqnarray}\label{eq:dis_ini_inversemap}
&&\ph_{1}\,=\,\frac{1}{x}p_{1},\;\ph_{i}\,=\,p_{i},\,\ph_{4}\,=\,\Lambda(\hat{K},K)p_{4}
\end{eqnarray}
where $\Lambda$ is defined in \eq (\ref{eq:LTini}) and 
\begin{\eqn*}
K\,=\,p_{1}+p_{i},\;\hat{K}\,=\,\frac{x+y'(1-x)}{x}p_{1}+\left[1-y'\,(1-x)\right]\,\ph_{i}\,+\,|k_\perp|\hat{k}_\perp
\end{\eqn*}
in terms of the $m$-particle phase space variables. Here,
\begin{\eqn*}
|k_\perp|\,=\,(1-x)\,\sqrt{\frac{(1-y')\,y'}{x}\,2\,p_{1}p_{i}},\;\hat{k}_\perp\,=\,\lb
\begin{array}{c}0\\2\,v-1\\-2\,\sqrt{v\,(1-v)}\\0 \end{array} \rb.
\end{\eqn*}

The complete subtraction term is given by
\begin{eqnarray}
\lefteqn{\frac{\al_{s}}{2\,\pi}\,C_{F}\,\frac{1}{\Gamma(1-\vareps)}\,\lb\frac{4\,\pi\,\mu^{2}}{2\,p_{1}\cdot p_{i}}\rb^{\vareps}}\nn\\
&&\,\int^{1}_{0}\,dx\,\left[\delta(x-1)\,I^{i}(\vareps)\,+\,\tilde{K}^{i}(x;\vareps)+P_\text{coll}(x,\vareps;\mu_{F})
  \right]|\M_\text{Born}(\tilde{p})|^{2}
\end{eqnarray}
with
\begin{\eqn*}
P_\text{coll}(x,\vareps;\mu_{F})\,=\,\frac{1}{\vareps}\,\frac{1}{x}\,\lb\frac{2 p_{a} \cdot p_{b}}{\mu^{2}_{F}}\rb^{\vareps}\,\lb\frac{1+x^{2}}{1-x}
\rb_{+}\,.
\end{\eqn*}
The integrated subtraction term contains two finite integrals
\begin{eqnarray}\label{eq:ifidis}
I^{0}_\text{fin}(\tilde{z}_{0})&=&\int^{1}_{0}\,dy'\,\lb\frac{\tilde{z}_{0}}{y'\,\sqrt{4\,y'^{2}\,(1-\tilde{z}_{0})
+\tilde{z}_{0}^{2}}}\nn\right.\\
&&\left. \hspace*{20mm}\times
  \ln\,\left[\frac{2\,\sqrt{4\,y'^{2}\,(1-\tilde{z}_{0})+\tilde{z}_{0}^{2}}\,
\sqrt{(1-y')}}{2\,y'+\tilde{z}_{0}-2\,y'\,\tilde{z}_{0}+
    \sqrt{4\,y'^{2}\,(1-\tilde{z}_{0})+\tilde{z}_{0}^{2}}}\right]\rb\,,\nonumber\\
I^{1}_\text{fin}(x,\tilde{z})&=&\frac{2}{(1-x)_{+}}\,\,\frac{1}{\pi}\int^{1}_{0}\,\frac{dy'}{y'}\,
\left[\int^{1}_{0}\,\frac{dv}{\sqrt{v\,(1-v)}}\,\frac{\tilde{z}}{N(x,y',\tilde{z},v)}-1\right],\nonumber\\
&&
\end{eqnarray}
which need to be integrated numerically. Here, 
\begin{\eqn*}
N\,=\,\frac{\ph_{3} \cdot \ph_{4}}{\ph_{4}\cdot \hat{Q}}\,\frac{1}{1-x}\,+\,y',
\end{\eqn*}
and $\ph_{3}$ needs to be reconstructed using \eqs (\ref{eq:pjsud}),
(\ref{eq:albe_sud}), and (\ref{eq:kperp}).  For the implementation of
the $+$ distribution, we also have to use
\begin{\eqn*}
N_{0}\,\equiv\,N(x\,=\,1)\,=\,\frac{(1-y')\,p_{1} \cdot
  p_{4}+y'\,p_{i} \cdot 
p_{4}-\sqrt{y'\,(1-y')\,2\,p_{1} \cdot p_{i}}\,\hat{k}_{\perp} \cdot p_{4}}{p_{4} \cdot Q}\,+\,y'.
\end{\eqn*}
If we expand in $\vareps$, we then obtain for the subtraction term
\begin{eqnarray*}
\lefteqn{\frac{\al_{s}}{2\,\pi}\,C_{F}\,\times}\\
&&\int^{1}_{0}\,dx\,\Bigg\{\delta(1-x)\,\left[\frac{1}{\vareps^{2}}\,+\,\frac{1}{\vareps}\,\lb
  \frac{3}{2}\,-\,\ln\tilde{z}_{0}\,+\,\ln\,A\,-\,\gamma
  \rb\,+\frac{1}{2}\,\text{Li}_{2}\,\lb 1-\tilde{z}_{0} \rb\right.\\
&&\left.
  \,+\,\frac{1}{2}\,\ln^{2}\tilde{z}_{0}\,+\,\frac{1}{2}\ln^{2}A\,-\,\gamma\,\ln\,A-\frac{\pi^{2}}{12}
\,+\,\frac{1}{2}\gamma^{2}\,+\,\lb\frac{3}{2}-\ln\,\tilde{z}_{0}
\rb\,(\ln\,A-\gamma) 
\,+\,2\,I_\text{fin}(\tilde{z}_{0})\right]\\&&+\frac{1}{x}\left[
2\,(1-x)\,\ln\,(1-x)
\,-\,\lb\frac{1+x^{2}}{1-x}\rb_{+}\,\ln\,x\,+\,4\,x\,\lb\frac{\ln(1-x)}{1-x} \rb_{+} \nn\right.\\
  &&\left. \hspace*{10mm}
    \,+\,\lb\frac{1+x^{2}}{1-x}\rb_{+}\,\ln\,B\right] 
\,+\,I_\text{fin}^{1}(\tilde{z},x) \Bigg\}|\M_\text{Born}(p)|^{2}
\end{eqnarray*}
with
\begin{\eqn*}
A\,=\,\frac{2\,\pi\,\mu^{2}}{p_{1} \cdot p_{i}},\;B\,=\,\frac{2 p_{1} \cdot p_{i}}{\mu_{F}^{2}},
\end{\eqn*}
and $\gamma$ being the Euler number.
\subsubsection*{Final-state subtraction}
The final-state integrated subtraction term is given by \eqs (\ref{eq:vqqg_fin}) and (\ref{eq:vif_finini}):
\begin{eqnarray*}
I^{f}(\vareps)&=&\frac{1}{\vareps^{2}}\,+\,\frac{1}{\vareps}\,\left[\frac{3}{2}\,+\,\ln(\tilde{a}_{0}+1)
  \right]
-1\,+\,\frac{\pi^{2}}{3}-2\,\ln\,(2)\ln(\tilde{a}_{0}+1)\,+\,\frac{1}{2}\,\ln^{2}(\tilde{a}_{0}+1)\\
&&
\,+\,\frac{5}{2}\text{Li}_{2}\lb
  \frac{\tilde{a}_{0}}{\tilde{a}_{0}+1}\rb\,-\,\frac{1}{2}\,\text{Li}_{2}\left[\lb\frac{\tilde{a}_{0}}{\tilde{a}_{0}+1} 
\rb^{2}\right]\,+\,I_\text{fin}^{0}(\tilde{a}_{0})\,+\,I^{1}_\text{fin}(\tilde{a})
 \end{eqnarray*}
where
\begin{eqnarray}\label{eq:iffdis}
I_\text{fin}^{0}(\tilde{a}_{0})&=&\int^{1}_{0}\,\frac{du}{u}\,\left\{2\,\ln\,2\,
+\,\frac{1}{\sqrt{1+4\,\tilde{a}_{0}(1+\tilde{a}_{0})\,u^{2}}}\,\nn\right.\\
&&\hspace*{20mm}\left.\times\ln\left[\frac{(1-u)}{\lb
      1+2\,\tilde{a}_{0}\,u\,
+\,\sqrt{1+4\,\tilde{a}_{0}\,(1+\tilde{a}_{0})\,u^{2}}
         \rb^{2}}
    \right]\right\},\nonumber\\
&&\nonumber\\
I_\text{fin}^{1}(\tilde{a})&=&2\int^{1}_{0}\,\frac{du}{u}\,\int^{1}_{0}\,\frac{dx}{x}\Bigg[\frac{x
  \lb 1-x+u\,x\,\left[(1-u\,x)\,\tilde{a}\,+\,2 \right]\rb}{k(u,x,\tilde{a})}\nn\\
&&\hspace*{35mm} -\frac{1}{\sqrt{1+4\,\tilde{a}_{0}\,u^{2}\,(1+\tilde{a}_{0})}}\Bigg]\,.
\end{eqnarray}
Here,
\begin{eqnarray*}
k^{2}(x,u,\tilde{a})&=& \left[(1+ux\,-x)(z-z')\,+\,ux\,\lb(1-ux)\,\tilde{a}+1 \rb \right]^{2}\nn\\
&&+\,4\,u\,x\,z'\,(1-z)\lb 1+u\,x-x \rb\,\lb (1-ux)\,\tilde{a}+1 \rb
\end{eqnarray*}
and
\begin{\eqn*}
z\,=\,\frac{x\,(1-u)}{1-u x},\;z'\,=\,u\,x\,\tilde{a},\;\tilde{a}\,=\,\frac{p_{1} \cdot p_{o}}{(1-y)p_{1} \cdot p_{4}\,+\,y\,p_{1} \cdot p_{i}}
\end{\eqn*}
in terms of the $m$-particle phase-space variables. The complete subtraction term is given by
\begin{\eqn*}
\frac{\al_{s}}{2\,\pi}\,C_{F}\,\frac{1}{\Gamma(1-\vareps)}\,\lb\frac{2\,\pi\,\mu^{2}}{p_{1}
 \cdot p_{i}}\rb^{\vareps}\,I^{f}(\vareps)|\M_\text{Born}(p)|^{2}\,.
\end{\eqn*}
Expanding the complete integrand in $\vareps$, we obtain
\begin{eqnarray*}
\lefteqn{\frac{\al_{s}}{2\,\pi}\,C_{F}\,\times}\\
&&\Bigg\{\frac{1}{\vareps^{2}}\,+\,\frac{1}{\vareps}\,\left[\frac{3}{2}\,+\,\ln(\tilde{a}_{0}+1)+\,\ln\,D\,-\,\gamma
  \right]
-1\,+\,\frac{\pi^{2}}{4}-2\,\ln\,(2)\ln(\tilde{a}_{0}+1)\nn\\
&&+\,\frac{1}{2}\,\ln^{2}(\tilde{a}_{0}+1) \,+\,\frac{5}{2}\text{Li}_{2}\lb
  \frac{\tilde{a}_{0}}{\tilde{a}_{0}+1}\rb
\,-\,\frac{1}{2}\,\text{Li}_{2}\,\left[\lb\frac{\tilde{a}_{0}}{\tilde{a}_{0}+1} \rb^{2} \right]
\,+\,\frac{\ln^{2}\,D+\gamma^{2}}{2}\,-\,\gamma\,\ln\,D \\
&&\,+\,\lb\frac{3}{2}+\,\ln(\tilde{a}_{0}+1)\rb\,\lb\ln\,D-\gamma \rb\,+\,I_\text{fin}^{0}(\tilde{a}_{0})\,+\,I_\text{fin}^{1}(\tilde{a})\Bigg\}|\M_\text{Born}|^{2}
\end{eqnarray*}
with
$D\,=\,{2\,\pi\,\mu^{2}}/{(p_{i} \cdot p_{1})}$.
\subsubsection*{Effective $I,\,K,\,P$ terms}
We can combine the initial- and final-state integrated subtraction
terms to an effective $I,\,K,\,P$ term which should be applied in the
spirit of an initial-state subtraction term (\ie the $I$ term is
multiplied with $\delta\,(1-x)$). We then obtain
\begin{eqnarray*}
\lefteqn{I^{\text{tot}}(\vareps)\,=\,\frac{\al_{s}}{2\,\pi}\,C_{F}\,\times\,
\Bigg\{\frac{2}{\vareps^{2}}\,+\,\frac{1}{\vareps}\left[3\,-\,2\,\ln\tilde{z}_{0}\,+\,2\,\ln\,A\,-2\,\gamma\right]}\\
&&+\,3\,\text{Li}_{2}\,(1-\tilde{z}_{0})\,+\,2\,\ln\,2\,\ln\tilde{z}_{0}\,-\,\frac{1}{2}\,\text{Li}_{2}\,\left[ (1-\tilde{z}_{0})^{2}\right]\,+\,\ln^{2}\tilde{z}_{0}\\
&&+\,\lb\ln\,A-\gamma
  \rb^{2}\,+\,\frac{\pi^{2}}{6}\,+\,\lb3\,-\,2\,\ln \tz_{0}
\rb\,\lb\ln\,A -\gamma\rb\,-\,1 +\,I^{\text{tot},0}_\text{fin}(\tz_{0})\,+\,I^{1
}_\text{fin}(\tilde{a})\Bigg\},
\end{eqnarray*}
\begin{eqnarray}\label{eq:dis_kp}
K^\text{tot}_\text{fin}(x;\tz)&=&\frac{\al_{s}}{2\,\pi}\,C_{F}  \Bigg\{\frac{1}{x}\left[
  2\,(1-x)\,\ln\,(1-x)\,-\,\lb\frac{1+x^{2}}{1-x}\rb_{+}\,\ln\,x\nn\right.\\
&&\hspace*{20mm} \left. \,+\,4\,x\,\lb\frac{\ln(1-x)}{1-x}\rb_{+}\right]\,+\,I_\text{fin}^{1}(\tilde{z},x)\Bigg\},\nonumber\\
P_\text{fin}^\text{tot}(x;\mu^{2}_{F})&=&\frac{\al_{s}}{2\,\pi}\,C_{F}\,\frac{1}{x}\lb\frac{1+x^{2}}{1-x}\rb_{+}\,\ln\,\lb\frac{2 p_{1} \cdot p_{i}}{\mu^{2}_{F}}\rb,
\end{eqnarray}
where
\begin{eqnarray*}
\lefteqn{I^\text{
  tot,0}_\text{fin}(\tz_{0})\,=\,}\\
&&2\,\int^{1}_{0}\,\frac{dy}{y}\,\Bigg\{\frac{\tilde{z}_{0}}{\sqrt{4\,y^{2}\,(1-\tilde{z}_{0})+\tilde{z}_{0}^{2}}}\nn\\
&&\hspace*{20mm}\times\,\ln\,\left[\frac{2\,z\,\sqrt{4\,y^{2}\,(1-\tilde{z}_{0})+\tilde{z}_{0}^{2}}\,(1-y)}{\lb 2\,y+\tilde{z}_{0}-2\,y\,\tilde{z}_{0}+
    \sqrt{4\,y^{2}\,(1-\tilde{z}_{0})+\tilde{z}_{0}^{2}}\rb^{2}}\right]+ \ln\,2\,\Bigg\}\,.
\end{eqnarray*}
$I^{1}_\text{fin}(\tilde{a})$ is given by \eq (\ref{eq:iffdis}) and
$I_\text{fin}^{1}(\tilde{z},x)$ by \eq(\ref{eq:ifidis}). We can
further simplify
\begin{\eqn*}
\ln\,\lb\frac{A}{\tz_{0}}\rb\,=\,\ln\,\lb\frac{2\,\pi\,\mu^{2}}{p_{i} \cdot p_{o}}\rb.
\end{\eqn*}
\subsubsection*{Combined two-particle phase-space contribution}
Adding the $I,\,K,\,P$ terms from the last section to the virtual
contribution, we have
\begin{eqnarray*}
\int^{1}_{0}\,dx\,|\M|^{2}_{2}&=&\int^{1}_{0}\,dx\,\Bigg\{\,\frac{\al_{s}}{2\,\pi}\,C_{F}\,\delta(1-x)\,\left[-9\,+\,\frac{1}{3}\pi^{2}\,-\,\frac{1}{2}\text{Li}_{2}[(1-\tz_{0})^{2}]\,\nn\right.\\
&&\left.+\,2\,\ln\,2\,\ln\tz_{0}\,+\,3\,\ln\tz_{0}\,+\,3\,\text{Li}_{2}(1-\tz_{0})\,+\,I^\text{tot,0}_\text{fin}(\tz_{0})\,+\,I^{\text 1
}_\text{fin}(\tilde{a}) \right. \bigg]\nn\\
&&\,+\,K_\text{fin}^\text{tot}(x;\tilde{z})\,+\,P^\text{tot}_\text{fin}(x;\mu^{2}_{F})  \Bigg\}|\M|^{2}_\text{Born}(x\,p_{1}),
\end{eqnarray*}
where $K_\text{fin}^\text{tot}(x;\tilde{z})$ and
$P^\text{tot}_\text{fin}(x;\mu^{2}_{F})$ are given by \eq
(\ref{eq:dis_kp}). Note that $I_\text{fin}^{1}(\tilde{z},x)$ contains
the four-vectors $\ph_{4}$ and $\ph_{3}$, which need to be
reconstructed using \eqs (\ref{eq:dis_ini_inversemap}) and
(\ref{eq:pjsud}), (\ref{eq:albe_sud}), (\ref{eq:kperp}), respectively.
\subsubsection{Results}
\begin{figure}
\centering
\includegraphics[width=3.5in]{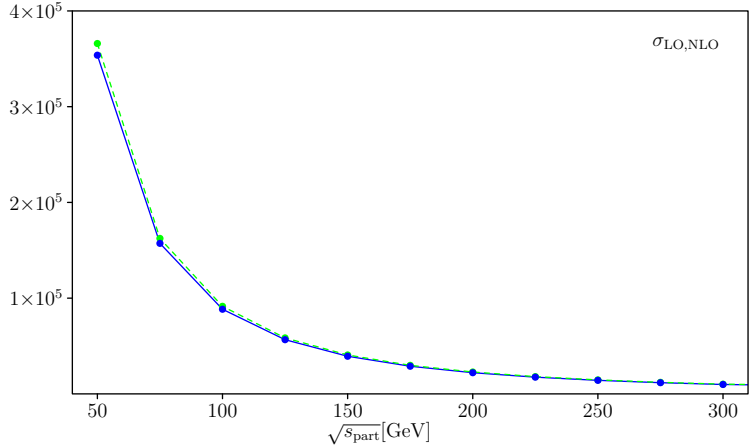}
\vspace{4mm}
\caption{\label{fig:dis_tot} NLO (green, dashed) and LO (blue, solid)
  partonic cross sections for DIS subprocess $e q\,\rightarrow\,e q
  (g)$, as a function of parton level (HERA-like) cm energies, with
  angular cuts $\cos\,\theta_{ee}<0.8$. Partonic cross section prior
  to convolution with PDFs. 
Relative NLO corrections are around $3.4\%$ }
\end{figure}

\begin{figure}
\centering
\includegraphics[width=3.5in]{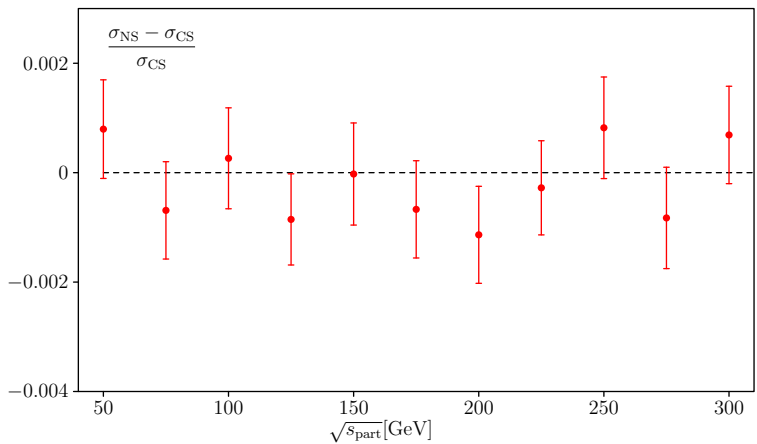}
\vspace{4mm}
\caption{\label{fig:dis_comp} As Figure \ref{fig:dis_tot}; relative difference between NLO contributions using Nagy-Soper (NS) and Catani Seymour (CS) subtraction terms. Errors are integration errors; results agree at the permil-level.}
\end{figure}

\begin{figure}
\centering
\includegraphics[width=3.5in]{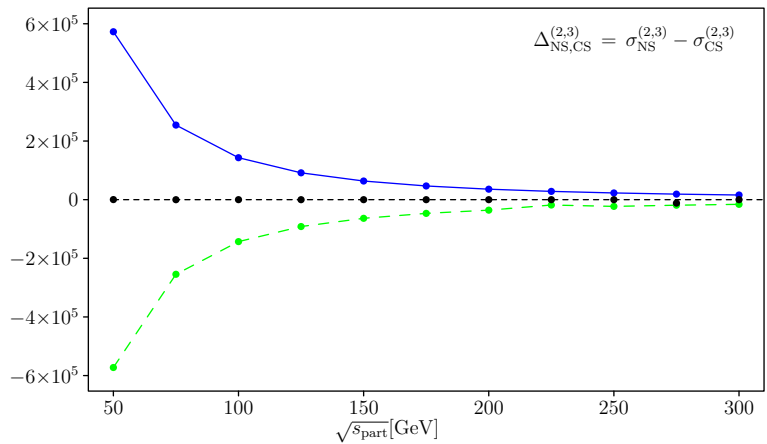}
\vspace{4mm}
\caption{\label{fig:dis_comp_diff} As Figure \ref{fig:dis_tot}. Behaviour of the {\sl difference} $\Delta^{(2,3)}_{\text{NS,CS}}\,=\,\sigma^{(2,3)}_{\text{NS}}-\sigma^{(2,3)}_{\text{CS}}$ between the two schemes for the two particle final state phase space (green, dashed) and three particle final state phase space (blue, solid), respectively. In the sum $\Delta^{(2)}_{\rm NS,CS}+\Delta^{(3)}_{\rm NS,CS}$ (black dots), the large differences cancel. In the new scheme, subtractions have been restricted to singular regions.
}
\end{figure}

We have compared the above results numerically with an implementation
of the Catani-Seymour dipole subtraction; the corresponding terms can
easily be obtained from \cite{Catani:1996vz} and are therefore omitted
here.  \Figs \ref{fig:dis_tot} and \ref{fig:dis_comp} show the
behaviour of the total partonic cross sections and the differences
between the application of the two schemes at parton level for varying
(HERA-like) center-of-mass energies, where we applied a cut of
$\cos\,\theta_{ee}<0.8$ to cut out terms for which
$t\,=\,(p_{i}-p_{o})^{2}\,=\,0$. We see that the results agree on the
per-mil level\footnote{Note that this result has been obtained with a
  relatively mild angular cut.}, therefore verifying the nontrivial
check of our subtraction prescription with a modified final-state
mapping. In \Fig \ref{fig:dis_comp_diff}, we show the behaviour of the
differences between the two- and three-particle phase-space
contributions for varying center-of-mass energies, again verifying
that their cancellation is non-trivial as the contributions from
different phase space integrations vary widely in magnitude for the
two schemes. The results for our new scheme have been obtained using
subtractions in the vicinities of the singular regions in phase space
only\footnote{This idea has been documented in \cite{Frixione:1995ms}
  in the context of FKS subtraction and in \cite{Nagy:2003tz} for
  Catani Seymour dipoles. Explicit expressions for our scheme will be
  presented elsewhere.}. A more detailed investigation for this
process, including the implementation in a parton level Monte Carlo
generator, is in the line of future work.  

\clearpage

\section{Conclusion and Outlook}
In this paper, we propose a new NLO subtraction scheme which is based
on the momentum mapping and on the splitting functions derived in the
context of an improved parton shower formulation \cite{Nagy:2007ty}.
One essential feature of our scheme is that we use a global momentum
mapping in which all of the partons participate.  As a result, the
number of momentum mappings needed to evaluate the subtraction terms
and the corresponding leading-order matrix elements is significantly
smaller than for standard subtraction schemes. A further important
feature of our scheme is that we derive the subtraction terms from the
splitting functions that describe an improved parton shower with
quantum interference.  The use of the shower splitting functions as
subtraction terms greatly simplifies the matching of a NLO calculation
with the corresponding parton shower.

To establish our scheme we have focused on the simple case of
collider processes with up to two massless particles in the final
state. We have presented formulae for all subtraction terms and
integrated counterparts needed to address such processes at NLO, and
have applied the results to a variety of basic lepton- and
hadron-collider processes.  In all cases, we have reproduced the
results from the literature and have shown that our implementation
agrees with results obtained using the Catani-Seymour dipole
subtraction. In the scheme proposed here, the mapping of parton
momenta is given by a generic description for initial- or final-state
emitters. The finite contributions in the integrated soft interference
terms between initial- and final-state partons thus depend on the
kinematics of the process and need to be integrated numerically.  The
implementation of initial-final and final-initial interference terms
therefore accounts for a non-trivial check of our subtraction
prescription.

The advantages of two main features of our scheme, \ie the global
mapping and the use of subtraction terms derived from the splitting
functions of an improved parton shower, become most apparent when
applying the scheme to NLO multi-parton processes and/or matching the
NLO calculation with the corresponding parton shower.  Both
applications are in the line of future work.\footnote{We note that
  work is underway ~\cite{NS-helac} to implement the scheme presented
  here into the Helac Event Generator framework \cite{Kanaki:2000ey}.}
In this paper, we have demonstrated in a first step that the splitting
functions of the improved shower, in combination with the
corresponding global mapping, can be used as local counterterms in a
subtraction scheme for processes with relatively simple final-state
kinematics. While the subtraction terms for the real-emission
contributions can readily be applied to generic multi-parton final
states, the mapping prescription for final-state emitters leads to
more involved finite parts of the integrated subtraction terms when
considering processes with three or more final-state particles.
Although the integrated subtraction terms for the general case have
been privately available for some time, and an example for the
final-state splitting function $g\,\rightarrow\,q\,\bar{q}$ has been
presented in \cite{Robens:2010zr}, a generic application to a more
challenging physical process is still work in progress. However, we
are confident that results for multi-parton final states, which will
allow for a more powerful test of the features of the new scheme, will
become available in the near future.

\section*{Acknowledgements}
This research was partially supported by the DFG SFB/TR9
``Computational Particle Physics'', the DFG Graduiertenkolleg
``Elementary Particle Physics at the TeV Scale'', the Helmholtz
Alliance ``Physics at the Terascale'', the BMBF, the STFC and the EU
Network MRTN-CT-2006-035505 ``Tools and Precision Calculations for
Physics Discoveries at Colliders''. We would like to thank Zolt\'{a}n
Nagy, Dave Soper and Zolt\'{a}n Tr\'{o}cs\'{a}nyi for many valuable
discussions, as well as Tobias Huber for help with integrations
including HypExp. In addition, TR thanks Adrian Signer for a
clarifying discussion about the FKS subtraction scheme, Rikkert
Frederix for discussions concerning the implementation of the FKS
scheme within the Madgraph framework, and David Miller and Chris White
for helpful comments regarding the manuscript. Finally, we want to
thank the hospitality of the Aspen Center of Physics, where parts of
this work were completed. Some of the plots were generated using the
gamelan graphics package \cite{Kilian:2007gr}.

\newpage

\begin{appendix}
\section{Splitting amplitudes}\label{app:splittings}
In Table \ref{tab:splitting}, we list the splitting amplitudes for
$q\bar{q}g$ splittings as given in \cite{Nagy:2007ty}.
\begin{table}
\begin{tabular}{cccccc}
\hline
\\
$\ell$&$f_\ell$&$\hat f_\ell$&$\hat f_{j}$\!\!\!\!&
$\displaystyle{v_\ell \times \frac{1}{\sqrt{4\pi\as}}}$
&colour \\
\\
$F$&
$q$&
$q$&
g&
$\displaystyle{
\varepsilon_\mu(\hat p_{j},\hat s_{j};\hat Q)^*\,
\frac{
\overline U({\hat p_\ell,\hat s_\ell})\gamma^\mu 
  [\s{\hat p}_\ell + \s{\hat p}_{j}] \s{n_\ell} U({p_\ell,s_\ell})}
{2p_\ell\!\cdot\! n_\ell\ [(\hat p_\ell + \hat p_{j})^2]}
}
$&
$t^a$\\
&&&&&\\
$I$&
$q$&
$q$&
g&
$\displaystyle{
-
\varepsilon_\mu(\hat p_{j},\hat s_{j};\hat Q)^*\,
\frac{\overline V({\hat p_\ell,\hat s_\ell})\gamma^\mu
(\s{\hat p}_\ell - \s{\hat p}_{j})
   \s{n}_\ell V({p_\ell,s_\ell})}
{2p_\ell\!\cdot\! n_\ell\ [(\hat p_\ell - \hat p_{j})^2]}
}
$&
$t^a$\\
&&&&&\\
$F$&
g&
$q$&
$\bar q$&
$\displaystyle{
-
\varepsilon^\mu(p_{\ell},s_{\ell};\hat Q)
D_{\mu\nu}(\hat p_\ell + \hat p_{j},n_\ell)
\frac{
\overline U({\hat p_\ell,\hat s_\ell})\gamma^\nu V({\hat p_{j},\hat s_{j})}
}{(\hat p_\ell + \hat p_{j})^2}
}$&
$t^a$\\
&&&&&\\
$I$&
g&
$\bar q$&
$q$&
$\displaystyle{
-
\varepsilon^\mu(p_\ell,s_\ell;\hat Q)^*
D_{\mu\nu}(\hat p_\ell - \hat p_{j};n_\ell)
\frac{
\overline U({\hat p_{j},\hat s_{j}})\gamma^\nu 
U({\hat p_\ell,\hat s_\ell})
}{(\hat p_\ell - \hat p_{j})^2}
}
$&
$t^a$\\
&&&&&\\
$I$&
$q$&
g&
$q$&
$\displaystyle{
-
\varepsilon_\mu(\hat p_\ell,\hat s_\ell;\hat Q)\,
\frac{
\overline U({\hat p_{j},\hat s_{j}})\gamma^\mu 
[\s{\hat p}_\ell - \s{\hat p}_{j}]\s{n_\ell}V({p_\ell,s_\ell})
}
{2p_\ell\!\cdot\! n_\ell\ [(\hat p_\ell - \hat p_{j})^2]}
}$&
$t^a$\\
&&&&&\\
\\  [10 pt]
\hline
\end{tabular}
\caption{Splitting amplitudes $v_\ell(\{\hat p, \hat f\}_{j},\hat
  s_{j},\hat s_{\ell},s_\ell)$ 
  involving a $q\bar{q}g$ splitting. We have removed a common factor 
  $\sqrt{4\pi\as}$ removed. The label $\ell$ denotes either
  initial-state indices  
  $I = \{\La,\Lb\}$ or final-state indices $F = \{1,\dots,m\}$. 
  The light-like vector $n_\ell$ is defined in Eq.~(\ref{eq:nldef}). Taken from \cite{Nagy:2007ty}.}
\label{tab:splitting}
\end{table}
For triple gluon splittings, we have for the final state
\begin{equation}
\begin{split}
\label{eq:VggF}
v_\ell(\{\hat p, \hat f\}_{m+1},&\hat s_{j},\hat s_{\ell},s_\ell)
\\ & =
\frac{\sqrt{4\pi\as}}{2 \hat p_{j}\!\cdot\! \hat p_\ell}\, 
\varepsilon_{\alpha}(\hat p_{j}, \hat s_{j};\hat Q)^*
\varepsilon_{\beta}(\hat p_{\ell}, \hat s_l;\hat Q)^*
\varepsilon^{\nu}(p_{\ell}, s_\ell;\hat Q)
\\&\quad\times
v^{\alpha \beta \gamma}(\hat p_{j},\hat p_\ell,-\hat p_{j}-\hat p_\ell)\,
D_{\gamma\nu}(\hat p_\ell + \hat p_{j};n_\ell)\,.
\end{split}
\end{equation}
For an initial state splitting, we have
\begin{equation}
\begin{split}
\label{eq:VggI}
v_\ell(\{\hat p, \hat f\}_{m+1},&\hat s_{j},\hat s_{\ell},s_\ell)
\\&=
-
\frac{\sqrt{4\pi\as}}{2 \hat p_{j}\!\cdot\! \hat p_\ell}\, 
\varepsilon_{\alpha}(\hat p_{j}, \hat s_{j};\hat  Q)^*
\varepsilon_{\beta}(\hat p_{\ell},\hat s_\ell; \hat Q)
\varepsilon^{\nu}(p_{\ell}, s_\ell; \hat Q)^*
\\&\quad\times
v^{\alpha \beta \gamma}(\hat p_{j}, -\hat p_\ell, \hat p_\ell - \hat p_{j})\,
D_{\gamma\nu}(\hat p_\ell - \hat p_{j};n_\ell)
\;\;.
\end{split}
\end{equation}
We use standard notation where
$U(p,s),\overline{U}(p,s),V(p,s),\overline{V}(p,s)$ denote spinors of
the fermions with a four-momentum $p$ and spin $s$, and
$\varepsilon_{\alpha}(p, s;Q)$ are the gluon polarisation vectors. The
$ggg$ vertex has the form \beq
\label{eq:vgg}
v^{\alpha \beta \gamma}(p_a, p_b, p_c)
= g^{\alpha\beta} (p_a - p_b)^\gamma
+ g^{\beta\gamma} (p_b - p_c)^\alpha
+ g^{\gamma\alpha} (p_c - p_a)^\beta\,.
\eeq
The transverse projection tensor $D_{\gamma\nu}(\hat p_\ell - \hat p_{j};n_\ell)$ is defined according to \eq (\ref{transverseprojectiontensor}).
The light-like vector $n_\ell$ is given by
\begin{equation}
\label{eq:nldef}
n_\ell = 
\begin{cases}
p_\LB \;\;,&  \ell = \La\;\;, \\
p_\LA \;\;,&  \ell = \Lb\;\;, \\
\displaystyle{
Q
-\frac{Q^2}
{Q\!\cdot\! p_\ell
+ \sqrt{(Q\!\cdot\! p_\ell)^2 }}\
p_\ell
}
\;\;,&  \ell \in \{1,\dots,m\}\;\;.
\end{cases}
\end{equation}

\section{Integration measures in terms of singular variables}\label{app:ints}
In this section, we give the integration measures for the initial- and
final-state splittings in terms of the singular variables, and relate
these to the four-momenta in the real-emission phase spaces.
\subsection{Initial-state integration measure}
The initial-state integration measure in $d\,=\,4-2\,\vareps$
dimensions is given by \eq (\ref{eq:ini_meas}) and related to the
variables in Section \ref{sec:inisq} via
\begin{eqnarray}
d\xi_{p}&=&\frac{d^{d}\ph_{j}}{(2\,\pi)^{d-1}}\,\delta_{+}\lb\ph_{j}^{2}\rb\nn\\
&=&dx\,dy'\,\frac{(2\,p_{a}\cdot p_{b})^{1-\vareps}\,x^{\vareps-1}}{\Gamma(1-\vareps)\,(4\,\pi)^{2-\vareps}}\,(1-x)^{1-2\,\vareps}\,\left[ y'\,(1-y') \right]^{-\vareps}\Theta\left[ (1-x)\,x\right]\,\Theta\left[(1-y')\,y'\right]\,,\nn\\&&
\end{eqnarray}
where the center of mass energy in the $m+1$ phase space is given by
$\hat{s}\,=\,\hat{\eta}_{a}\,\eta_{b}\,s\,=\,(2 p_{a} \cdot p_{b})/x$.
Here, $x\rightarrow\,1$ corresponds to the soft and
$y'\,\rightarrow\,0$ to the collinear singular limit. For the
integration of the interference terms, we additionally have to
parametrise the azimuthal angle of $\ph_{j}$ in the integration
measure; this parametrisation is frame dependent. We obtain
\begin{eqnarray}
d\xi_{p}&=&dx\,dy'\,\,dv\frac{(2\,p_{a}\cdot p_{b})^{1-\vareps}\,x^{\vareps-1}}{(4\,\pi)^{2}}\,\frac{\pi^{\vareps-\frac{1}{2}}}{\Gamma\lb\frac{1-2\vareps}{2}\rb}\,(1-x)^{1-2\,\vareps}\,\left[ y'\,(1-y')\right]^{-\vareps}\,\left[ v\,(1-v)\right]^{-\frac{1+2\,\vareps}{2}}\nn\\
&&\times\,\Theta\left[ (1-x)\,x\right]\,\Theta\left[(1-y')\,y'\right] \,\Theta\left[ v\,(1-v)\right]
\end{eqnarray}
where $v\,=\,\frac{1}{2}\lb 1-\cos\varphi \rb$ is related to the azimuthal angle of $\ph_{j}$ in the center for mass frame of $\ph_{a},\ph_{b}$ and $\ph_{k}$ defines the $x,z$ plane.
\subsection{Final-state integration measure}
The final-state integration measure in $d\,=\,4-2\,\vareps$ dimensions
is given by \eq (\ref{eq:meas_fin}). In terms of integration variables
which parametrise the singularities of the integrands, we obtain
\begin{eqnarray}
d\xi_p&=&
dy\,\theta\lb y(1-y)\rb\,(1-y)^{d-3}\,\frac{p_{\ell}\cdot Q}{\pi}\,\frac{d^d\hat p_{\ell}}{(2\,\pi)^{d}}\,
2\,\pi\,\delta^{+}(\hat p_{\ell}^{2} )\,
\frac{d^{d}\hat p_{j}}{(2\,\pi)^{d}}\,2\,\pi\,\delta^{+}(\hat p_{j}^{2} )\, \nn
\\
&&\times
(2\,\pi)^{d}\,\delta^{(d)}\left(\hat p_\ell + \hat p_{j}- (1-y) p_\ell - y Q
\right)\nn\\
&=&\,\frac{(2\,p_{\ell} \cdot Q)^{1-\vareps}}{16\,\pi^{2}}\,\frac{(4\,\pi)^{\vareps}}{\Gamma(1-\vareps)}\,\int^{1}_{0}\,du\,u^{-\vareps}\,(1-u)^{-\vareps}\,\int^{1}_{0}\,dx\,x^{1-2\,\vareps}\,(1-x)^{-\vareps},
\end{eqnarray}
where $u$ and $x$ are related to the $m+1$ particle kinematics via
\begin{\eqn*}
x\,=\,\frac{\ph_{j}\cdot Q}{p_{\ell} \cdot Q},\;u\,=\,\frac{\ph_{\ell}\cdot\ph_{j}}{\ph_{j}\cdot Q}\,.
\end{\eqn*}
Here, $x\,\rightarrow\,0$ corresponds to the soft and
$u\,\rightarrow\,0$ to the collinear limit of the integration.  While
the above parametrisation suffices in the integration of the collinear
subtraction terms, we need to introduce an additional angle between
emitted parton and spectator in the interference terms. In contrast to
the above parametrisation, the additional integration variable is
frame dependent; in the center of mass system of $\ph_{\ell},\ph_{j}$,
where $\ph_{k}$ defines the $x,z$ plane, we can write
 \begin{eqnarray}
d\xi_{p}&=&\frac{(2\,p_{\ell} \cdot Q)^{1-\vareps}}{16}\,\frac{\pi^{-\frac{5}{2}+\vareps}}{\Gamma\lb \frac{1}{2}-\vareps\rb}\,\nn\\
&\times&\!\!\!
\int^{1}_{0}\,du\,u^{-\vareps}\,(1-u)^{-\vareps}\,\int^{1}_{0}\,dx\,x^{1-2\,\vareps}\,(1-x)^{-\vareps}\, \int^{1}_{0}\,dv\,\left[v\,(1-v)\right]^{-\frac{1+2\,\vareps}{2}},
\end{eqnarray}
with $v\,=\,\frac{1}{2}\lb 1-\cos\varphi\rb$ is related to the
azimuthal angle of $\ph_{j}$.  The above parametrisations were used in
all integrations presented in this paper.
\section{Altarelli-Parisi splitting functions}\label{sec:app_splittings}
This section contains a list of the well known Altarelli-Parisi
splitting functions \cite{Altarelli:1977zs}, which are evolution
kernels of the DGLAP equation \cite{Gribov:1972ri,
  Lipatov:1974qm,Altarelli:1977zs,Dokshitzer:1977sg}. They also
describe the behaviour of parton splittings by giving the probability
of finding a parton of type $b$ with momentum fraction $x$ in a parton
of type $a$ in the collinear limit:
\begin{\eqn}
a(p)\,\longrightarrow\,b\lb x\,p+k_{\perp}+\,\mO(k_{\perp}^{2})\rb\,+\,c\,\lb (1-z)\,p-k_{\perp}+\,\mO(k_{\perp}^{2}) \rb.
\end{\eqn}
At leading order, the splitting functions are given by
\begin{eqnarray}
P^{qq}(x)&=&C_{F}\,\left[ \frac{1+x^{2}}{(1-x)_{+}}+\frac{3}{2}\,\delta(1-x)\right],\nn\\
P^{gq}(x)&=&T_{R}\,\left[x^{2}+(1-x)^{2} \right],\;T_{R}\,=\,\frac{1}{2},\nn\\
P^{qg}(x)&=&C_{F}\,\left[\frac{1+(1-x)^{2}}{x}\right],\nn\\
P^{gg}(x)&=&2\,C_{A}\,\left[ \frac{x}{(1-x)_{+}}\,+\,\frac{1-x}{x}+x\,(1-x) \right]+\delta(1-x)\,\frac{11\,C_{A}-4\,n_{f}\,T_{R}}{6}\,,\nn\\&&
\end{eqnarray}
where $n_{f}$ is the number of quark flavours in the theory. The $+$
distribution is defined in the standard way
\begin{\eqn}\label{eq:plusdef}
\int^{1}_{0}\,f(x)\,g_{+}(x)\,dx\,=\,\int^{1}_{0}\,g(x)\,(f(x)-f(1))\,dx\,=\,\int^{1}_{0}\,g(x)\,f(x)\,dx\,-\,f(1)\,\int^{1}_{0}g(x)\,dx
\end{\eqn}
for the convolution with a test function $f(x)$.
\end{appendix}
\bibliographystyle{utcaps}
\bibliography{NS_subtraction}
\end{document}